\documentclass{cernyrep}
\def\const{\mbox{const}}
\def\e{{\rm e}}

\def\d{\partial}
\def\l{\left(}
\def\r{\right)}
\def\la{\langle }
\def\ra{\rangle }
\newcommand{\be}{\begin{equation}}
\newcommand{\ee}{\end{equation}}
\newcommand{\bea}{\begin{eqnarray}}
\newcommand{\eea}{\end{eqnarray}}

\renewcommand{\ln}{\mathop{\rm ln}\nolimits}
\newcommand{\sm}[1]{{\scriptscriptstyle \rm #1}}

\usepackage{varwidth}
\usepackage{xcolor}

\begin{document}

\title{Cosmology}


\author{V.A. Rubakov}

\institute{Institute for Nuclear Research
of the Russian Academy of Sciences, \\
Moscow, Russia\\
and\\
Department of Particle Physics and Cosmology,
Physics Faculty, Moscow State University, \\ Moscow, Russia}

\maketitle

\begin{abstract}Cosmology and particle physics are
deeply interrelated. Among the common problems are
dark energy, dark matter and baryon asymmetry of the Universe.
We discuss these problems in general terms, and concentrate
on several particular hypotheses. On the dark matter side, we consider
weakly interacting massive particles and axions/axion-like particles
as cold dark matter, sterile neutrinos and
gravitinos as warm dark matter. On the baryon asymmetry side,
we discuss electroweak baryogenesis as a still-viable
mechanism. We briefly describe diverse experimental and
observational approaches towards checking these hypotheses.
We then turn to the earliest cosmology.
We give arguments showing that the hot stage was preceded
by another epoch at which density perturbations and possibly
primordial gravity waves were generated.
The best guess here is inflation, which is consistent with
everything we know of density perturbations,
but there are alternative scenarios.
Future measurements of the properties of density perturbations
and possible discovery of primordial gravity waves
have strong potential in this regard.
\end{abstract}


\section{Introduction}
\label{Intro}
Cosmology is one of the major sources of inspiration---and confusion---for particle physicists. It gives direct evidence
for the necessity to extend the Standard Model of particle physics,
possibly at an energy scale that can be probed by collider experiments.
Indeed, there is no doubt that most part of the mass in the present
Universe is in the form of mysterious dark matter particles
which are not present  in the Standard Model.
Also, the very existence of conventional matter
in our
Universe (i.e., matter--antimatter asymmetry) calls
for processes with baryon number violation and substantial
charge parity (CP)-violation, 
which have not been observed
in experiments.
These processes
had to be rapid in the early Universe   and, furthermore,
the asymmetry between matter and antimatter had to be generated
in a fairly turbulent cosmological epoch. Again,
the conditions necessary for
the generation of this asymmetry
are not present in the Standard Model.
Solving the problems of dark matter and matter--antimatter
asymmetry are the two immediate challenges for particle physics.

Going very much back into the cosmological history, we encounter
another challenging issue. It is very well known that
matter in the Universe was very hot and dense early on.
It is less known that the properties of the matter distribution in
the past and present Universe, reflected in the properties of the
cosmic microwave background (CMB), galaxy distribution etc,
unambiguously tell us that the hot epoch was not the earliest.
It was preceded by another, completely different epoch
responsible for the generation of inhomogeneities which in the end
have become galaxies and their clusters, stars and ourselves.
Obviously, the very fact that we are
confident
about the existence of such an
epoch
is a fundamental result of theoretical and observational
cosmology.
The most plausible
hypothesis on that epoch is cosmological inflation,
though the observational support of this scenario
is presently not overwhelming, and alternative possibilities
have not been ruled out. For the time being it appears
unlikely that we will be able to probe the physics behind that
epoch in terrestrial experiments, but there is no doubt that
this physics belongs to the broad domain of `particles and fields'.

After this brief introduction, the scope of these lectures
must be clear. To set the stage, we briefly consider the basic notions of
cosmology.
We then discuss several  dark matter particle
candidates and mechanisms
for dark matter generation. Needless to say, these candidates do not
exhaust the long list of the candidates proposed; our choice
is based on a personal view of what candidates are more plausible.
Our next topic is the matter--antimatter
asymmetry of the Universe, and we present electroweak baryogenesis
as a mechanism particularly
interesting from the viewpoint of the LHC experiments.
The last part of these lectures deals with cosmological perturbations,
inflation (and its alternatives) and
the potential of
future observational data.

These lectures are meant to be self-contained, but
we necessarily omit numerous details,
while trying to make clear the basic ideas and results.
More complete accounts of cosmology and its
particle-physics aspects
may be found in various books~\cite{books1,books2,books3,books4,books5,books6}. Dark matter
candidates we consider in these lectures are
reviewed in Refs.~\cite{dark-rev1,dark-rev2,dark-rev3,dark-rev4}. Electroweak baryogenesis
is presented in detail in reviews~\cite{ew-rev,ew-rev-2-1,ew-rev-2-2}; for reference,
a plausible alternative scenario, leptogenesis, is
discussed in reviews~\cite{leptogen-1,leptogen-2}.
Aspects of inflation and its alternatives are reviewed
in Refs.~\cite{pert-rev-1,pert-rev-2,pert-rev-3,pert-rev-4,pert-rev-5}.

\section{Expanding universe}

\subsection{Friedmann--Lema\^itre--Robertson--Walker metric}
\label{sub:FLRW}

Our Universe (more precisely, its visible part) is
{\it homogeneous and isotropic}.
Clearly, this does not apply to relatively small spatial scales:
there are galaxies, clusters of galaxies and giant voids.
But boxes of sizes exceeding about 200~Mpc all look the same.
Here the Mpc is the distance unit conventionally used in cosmology,
\[
1~\mbox{Mpc} \approx 3\times 10^6~\mbox{light~years} \approx
3\times 10^{24}~\mbox{cm} \; .
\]
There are three types of homogeneous and isotropic three-dimensional
spaces, labelled by an integer parameter $\varkappa$.
These are
three-sphere (closed model, $\varkappa=+1$), flat
(Euclidean) space (flat model, $\varkappa = 0$)
and three-hyperboloid (open model, $\varkappa = -1$).
We will see that the parameter $\varkappa$ enters
the dynamical equations governing the space--time fabric
of the Universe.

Another basic property of our Universe is that
it {\it expands}.
This is encoded in the space--time metric
\begin{equation}
 {\rm d}s^2 = {\rm d}t^2 - a^2(t) {\rm d}{\bf x}^2 \; ,
\label{FRW}
\end{equation}
where ${\rm d}{\bf x}^2$ is the distance on a unit
three-sphere, Euclidean space or hyperboloid. The metric (\ref{FRW})
is called the Friedmann--Lema\^itre--Robertson--Walker (FLRW) metric,
and $a(t)$ is the scale factor. In these lectures we use natural units,
setting the speed of light and Planck and Boltzmann constants equal to 1,
\[
c=\hbar = k_{\rm B} =1 \; .
\]
In these units, Newton's gravity constant is
$G=M_{\rm Pl}^{-2}$, where $M_{\rm Pl} = 1.2\times 10^{19}$~GeV is the Planck mass.

The meaning of Eq.~\eqref{FRW} is as follows. One can check that
a free mass put at a certain ${\bf x}$ at zero velocity will stay at the same
${\bf x}$ forever. In other words, the coordinates ${\bf x}$ are comoving.
The scale factor $a(t)$ increases in time, so
the distance between free masses of fixed spatial
coordinates
${\bf x}$ grows,
$  {\rm d}l^2 = a^{2}(t) {\rm d}{\bf x}^2$. The space stretches out;
the galaxies run away from each other.

This expansion manifests itself as a red shift. Red shift is often
interpreted
as the Doppler effect for a source running away from us with velocity
$v$:
if the wavelength at emission is   $\lambda_{\rm e}$, then the wavelength
we measure is $\lambda_0 = (1 + z) \lambda_{\rm e}$, where $z = v/c$
(here we temporarily restore the speed of light). This interpretation is useless and
rather misleading in cosmology
(with respect to which reference frame does the source move?).
The correct interpretation is that as the Universe expands, space stretches
out and the photon wavelength increases proportionally to the
scale factor $a$. So, the relation between
the wavelengths is
\[
\lambda_0 = (1 + z) \lambda_{\rm e} \; , \;\;\;\;
\mbox{where} \; z = \frac{a(t_0)}{a(t_{\rm e})} - 1 \; ,
\]
where $t_{\rm e}$ is the emission time.
For $z \ll 1$, this relation reduces to  the Hubble law,
\be
z = H_0 r \; ,
\label{mar22-15-1}
\ee
where $r$ is the physical distance to the source and $H_0 \equiv
H(t_0)$ is the present value of the Hubble parameter
\[
H (t) = \frac{\dot{a}(t)}{a(t)} \; .
\]
In the formulas above, we
label the present values of time-dependent quantities
by subscript $0$; we will always do so in these lectures.


\vspace{0.3cm}
\noindent
{\it Question.} Derive the Hubble law \eqref{mar22-15-1} for $z \ll 1$.
\vspace{0.3cm}

\begin{figure}[htb!]
\begin{center}
\includegraphics[width=0.5\textwidth,angle=0]{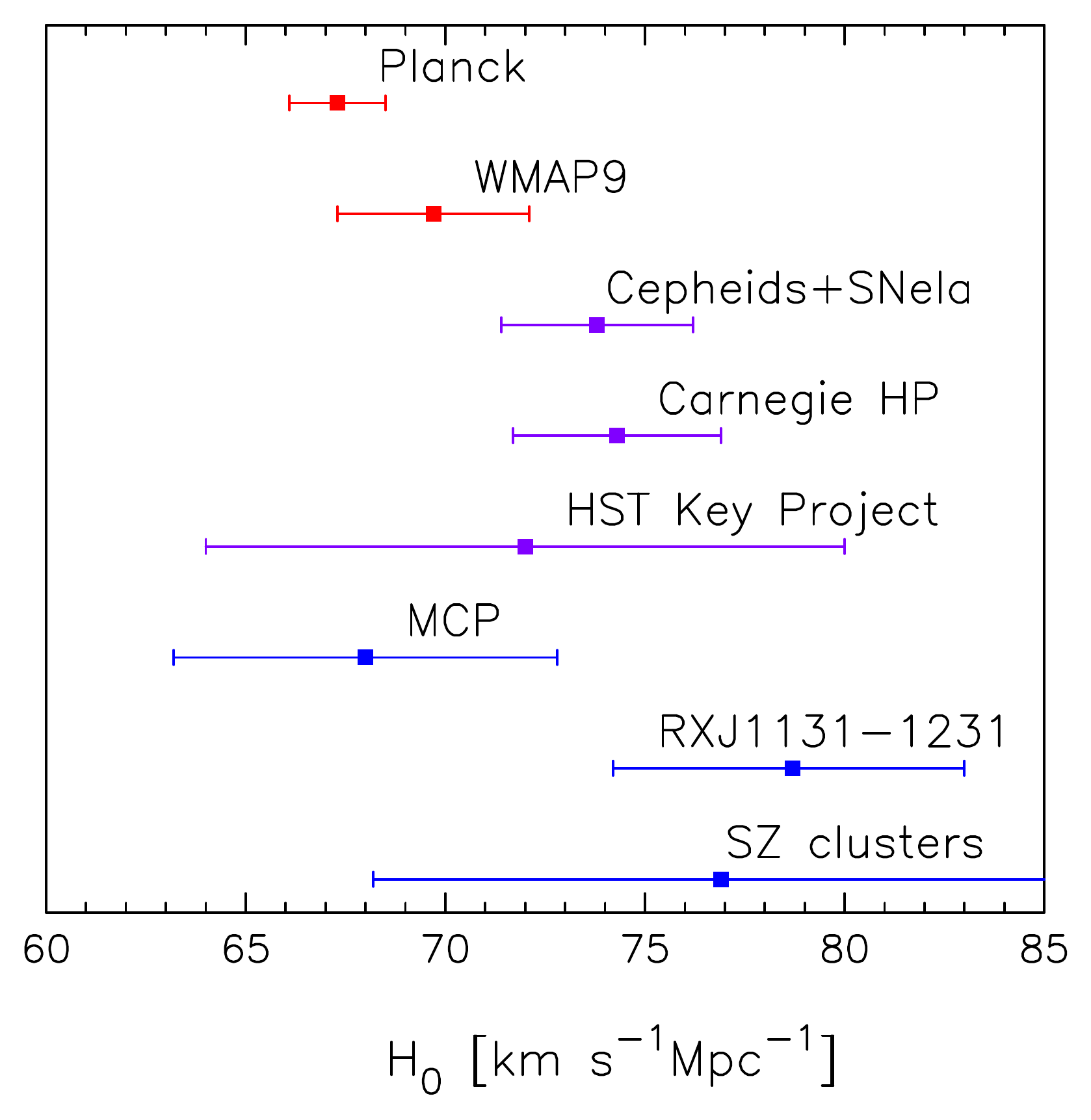}
\end{center}
\caption{Recent determinations of the Hubble parameter
$H_0$~\cite{Ade:2013zuv}
 \label{H0-syst}
 }
\end{figure}

The red shift of an object
is directly measurable.  The wavelength $\lambda_{\rm e}$ is fixed by
physics of the source, say, it is the wavelength of a photon
emitted by an excited hydrogen atom. So,
one identifies a series of emission
or absorption lines, thus determining $\lambda_{\rm e}$,
and measures their actual wavelengths $\lambda_0$.
These spectroscopic measurements give accurate values of $z$
even for distant sources. On the other hand, the red shift
is related to
the time of emission and hence to the distance to the source.
Absolute distances to astrophysical sources have a lot more systematic
uncertainty, and so do the direct measurements of the Hubble
parameter $H_0$.
According to the Planck Collaboration~\cite{Planck:2015xua},
the combination of observational data
gives
\be
     H_0 = (67.8 \pm 0.9)~\frac{\mbox{km}}{\mbox{s Mpc}} \approx
(14.4 \times 10^9~\mbox{yr})^{-1} \; ,
\label{H00}
\ee
where the unit used in the first expression reflects the interpretation
of red shift in terms of the Doppler shift. The fact that the
systematic uncertainties in the determination of $H_0$ are pretty large
is illustrated in Fig.~\ref{H0-syst}.

Traditionally, the present value of the Hubble parameter is written as
\begin{equation}
   H_0 = h \times 100 ~\frac{\mbox{km}}{\mbox{s Mpc}} \; .
\label{H0}
\end{equation}
Thus, $   h \approx 0.7$.
We will use this value in further estimates.

\subsection{Hot Universe: recombination, Big Bang nucleosynthesis and neutrinos}

\begin{figure}[htb!]
\begin{center}
\includegraphics[width=0.5\textwidth,angle=0]{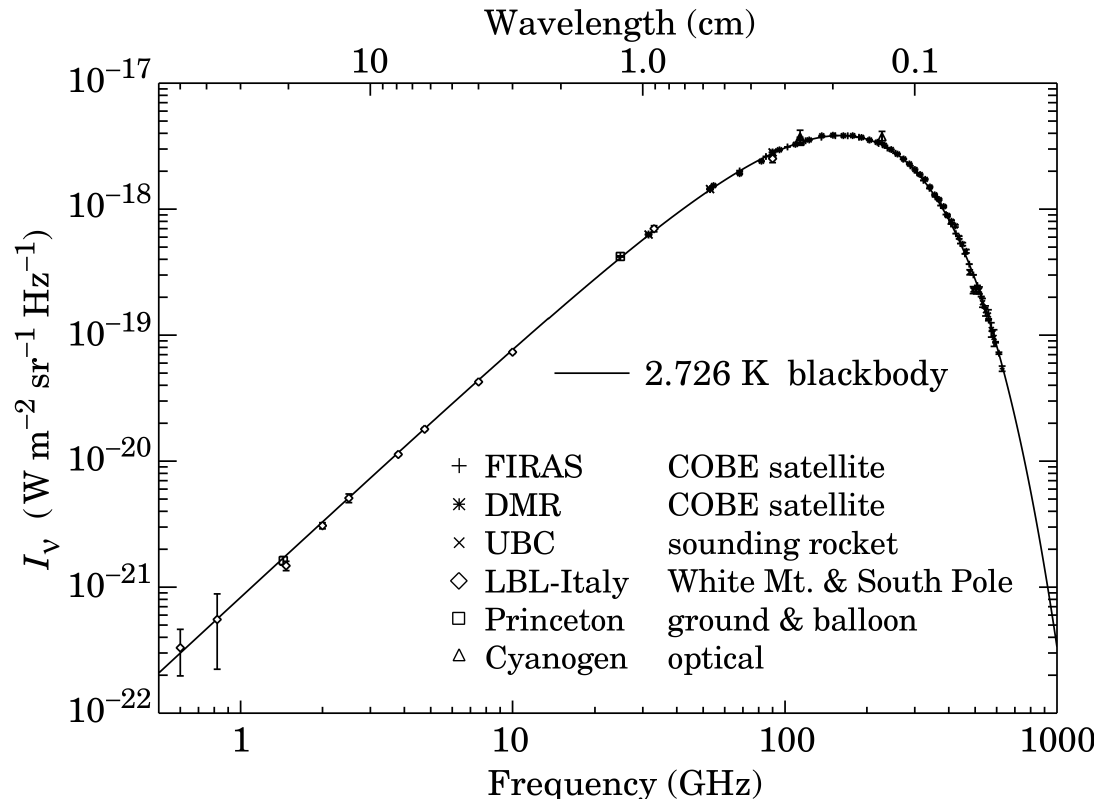}
\end{center}
\caption{Measured CMB energy spectrum as compiled in
Ref.~\cite{Hagiwara:2002fs}
 \label{bbody}
 }
\end{figure}

Our Universe is filled with CMB.
The CMB as observed today
consists of photons with an excellent black-body spectrum of temperature
\be
   T_0 = 2.7255 \pm 0.0006~\mbox{K} \; .
\label{temperature}
\ee
The spectrum has been precisely measured by various instruments,
see Fig.~\ref{bbody},
and does not show any deviation from the Planck
spectrum (see Ref.~\cite{black-body} for a detailed review).


Once the present photon temperature is known, the number density
and energy density of CMB photons are known from the Planck distribution
formulas,
\be
n_{\gamma, 0} = 410~~\mbox{cm}^{-3} \; , \;\;\;\;\;\;
  \rho_{\gamma, 0} = \frac{\pi^2}{15} T_0^4 =
2.7 \times 10^{-10} ~\frac{\mbox{GeV}}{\mbox{cm}^3} \;
\label{rhogamma} \;
\ee
(the second expression is the Stefan--Boltzmann formula).

The CMB is a remnant of an earlier cosmological epoch.
The Universe was hot at early times and, as it expands,
the matter in it cools down.
Since the wavelength of a photon evolves in time as $a(t)$,
its energy and hence temperature scale as
\[
\omega(t) \propto a^{-1}(t) \; , \;\;\;\;\;\;\;
T(t) = \frac{a_0}{a(t)} T_0 = (1+z) T_0 \; .
\]
When the Universe was hot, the usual matter (electrons and protons
with a rather small admixture of light nuclei, mainly $^4$He)
was in the plasma phase.
At that time photons strongly interacted with electrons
due to the Thomson scattering
and protons interacted with electrons via the Coulomb force,
so all these particles were in thermal equilibrium.
As the Universe cooled down, electrons `recombined' with protons into
neutral hydrogen atoms (helium recombined earlier),
and the Universe became transparent to photons: at that time,
the density of hydrogen atoms was quite small,
$250~\mbox{cm}^{-3}$.
The photon last scattering occurred at temperature and red shift
\[
T_{\rm rec} \approx 3000~{\mbox K} \; , \;\;\;\;\;\; z_{\rm rec} \approx 1090 \; ,
\]
when the age of the Universe was about $t \approx 380$~thousand years
(for comparison, its present age is about 13.8 billion years).
Needless to say, CMB photons got red shifted since the last scattering,
so their present temperature is $T_0 = T_{\rm rec}/(1+z_{\rm rec})$.

The photon last scattering epoch is an important cornerstone in the
cosmological history. Since after that CMB photons travel freely
through the Universe, they give us
a photographic picture of the Universe at that
epoch.
Importantly, the
duration of the last scattering epoch was considerably shorter
than the Hubble time $H^{-1}(t_{\rm rec})$; to a reasonable approximation,
recombination occurred instantaneously. Thus, the photographic picture
is only slightly washed out due to the finite thickness  of
the last scattering surface.

At even earlier times, the temperature of the Universe was even
higher. We have direct evidence that at some point the temperature in
the Universe was in the MeV range. A traditional source of evidence
is
the Big Bang nucleosynthesis (BBN). The story begins
at a temperature of about 1~MeV, when the age of the Universe
was about 1~s. Before that time neutrons were rapidly created and
destroyed in
weak processes like
\be
{\rm e}^- + {\rm p} \longleftrightarrow {\rm n} + \nu_{\rm e} \; ,
\label{mar22-15-2}
\ee
while at $T_{\rm n} \approx 1$~MeV these processes
switched off, and the comoving number density of neutrons
froze out. The neutron-to-proton ratio at that time was given by
the Boltzmann factor,
\[
\frac{n_{\rm e}}{n_{\rm p}} = \e^{-\frac{m_{\rm n} - m_{\rm p}}{T_{\rm n}}} \; .
\]
Interestingly,
$m_{\rm n} - m_{\rm p} \sim T_{\rm n}$, so the neutron--proton ratio
at neutron freeze-out and later was neither equal to 1, nor very small.
Were it equal to 1, protons would combine with neutrons into
$^4$He at a somewhat later time, and there would remain no hydrogen in the
Universe. On the other hand, for very small $n_{\rm n}/n_{\rm p}$, too few light nuclei
would be formed, and we would not have any observable
remnants of the BBN epoch. In either case, the Universe would be quite
different from what it actually is. It is worth noting that the approximate
relation $m_{\rm n} - m_{\rm p} \sim T_{\rm n}$ is a coincidence:
$m_{\rm n} - m_{\rm p}$ is determined by light quark masses and electromagnetic
coupling, while $T_{\rm n}$ is determined by the strength of
weak interactions (which govern the rates of the processes \eqref{mar22-15-2})
and gravity (which governs the expansion of the Universe).
This is one of numerous coincidences we encounter in cosmology.

\begin{figure}[htb!]
\begin{center}
\includegraphics[width=0.5\textwidth,angle=0]{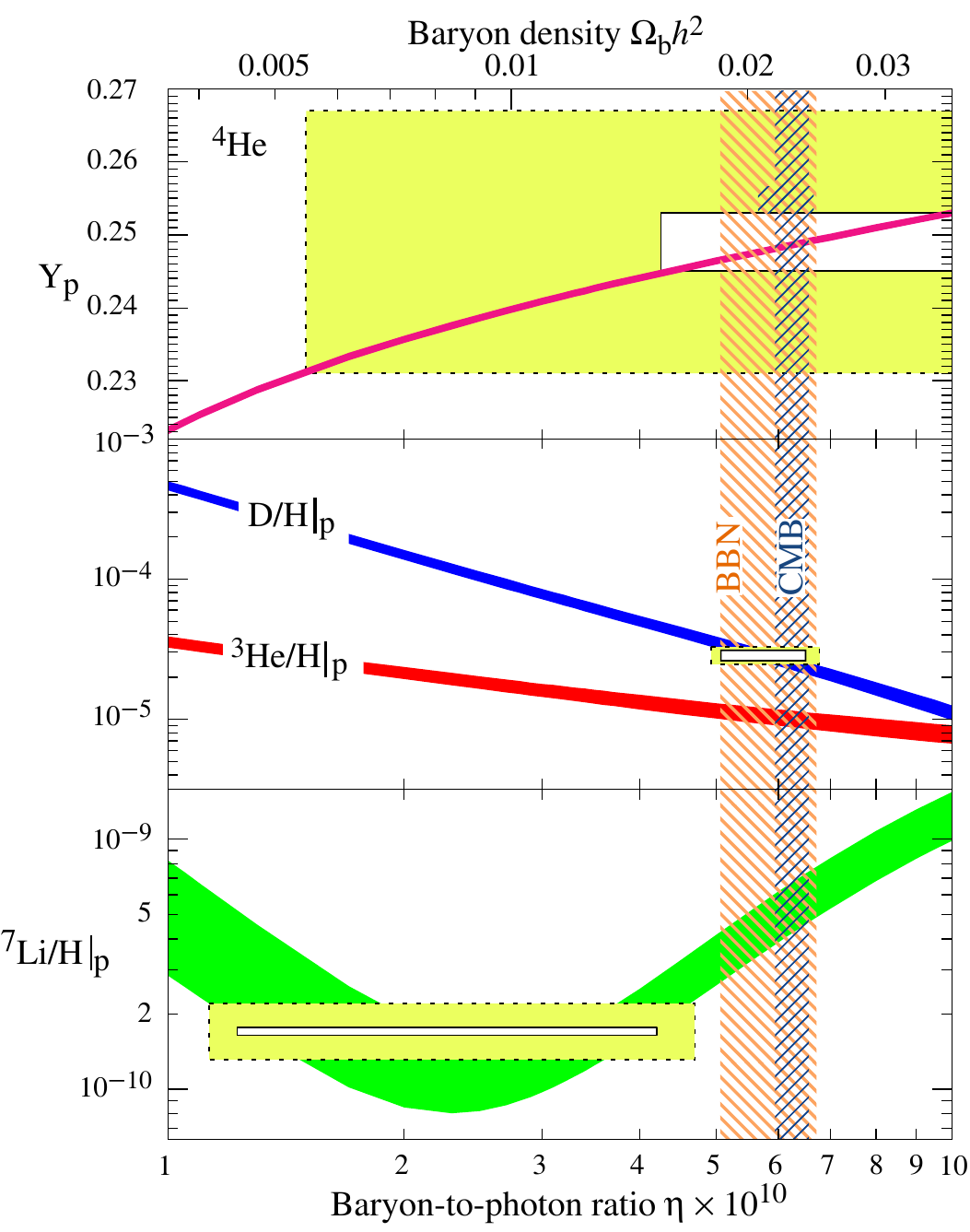}
\end{center}
\caption{Abundances of light elements, measured (boxes;
larger boxes include systematic uncertainties) and calculated
as functions of baryon-to-photon ratio $\eta$~\cite{Agashe:2014kda}.
The determination of $\eta \equiv \eta_{\rm B}$ from BBN (vertical range marked BBN) is in excellent agreement with the determination from the analysis of CMB temperature fluctuations  (vertical range marked CMB).
 \label{bbn09} }
\end{figure}

At temperatures somewhat below $T_{\rm n}$,
the neutrons combined with protons into
light elements in thermonuclear reactions like
\begin{eqnarray}
  {\rm p} + {\rm n} &\to& ^2\mbox{H} + \gamma\; ,
\nonumber\\
  ^2\mbox{H} + {\rm p} &\to& ^3\mbox{He} + \gamma \; ,
\nonumber\\
   ^3\mbox{He} + ^2\mbox{H} &\to&  ^4\mbox{He} + {\rm p}  \; ,
\end{eqnarray}
etc, up to $^7$Li. The abundances of light elements
have been measured; see Fig.~\ref{bbn09}.
On the other hand, the only parameter relevant for calculating
these abundances (assuming negligible neutrino--antineutrino asymmetry)
is the baryon-to-photon ratio
\be
\eta_{\rm B} \equiv \eta = \frac{n_{\rm B}}{n_\gamma} \; ,
\label{mar23-15-10}
\ee
 characterizing the number
density of baryons.
Comparison of the BBN
theory with the observational determination of the composition
of the cosmic medium enables one to determine $\eta_{\rm B}$ and check the
overall consistency of the BBN picture. It is even more reassuring
that a completely
independent measurement of $\eta_{\rm B}$ that makes use of the CMB
temperature fluctuations is in excellent agreement with BBN.
Thus, BBN
gives us confidence that we understand the
Universe at $T\sim 1$~MeV, $t \sim 1$~s.
In particular, we are convinced that
the cosmological expansion was
governed by general relativity.

Another class of processes of interest at temperatures in the
MeV range is neutrino production, annihilation and scattering,
\[
\nu_\alpha + \bar{\nu}_\alpha \longleftrightarrow {\rm e}^+ + {\rm e}^-
\]
and crossing processes. Here the subscript $\alpha$ labels neutrino flavours.
These processes switch off at $T \sim 2$--3~MeV, depending on
neutrino flavour. Since then neutrinos do not interact with the cosmic medium
other than gravitationally,
but they do affect the properties of CMB and distribution of galaxies
through their gravitational interactions. These effects are not negligible,
since the energy density of relativistic neutrinos is almost
the same as that of photons and, at temperature $T_{\rm rec} \simeq 3000$~K,
the energy density of these relativistic species is only three times smaller
than the energy density of non-relativistic particles (dark matter
and baryons). Thus, observational
data can be used to establish, albeit somewhat indirectly,
the existence of relic neutrinos and set limits on neutrino masses.
\begin{figure}[htb!]
\begin{center}
\includegraphics[width=0.5\textwidth,angle=0]{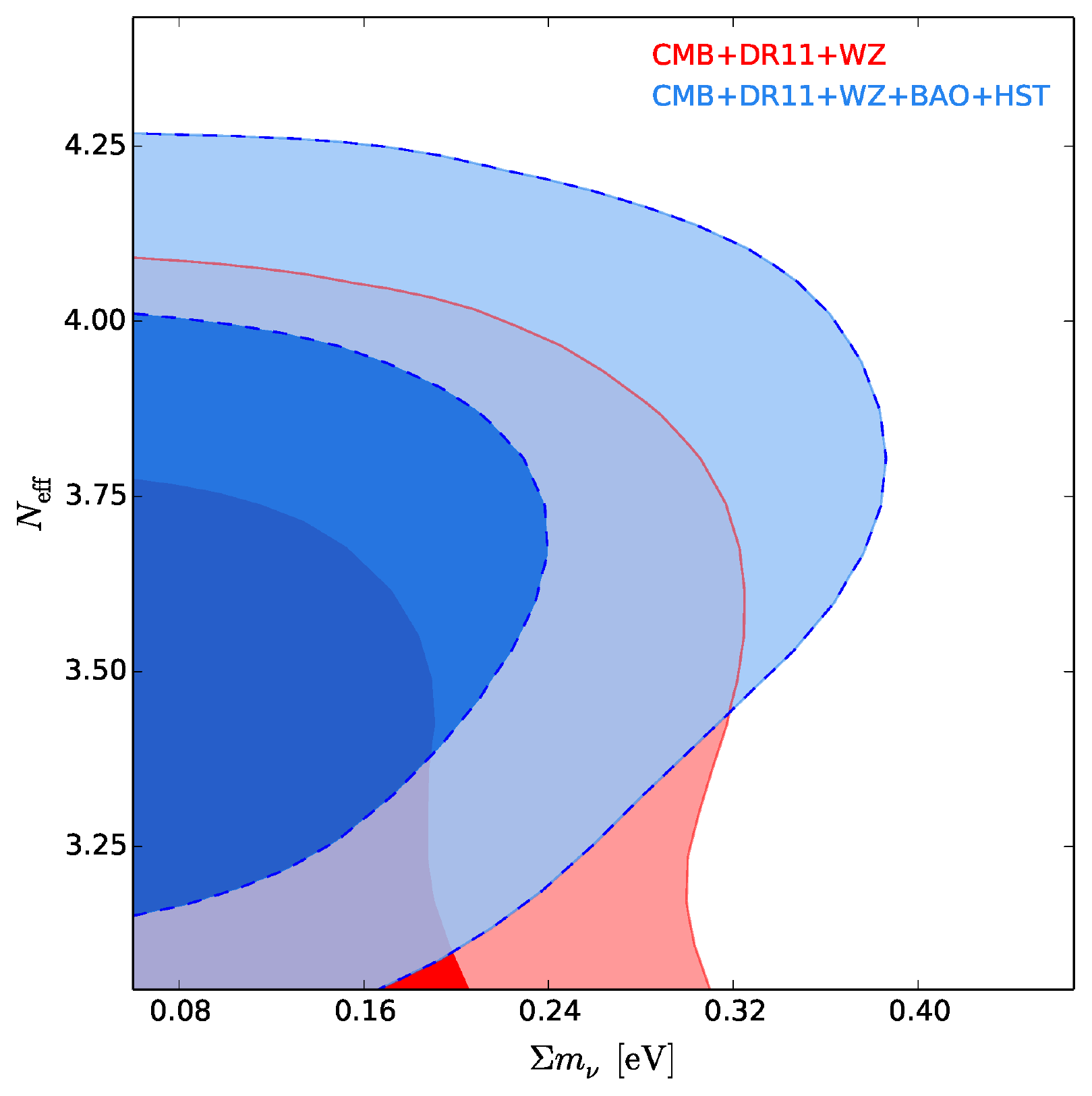}
\end{center}
\caption{Effective number of neutrino species and sum of neutrino masses
allowed by cosmological observations~\cite{Giusarma:2014zza}.
 \label{nmass}
 }
\end{figure}
An example is shown in Fig.~\ref{nmass}, where the number of neutrino flavours
$N_{\rm eff}$ and the sum of
neutrino masses
 are taken as free parameters. We see that cosmology {\it requires}
relic neutrinos of at least three flavours and sets the limit on neutrino mass
$m_\nu \lesssim 0.1$~eV  (neutrino oscillation data tell that
neutrinos with masses
above 0.1~MeV are degenerate in mass). The latest Planck analysis
gives~\cite{Planck:2015xua}
\[
\sum_i m_{\nu_i} < 0.23~\mbox{eV} \; , \;\;\;\;\;\;
N_{\rm eff} = 3.15 \pm 0.23 \; .
\]

\subsection{Dynamics of expansion}
\label{subsec:Friedeq}

The basic equation governing the expansion rate of the Universe
is the Friedmann equation,
\begin{equation}
    H^2 \equiv \left( \frac{\dot{a}}{a} \right)^2
= \frac{8\pi}{3M_{\rm Pl}^2}  \rho - \frac{\varkappa}{a^2}\; ,
\label{FriedmannC}
\end{equation}
where the dot denotes derivative with respect to time $t$, $\rho$ is the
{\it total} energy density in the Universe
and $\varkappa= 0, \pm 1$ is the parameter,
introduced in Section~\ref{sub:FLRW},  that discriminates
the  Euclidean 3-space
($\varkappa =0$) and curved 3-spaces.
The Friedmann equation is nothing but the $(00)$-component
of the Einstein equations of general relativity,
$R_{00} - \frac{1}{2}g_{00}R = 8\pi T_{00}$,
specified to the FLRW metric.  Observationally,
the spatial curvature of the Universe is very small:
the last,
curvature term in the right-hand side of
Eq.~\eqref{FriedmannC} is small compared to
the energy density term~\cite{Planck:2015xua},
\[
\frac{1/a^2}{8 \pi \rho/(3M_{\rm Pl}^2)} < 0.005 \; ,
\]
while the theoretical expectation is that the spatial curvature
is completely negligible. Establishing that the three-dimensional
space is (nearly) Euclidean is one of the profound results of CMB
observations.

In what follows we set $\varkappa =0$ and write the Friedmann equation as
\begin{equation}
    H^2 \equiv \left( \frac{\dot{a}}{a} \right)^2
= \frac{8\pi}{3M_{\rm Pl}^2}  \rho \; .
\label{Friedmann}
\end{equation}

The standard parameter used in cosmology is the critical density,
\be
\rho_{\rm c} = \frac{3}{8\pi} M_{\rm Pl}^2 H_0^2 \approx 5\times 10^{-6}\,
\frac{\mbox{GeV}}{\mbox{cm}^3} \; .
\label{rhoc-new}
\ee
According to Eq.~(\ref{Friedmann}), it is equal to the sum of all forms
of energy density
in the present Universe. There are at least three of such forms:
relativistic matter, or {\it radiation}, non-relativistic {\it matter, M}
and {\it dark energy, $\Lambda$}.  For every form $\lambda$ with
the {\it present} energy density $\rho_{\lambda,0}$, one defines the parameter
\be
   \Omega_\lambda = \frac{\rho_{\lambda,0}}{\rho_{\rm c}} \; .
\nonumber
\ee
One finds from Eq.~(\ref{Friedmann}) that
\[
\sum_\lambda \Omega_\lambda = 1\; .
\]
The $\Omega$ are important cosmological
parameters characterizing the energy balance in the present Universe.
Their numerical values are
\begin{subequations}
\begin{align}
\Omega_{\rm rad} &= 8.7 \times 10^{-5} \; ,
\label{mar23-15-1}
\\
\Omega_{\rm M} &= 0.31 \; ,
\\
\Omega_\Lambda &= 0.69 \; .
\label{mar13-15-11}
\end{align}
\end{subequations}
The value of $\Omega_{\rm rad}$ needs qualification.
At early times, when the temperature exceeds the masses of all neutrino
species, neutrinos are relativistic. The value of $\Omega_{\rm rad}$ in
Eq.~\eqref{mar23-15-1} is calculated for the unrealistic case in which
{\it all neutrinos are relativistic today}, so the radiation component
even at present consists of CMB photons and three neutrino species.
This prescription is convenient
for studying the energy (and entropy) content in the early Universe,
since it enables one to scale the energy density (and entropy) back in
time in a simple way, see below. For future reference, let us give
the value of the present entropy density in the Universe, pretending
that neutrinos are relativistic,
\be
s_0 \approx 3000~\mbox{cm}^{-3} \; .
\label{e-density}
\ee

\vspace{0.3cm}
\noindent
{\it Question.} Calculate the numerical value of $\Omega_\gamma$ and the entropy density of CMB photons.
\vspace{0.3cm}

Non-relativistic matter consists of baryons and dark matter.
The contributions of each of these fractions are~\cite{Planck:2015xua}
\begin{align*}
\Omega_{\rm B} &= 0.048 \; ,
\\
\Omega_{\rm DM} &=0.26 \; .
\end{align*}

Different components of the energy density evolve differently
in time. The energy of a given
photon or massless neutrino scales as $a^{-1}$,
and the number density of these species scales as $a^{-3}$.
Therefore, the energy density of radiation scales as
$\rho_{\rm rad} \propto a^{-4}$ and
\be
\rho_{\rm rad} (t) = \left(\frac{a(t)}{a_0}\right)^4 \rho_{\rm rad, 0}
= (1+z)^4 \, \Omega_{\rm rad} \rho_{\rm c} \; .
\label{mar23-15-3}
\ee
The energy of non-relativistic matter is dominated by the mass of its
particles, so the energy density scales as the number density, i.e.,
\be
\rho_{\rm M} (t) = \left(\frac{a(t)}{a_0}\right)^3 \rho_{\rm M, 0}
= (1+z)^3 \, \Omega_{\rm M} \rho_{\rm c} \; .
\label{mar23-15-4}
\ee
Finally, the energy density of dark energy does not change in time,
or changes very slowly. We assume for definiteness that $\rho_\Lambda$
stays constant in time,
\be
\rho_\Lambda = \Omega_\Lambda \rho_{\rm c} = \mbox{const} \; .
\label{mar23-15-5}
\ee
In fact, whether or not $\rho_\Lambda$ depends on time (even slightly)
is a very important question. If dark energy is a cosmological constant
(or, equivalently, vacuum energy), then it does not depend on time at all.
Even a slight dependence of $\rho_\Lambda$ on time would mean that we
are dealing with something different from the cosmological constant,
like, e.g., a new scalar field with a very flat scalar potential.
The existing limits on the time evolution of dark energy correspond,
roughly speaking,
to the variation of $\rho_\Lambda$ by not more than 20\% in the last
8 billion years (from the time corresponding to $z \approx 1$);
usually these limits are expressed in terms of the equation-of-state
parameter relating energy density and effective pressure
$p_\Lambda = w_\Lambda \rho_\Lambda$:
\be
w_\Lambda \approx 1.0 \pm 0.1 \; .
\label{mar23-15-12}
\ee
The relevance of the effective pressure is seen from the
covariant conservation equation for the energy--momentum tensor,
$\nabla_\mu T^{\mu \nu} =0$, whose $\nu=0$ component
reads
\[
\dot{\rho} = -3 \frac{\dot{a}}{a} (\rho + p) \; .
\]
It shows that the energy density of a
 component with equation of state $p=w\rho$, $w=\mbox{const}$
scales as $\rho \propto a^{-3(1+w)}$. As pointed out above,
radiation ($w_{\rm rad} = 1/3$) and matter
($w=0$) scale as $\rho_{\rm rad} \propto a^{-4}$ and $\rho_{\rm M} \propto a^{-3}$,
respectively,
while the cosmological constant case corresponds to $w_\Lambda = -1$.

\vspace{0.3cm}
\noindent
{\it Question.} Show that for a gas of relativistic particles, $p=\rho/3$.
\vspace{0.3cm}

According to Eqs.~\eqref{mar23-15-3}, \eqref{mar23-15-4}
and \eqref{mar23-15-5}, different forms of energy dominate at different
cosmological epochs. The present Universe is at the end of the
transition from matter domination to $\Lambda$ domination:
the dark energy will `soon' completely dominate over non-relativistic
matter because of the rapid decrease of the energy density of the latter.
Conversely, the matter energy density increases as we go backwards in time,
and
until relatively recently ($z \lesssim 0.3$) it dominated over dark energy
density.
At even more distant past, the radiation energy density
was the highest, as it increases most rapidly  backwards in time.
The red shift at radiation--matter equality, when the energy densities of
radiation and matter were equal, is
\[
1+ z_{\rm eq} = \frac{a_0}{a(t_{\rm eq})} = \frac{\Omega_{\rm M}}{\Omega_{\rm rad}}
\approx 3500 \;
\]
and, using the Friedmann equation, one finds the age of the Universe at
equality
\[
t_{\rm eq} \approx 50~000~\mbox{years} \; .
\]
Note that recombination occurred at matter domination,
but rather soon after equality. So, we have the following sequence
of the regimes of evolution:
\[
\dots \Longrightarrow \mbox{Radiation~domination}
\Longrightarrow \mbox{Matter~domination}\Longrightarrow
\Lambda\mbox{ domination} \; .
\]
The dots here denote some cosmological
epoch preceding the hot stage of the evolution; as we mentioned in
Section~\ref{Intro}, we are confident that such an epoch
existed, but do not quite know what it was.

\subsection{Radiation domination}
\label{sub:RD}

The epoch of particular interest for our purposes is radiation
domination.
By inserting $\rho_{\rm rad} \propto a^{-4}$ into the Friedmann equation
\eqref{Friedmann}, we obtain
\[
\frac{\dot{a}}{a} = \frac{\mbox{const}}{a^2} \; .
\]
This gives the evolution law
\be
a (t) = \mbox{const} \cdot \sqrt{t} \; .
\label{sep13-11-5}
\ee
The constant here does not have physical significance,
as one can rescale the coordinates ${\bf x}$ at some fixed moment
of time, thus changing the normalization of $a$.

There are several points to note regarding the result
\eqref{sep13-11-5}. First, the expansion {\it decelerates}:
\[
\ddot{a} < 0 \; .
\]
This property holds also for the matter-dominated epoch, but
it does not hold for the domination of the dark
energy.

\vspace{0.3cm}
\noindent
{\it Question.} Find the evolution laws, analogous to Eq.~\eqref{sep13-11-5},
for matter- and $\Lambda$-dominated Universes. Show that the expansion
decelerates, $\ddot{a} < 0$, at matter domination and accelerates,
$\ddot{a} >0$, at $\Lambda$ domination.
\vspace{0.3cm}

Second, time $t=0$ is the Big Bang singularity
(assuming erroneously
that the Universe starts being radiation dominated).
The expansion rate
\[
H(t) = \frac{1}{2t}
\]
diverges as $t \to 0$, and so do the energy
density $\rho(t) \propto H^2 (t)$ and temperature
$T \propto \rho^{1/4}$. Of course, the classical
general relativity and usual notions of statistical
mechanics (e.g., temperature itself) are not applicable
very near the singularity, but our result
suggests that in the picture we discuss
(hot epoch right after the Big Bang), the Universe
starts its classical evolution in a very hot and dense state, and its expansion
rate is very high in the beginning. It is customary to
consider for illustrational purposes
that the relevant quantities in the beginning of
the classical expansion take the Planck values,
$\rho \sim M_{\rm Pl}^4$, $H \sim M_{\rm Pl}$ etc.

Third, at a given moment of time the size of a causally connected
region is finite. Consider signals emitted right after the Big Bang
and travelling with the speed of light. These signals travel along the
light cone with
${\rm d}s=0$ and hence $a(t) {\rm d}x = {\rm d}t$. So, the coordinate distance
that a signal travels from the Big Bang to time $t$ is
\be
x = \int_0^{t} \frac{{\rm d}t}{a(t)} \equiv \eta \; .
\label{sep13-11-6}
\ee
In the radiation-dominated Universe,
\[
\eta = \mbox{\const} \cdot \sqrt{t} \; .
\]
The physical distance from the emission point to
the position of the signal is
\[
l_{\rm H}( t) = a(t) x = a(t) \int_0^{t} \frac{{\rm d}t}{a(t)} =2t \; .
\]
As expected, this physical distance is finite, and it
gives the size of a causally connected region at time $t$.
It is called the horizon size (more precisely, the
size of the particle horizon). A related property is that
an observer at time $t$ can see only the part
of the Universe whose current physical size is $l_{\rm H} (t)$.
Both at radiation and matter
domination one has, modulo a numerical constant of order 1,
\be
l_{\rm H}(t) \sim H^{-1}(t) \; .
\label{mar23-15-7}
\ee
To give an idea of numbers, the horizon size at
the present epoch is
\[
l_{\rm H}(t_0) \approx 15~\mbox{Gpc} \simeq 4.5 \times 10^{28}~\mbox{cm}
\; .
\]

\vspace{0.3cm}
\noindent
{\it Question.} Find the proportionality constant
in Eq.~\eqref{mar23-15-7} for a matter-dominated Universe.
Is there a particle horizon in a Universe without matter but with
positive cosmological constant?
\vspace{0.3cm}

It is convenient to express the Hubble parameter at radiation
domination in terms of temperature. The Stefan--Boltzmann law gives
for the energy density of a gas of relativistic particles in thermal
equilibrium at zero chemical potentials (chemical potentials in the Universe
are indeed small)
\be
\rho_{\rm rad} = \frac{\pi^2}{30} g_* T^4 \; ,
\label{sep13-11-2}
\ee
with $g_*$ being the effective number of degrees of freedom,
\[
g_* = \sum_{\rm bosons} g_i + \frac{7}{8}\sum_{\rm fermions} g_i \; ,
\]
where $g_i$ is the number of spin states and the factor $7/8$ is due
to Fermi statistics. Hence, the Friedmann equation \eqref{Friedmann}
gives
\be
H = \frac{T^2}{M_{\rm Pl}^*} \; , \;\;\;\;\; M_{\rm Pl}^* = \frac{M_{\rm Pl}}{1.66 \sqrt{g_*}}
\; .
\label{mar25-15-10}
\ee
One more point has to do with entropy: the cosmological expansion
is slow, so that the entropy is conserved (modulo exotic scenarios with
large entropy generation).
The entropy density in thermal equilibrium is given by
\[
s = \frac{2\pi^2}{45} g_* T^3 \; .
\]
The conservation of entropy means that
the entropy density scales {\it exactly} as $a^{-3}$,
\be
sa^3 = \mbox{const} \; ,
\label{sep13-11-1}
\ee
while
temperature scales {\it approximately} as $a^{-1}$.
The temperature would scale as $a^{-1}$ if the number
of relativistic degrees of freedom would be independent of
time. This is not the case, however.
Indeed, the value of $g_*$ depends on temperature:
at $T \sim 10$~MeV relativistic species are
photons, neutrinos, electrons and positrons, while
at $T \sim 1$~GeV four flavours of quarks, gluons, muons
and $\tau$-leptons
are relativistic too. The number of degrees of freedom
in the Standard Model at $T \gtrsim 100$~GeV
is
\[
g_* (100~\mbox{GeV}) \approx 100 \; .
\]

If there are conserved quantum numbers, such as the baryon number
after baryogenesis,
their density also scales as $a^{-3}$. Hence, the time-independent
characteristic of, say, the baryon abundance is the baryon-to-entropy ratio
\[
\Delta_{\rm B} = \frac{n_{\rm B}}{s} \; .
\]
The commonly used baryon-to-photon ratio $\eta_{\rm B}$, Eq.~\eqref{mar23-15-10},
is related to $\Delta_{\rm B}$ by a numerical factor, but this factor
depends on time through $g_*$ and stays constant only after
${\rm e}^+ {\rm e}^-$ annihilation, i.e., at $T \lesssim 0.5$~MeV. Numerically,
\be
\Delta_{\rm B} = 0.14 \eta_{\rm B, 0} = 0.86 \times 10^{-10} \; .
\label{mar28-15-1}
\ee
\section{Dark energy}
\label{sec:de}

\begin{figure}[!htb]
\centering
\includegraphics[width=0.5\textwidth, angle=90]{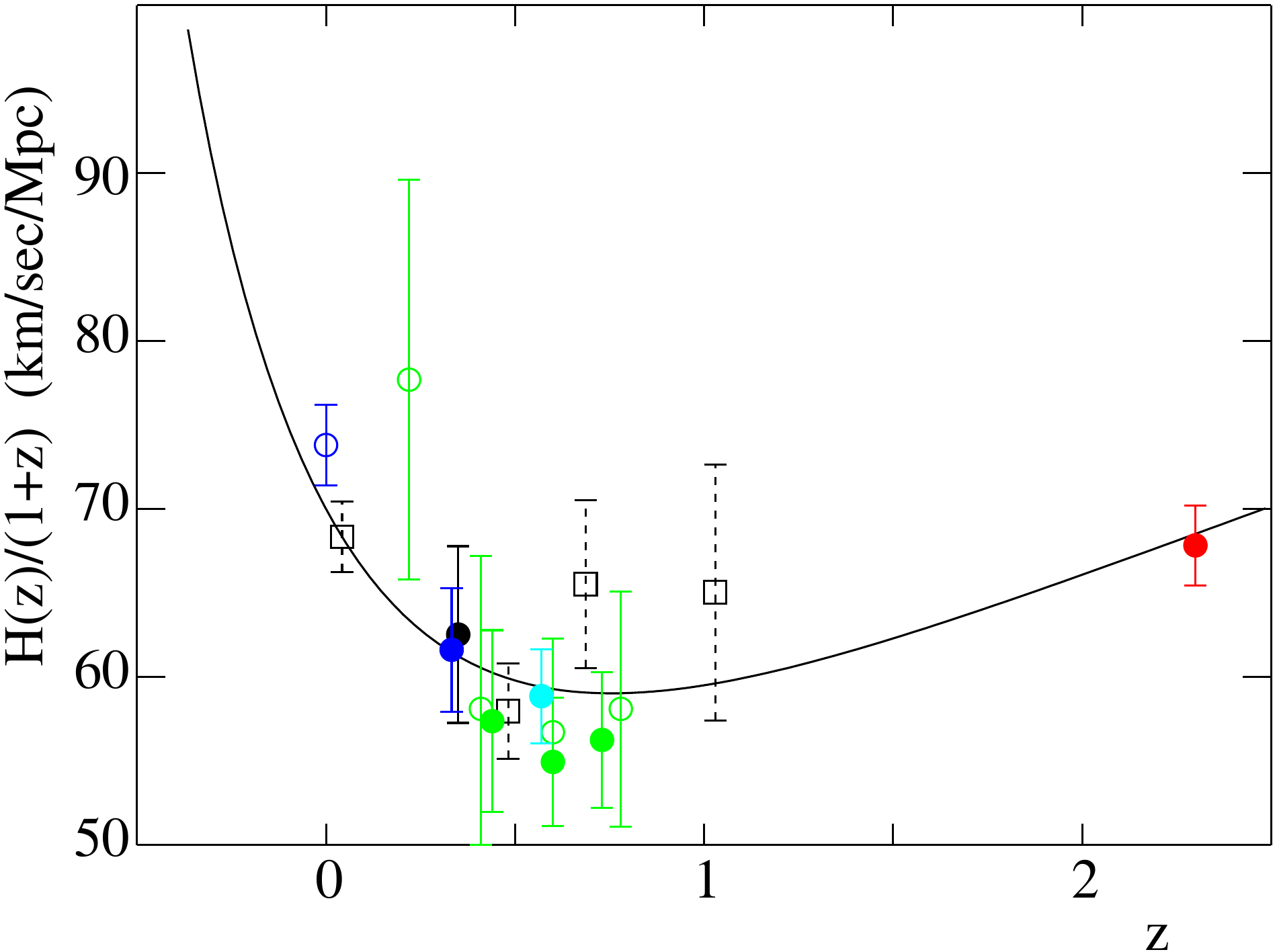}
\caption{Observational data on the time derivative of the scale factor
as function of red shift $z$~\cite{Busca:2012bu}. The change of the behaviour from decreasing to increasing with decreasing $z$ means the change from decelerated to accelerated expansion. The theoretical curve corresponds to
a spatially flat Universe with $h=0.7$ and $\Omega_\Lambda = 0.73$.
\label{adot}
}
\end{figure}

Before turning to our main topics, let us briefly discuss
dark energy. We know very little
about this `substance': our knowledge is summarized in
Eqs.~\eqref{mar13-15-11} and \eqref{mar23-15-12}. We also know that
dark energy does not clump,
unlike dark matter and baryons.  It gives rise to the
accelerated expansion of the Universe.  Indeed, the solution
to the Friedmann equation \eqref{Friedmann} with constant
$\rho = \rho_\Lambda$ is
\[
a (t) = \e^{H_\Lambda t} \; ,
\]
where $H_\Lambda = (8\pi \rho_\Lambda/3M_{\rm Pl}^2)^{1/2} = \mbox{const}$.
This gives $\ddot{a} > 0$, unlike at radiation or matter domination.
The observational discovery of the
accelerated expansion of the Universe was the discovery of dark energy.
Recall that early on (substantial $z$), the Universe was matter dominated,
so its expansion was decelerating. The transition from
decelerating to accelerating expansion is confirmed by combined observational
data, see Fig.~\ref{adot}, which shows the dependence on red shift of the quantity
$H(z)/(1+z)=\dot a(t)/a_0$.

\vspace{0.3cm}
\noindent
{\it Question.} Find the red shift $z$ at which decelerated
expansion turned into an accelerated one.
\vspace{0.3cm}

As a remark, the effective pressure of dark energy or any other component
is defined as the (possibly time-dependent) parameter determining the
spatial components of the energy--momentum tensor in a locally Lorentz frame
($a=1$ in the FLRW context),
\[
T_{\mu \nu} = \mbox{diag} ~(\rho, p,p,p) \; .
\]
In the case of the cosmological constant,
the dark energy density does not depend on time at all:
\[
T_{\mu \nu} = \rho_\Lambda \eta_{\mu \nu} \; ,
\]
where $\eta_{\mu \nu}$ is the Minkowski tensor. Hence, $w_\Lambda = -1$.
One can view this as the
characteristic of vacuum, whose energy--momentum tensor must
be Lorentz-covariant. As we pointed out above,
any deviation from $w=-1$ would mean that we are dealing
with something other than vacuum energy density.

The problem with dark energy is that
its present value is extremely small by particle-physics standards,
\begin{eqnarray*}
\rho_{\rm DE}\approx 4~\mbox{GeV m$^{-3}$}=(2\times10^{-3} \mbox{ eV})^4\,.
\end{eqnarray*}
In fact, there are two hard problems.
One is that particle-physics scales are much larger
than the scale relevant to the dark energy density,
so the dark energy density is zero to an excellent approximation.
Another is that
it is non-zero nevertheless, and one has to understand its energy scale.
To quantify the first problem, we recall the known scales of
particle physics and gravity,
\begin{eqnarray*}
\mbox{Strong interactions}:&\quad&\Lambda_{\rm QCD}\sim 1\,\mbox{GeV}\,,
\cr
\mbox{Electroweak}:&&M_{\rm W}\sim 100\,\mbox{GeV}\,,
\cr
\mbox{Gravitational}:&&M_{\rm Pl}\sim 10^{19}\, \mbox{GeV}\,.
\end{eqnarray*}
Off hand, physics at scale $M$ should contribute to
the vacuum energy density
as $\rho_\Lambda \sim M^4$, and
there is absolutely no reason for vacuum to be as light as it is.
The discrepancy here is huge, as one sees from the above numbers.

To elaborate on this point, let us note that the action
of gravity plus, say, the Standard Model has the general form
\[
S = S_{\rm EH} + S_{\rm SM} - \rho_{\Lambda, 0} \int~\sqrt{-g}~{\rm d}^4x \; ,
\]
where $S_{\rm EH} = -(16 \pi G)^{-1} \int ~R~\sqrt{-g}~{\rm d}^4x$ is the
Einstein--Hilbert action of general relativity, $S_{\rm SM}$ is the action
of the Standard Model and   $\rho_{\Lambda, 0}$ is the bare
cosmological constant. In order that the vacuum energy density be
almost zero, one needs fantastic
cancellations between the contributions of the Standard Model fields
into the vacuum energy density, on the one hand, and  $\rho_{\Lambda, 0}$
on the other.
For example, we know that quantum chromodynamics (QCD) has a complicated vacuum structure, and one would expect that
the energy density of QCD
should be of the order of $(1~\mbox{GeV})^4$.
 At least for QCD, one needs a cancellation of the order of $10^{-44}$.
 If one goes further and considers other interactions,
the numbers get even worse.

What are the hints from this `first' cosmological constant problem?
There are several options, though not many.
 One is that the Universe could have a very long prehistory: extremely long.  This option has to do with relaxation mechanisms.
Suppose that the original vacuum energy density is indeed large, say,
comparable to the particle-physics scales. Then there must be a
mechanism which can  relax this value down to an
acceptably small number.  It is easy to
convince oneself that this relaxation could not happen in the
history of the Universe we know of.  Instead,
the Universe should have a very long prehistory
during which this relaxation process might occur.  At that
prehistoric time, the vacuum in the
Universe must have been exactly the same
as our vacuum, so the Universe in its prehistory
must have been exactly like ours, or almost exactly like ours.
Only in that case could a relaxation mechanism  work.
There are concrete scenarios of this sort~\cite{prehistory-1,prehistory-2}.
However, at the moment it seems that these scenarios are hardly testable,
since this is prehistory.

Another possible hint is towards anthropic selection.
The argument that goes back to Weinberg
and Linde \cite{Weinberg:1987dv,Linde:1986dq}
is that if the
cosmological constant were larger, say, by a factor of 100, we simply would
not exist: the stars would not have formed because of the
fast expansion of the Universe.
So, the vacuum energy density may be selected anthropically.
The picture is that the Universe may be much, much larger
than what we can see, and different large regions of the Universe
may have different properties. In particular, vacuum energy
density may be different in different regions.
Now, we are somewhere in the place where one can live.
All the rest is empty of observers, because
there the parameters such as vacuum
energy density are not suitable for their existence.
This is disappointing for a theorist, as this point of view allows for
arbitrary tuning of fundamental parameters.  It is
hard to disprove this option,
on the other hand.
We do exist, and this is an experimental fact.
The anthropic viewpoint may, though hopefully  will not,
get more support from the LHC, if no or insufficient
new physics is found there.  Indeed, another candidate for
an environmental quantity is the electroweak scale, which is fine tuned
in the Standard Model in the same sense as the cosmological constant is fine
tuned in gravity (in the Standard Model context, this fine tuning
goes under the name of the gauge hierarchy problem).

Turning to the `second' cosmological constant problem,
we note that the scale $10^{-3}$~eV may be associated with some new
light field(s), rather than with vacuum. This implies
that $\rho_\Lambda$ depends on time, i.e., $w_\Lambda \neq -1$
and $w_\Lambda$ may well depend on time itself.
Current data are compatible with time-independent $w_\Lambda$ equal to
$-1$,  but their precision is not particularly high.
We conclude that
 future cosmological
observations may shed new light on the field content of
the fundamental theory.

\section{Dark matter}
\label{sec:dm}


Unlike dark energy,
dark matter experiences the same gravitational force as
baryonic matter. It
consists presumably of new stable massive
particles. These make clumps of mass
which constitute
most of the mass of galaxies and
clusters of galaxies.
There are various ways of measuring the contribution of
non-baryonic dark matter into the total energy density of the
Universe (see Refs.~\cite{dark-rev1,dark-rev2,dark-rev3,dark-rev4} for details).

\begin{enumerate}
\item The composition of the Universe affects the angular anisotropy
and polarization of
CMB. Quite accurate CMB
measurements available today enable one to measure the total mass
density
of  dark matter.

\item  There is direct evidence that dark matter exists in the largest
gravitationally bound objects---clusters of
galaxies. There are various methods to determine the gravitating mass
of a cluster, and even the mass distribution in a cluster, which give
consistent results. As an example, the total gravitational field of a cluster, produced by both dark matter and baryons, acts as a gravitational lens
for extended light sources behind the cluster. The images of these sources
enable one to reconstruct the mass distribution in the cluster. This is
shown in Fig.~\ref{grav-lens-dark}. These determinations show that
baryons (independently measured through their X-ray emission)
make less than $1/3$ of total mass in clusters. The rest is dark matter.
\begin{figure}[htb]
\centerline{\includegraphics[width=0.3337667\textwidth]{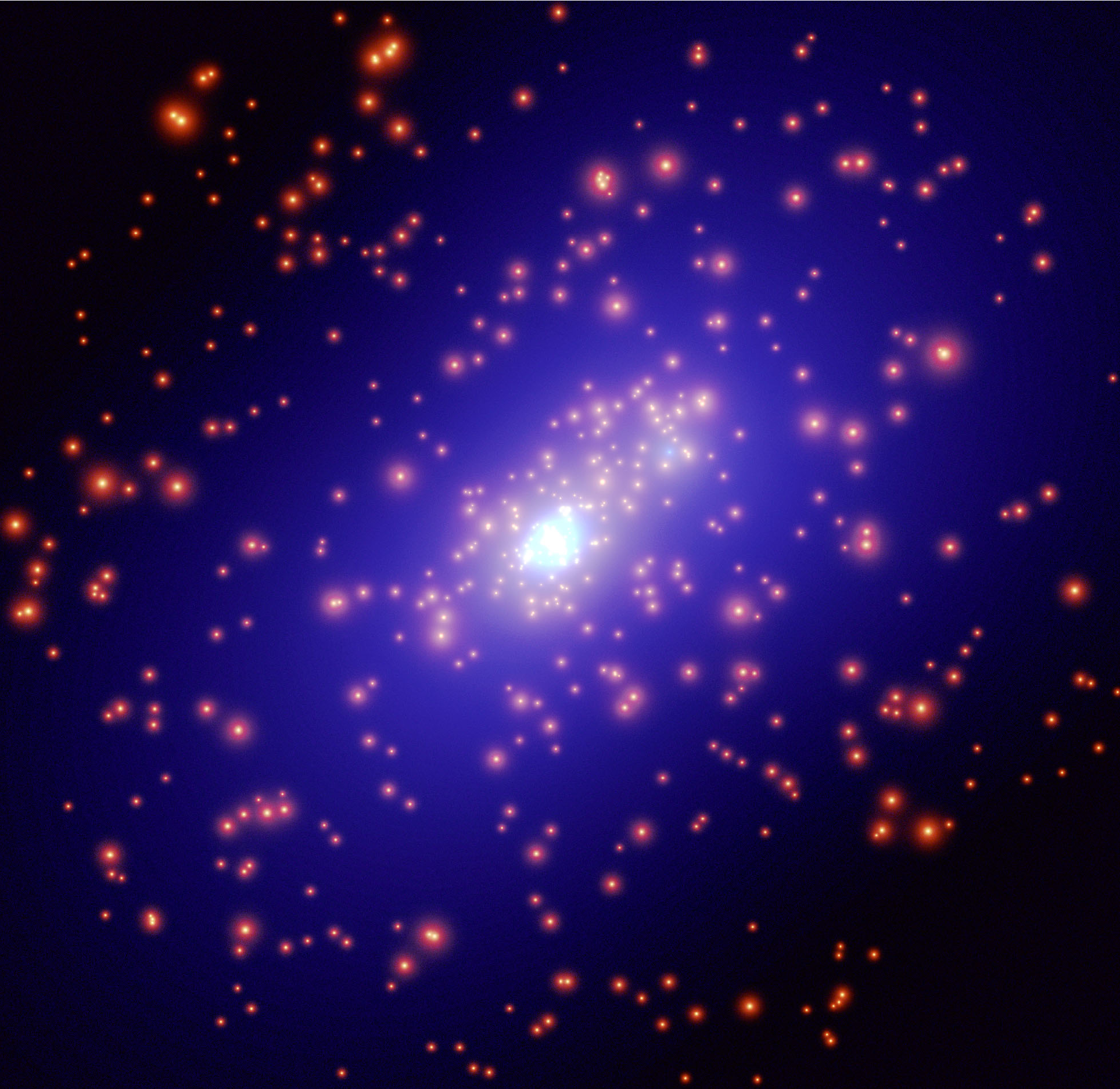}
\includegraphics[width=0.329\textwidth]{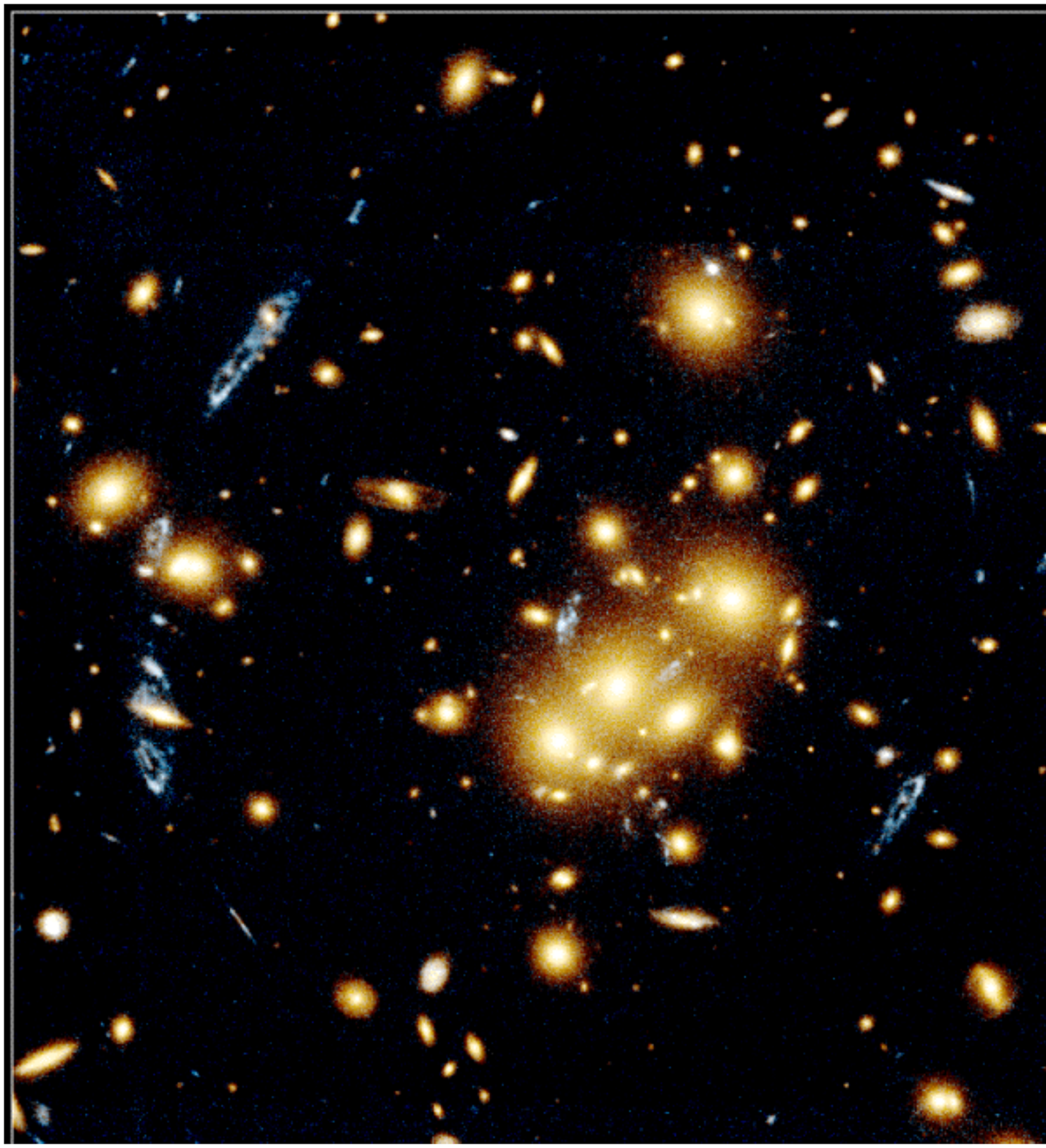}}
\caption{\label{grav-lens-dark}
Cluster of galaxies
CL$0024+1654$~\cite{CL0024+1654}, acting as gravitational lens.
Right-hand panel: cluster in visible light.
Round yellow spots are galaxies in the cluster.
Elongated blue images are those of one and the same galaxy beyond the
cluster. Left-hand panel: reconstructed distribution of gravitating mass in
the cluster; brighter regions have larger mass density.}
\end{figure}

A particularly convincing case is the Bullet Cluster,
Fig.~\ref{colliding-clusters}. Shown are two galaxy clusters that
passed through each other. The dark matter and galaxies do not
experience friction and thus do not lose their velocities.
On the contrary, baryons in hot, X-ray-emitting gas do experience
friction and hence get slowed down and lag behind dark matter and galaxies.
In this way the baryons (which are mainly in hot gas) and dark matter
are separated in space.
\begin{figure}[htb!]
\centerline{
\includegraphics[width=0.4\textwidth]{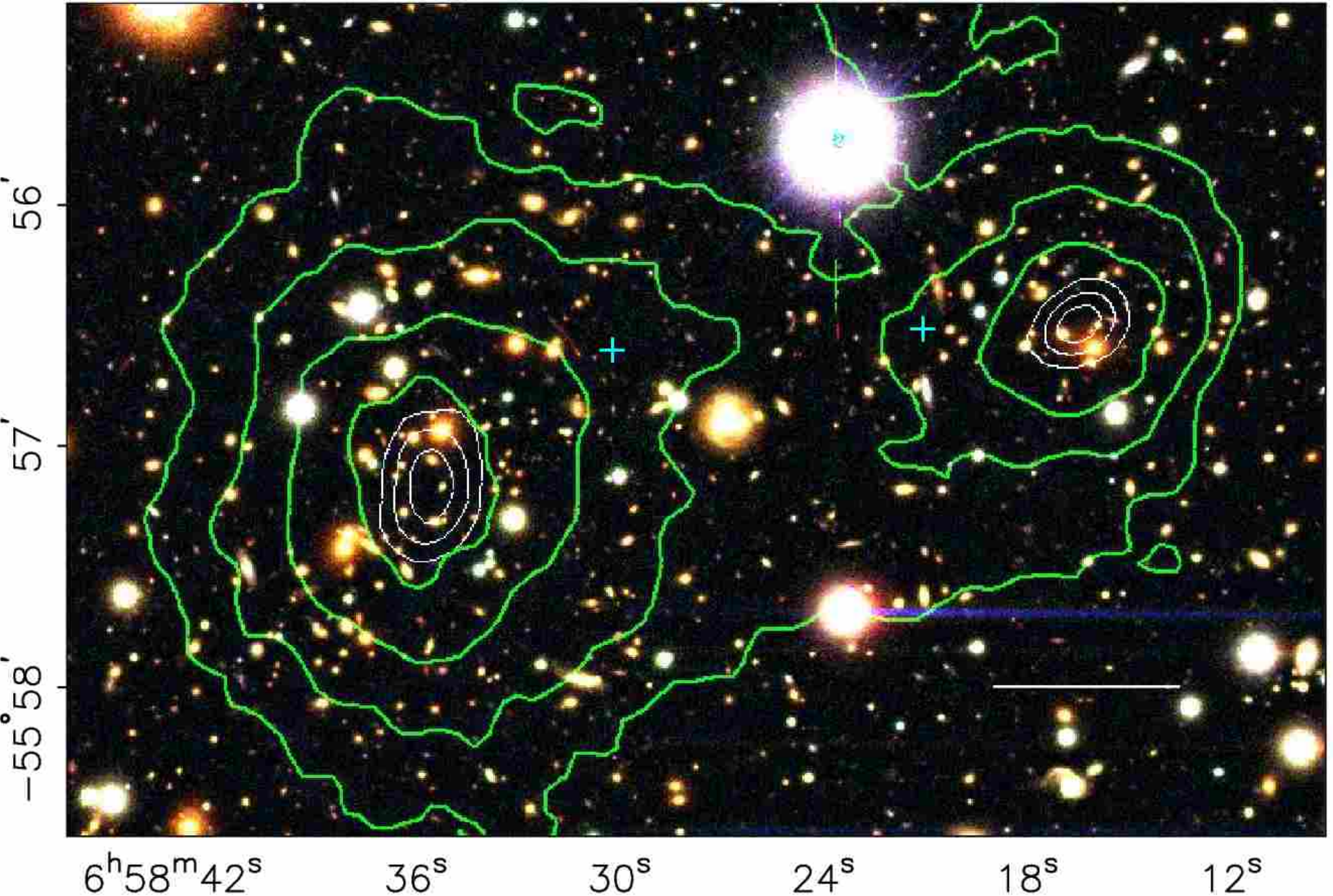}
\includegraphics[width=0.4\textwidth]{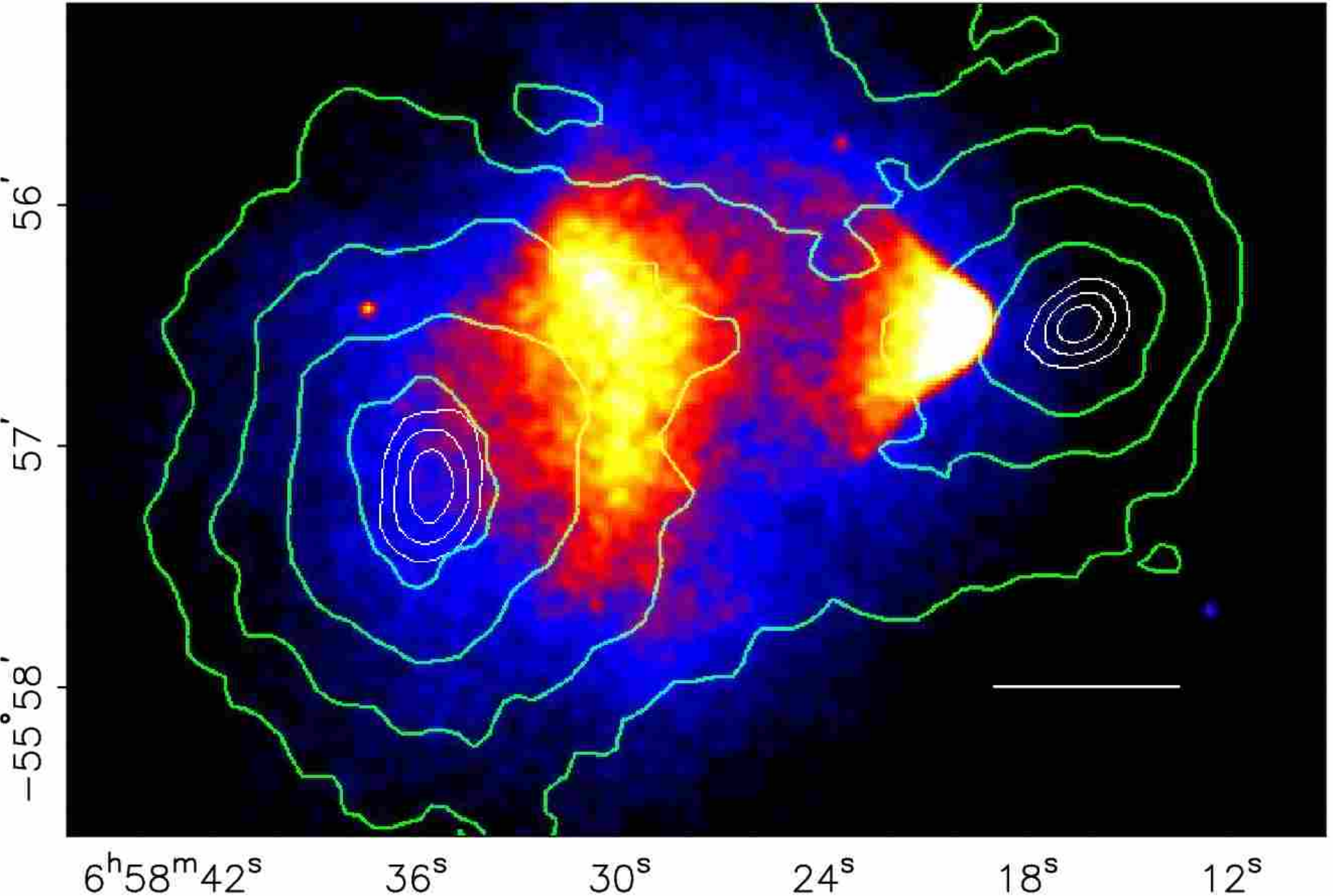}
}
\caption{Observation~\cite{Clowe:2006eq} of the Bullet Cluster
1E0657\hbox{-}558 at $z=0.296$. Closed lines show the gravitational potential
produced mainly by dark matter and measured through gravitational
lensing. Bright regions show X-ray emission of hot baryon gas,
which makes most of the baryonic matter in the clusters. The length
of the white interval is 200~kpc in the comoving frame.
\label{colliding-clusters}
}
\end{figure}

\item Dark matter exists also in galaxies. Its distribution is
measured by the observations of rotation velocities of distant stars
and gas clouds around a galaxy, Fig.~\ref{rotation-curve}.
Because of the existence of
dark matter away from the luminous regions, i.e., in
halos, the rotation velocities do not decrease with the distance from
the galactic centres; rotation curves are typically flat up to distances
exceeding the size of the bright part by a factor of 10 or so.
The fact that dark matter halos are so large is explained by
the defining property of dark matter particles: they do not lose their
energies by emitting photons and, in general, interact with conventional
matter very weakly.
\begin{figure}[b!]
\begin{center}
\includegraphics[width=0.5\textwidth]{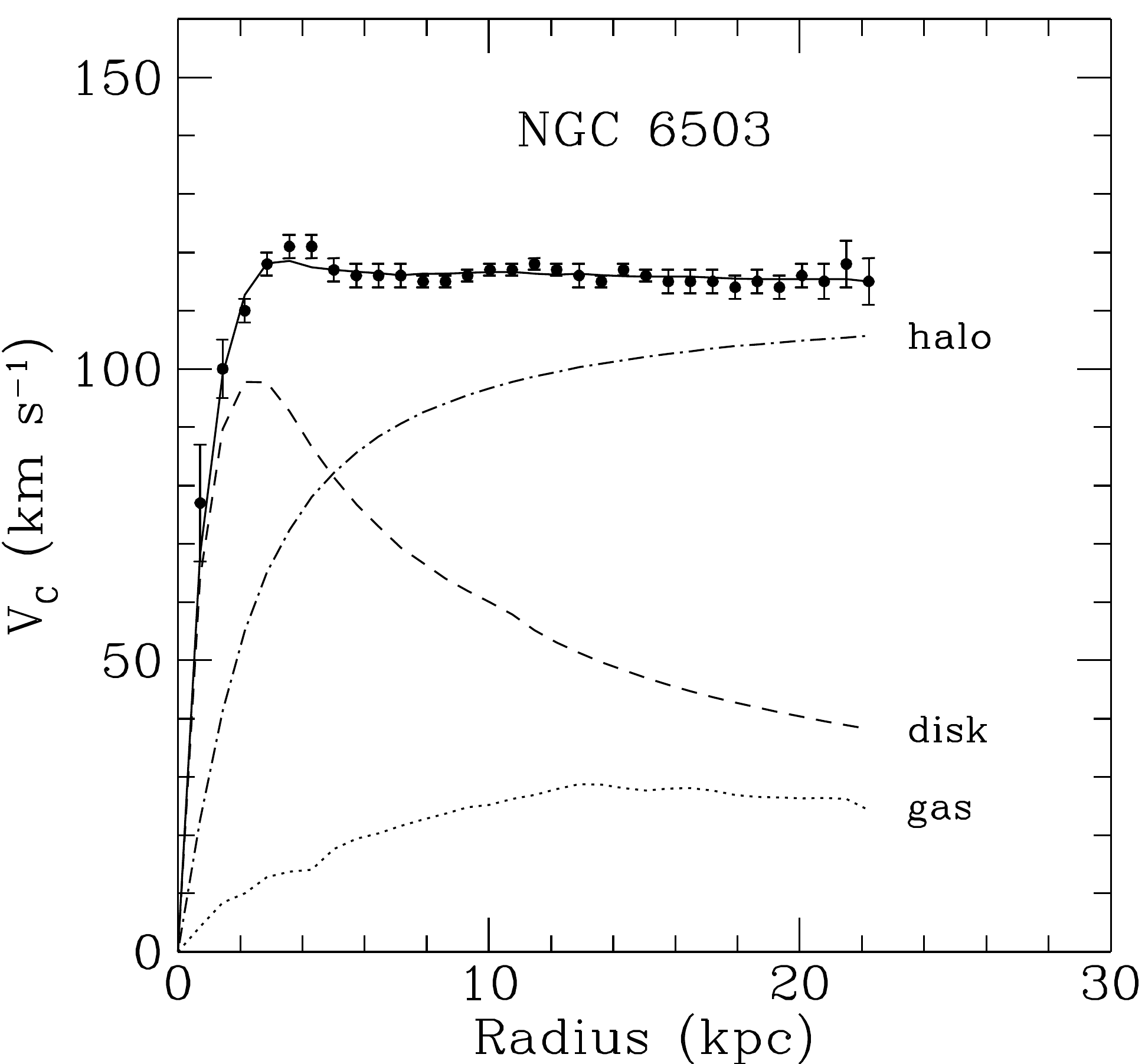}
\end{center}
\caption{\label{rotation-curve}
Rotation velocities of hydrogen gas clouds around the galaxy
NGC 6503~\cite{Extended-curves}. Lines show the
contributions of the three main components that produce the
gravitational potential. The main contribution at large distances
is due to dark matter, labelled `halo'.
}
\end{figure}

\end{enumerate}


Dark matter  is characterized by the mass-to-entropy ratio,
\be
   \left( \frac{\rho_{\rm DM}}{s} \right)_0 = \frac{\Omega_{\rm DM} \rho_{\rm c}}{s_0}
\approx
\frac{0.26 \times 5 \times 10^{-6}~ \mbox{GeV cm}^{-3}}{3000
~\mbox{cm}^{-3}} = 4 \times 10^{-10}~ \mbox{GeV} \; .
\label{10p*}
\ee
This ratio is constant in time since the freeze out of
dark matter density: both number density of dark matter particles
$n_{\rm DM}$ (and hence their mass density
$\rho_{\rm DM}=m_{\rm DM} n_{\rm DM}$) and entropy density
get diluted exactly as $a^{-3}$.

Dark matter is crucial for our existence, for the following reason.
Density perturbations in baryon--electron--photon plasma before recombination
do not grow because of high pressure, which is mostly due to photons;
instead,  perturbations are sound waves
propagating in plasma with time-independent
amplitudes.
Hence, in a Universe without dark matter, density perturbations
in the baryonic component would start to grow only after baryons decouple from photons, i.e., after recombination. The mechanism of the growth is
pretty simple: an overdense region gravitationally attracts
surrounding matter; this matter falls into the overdense region,
and the density contrast increases. In the expanding matter-dominated
Universe this gravitational instability results in the density contrast
growing like $(\delta \rho/\rho) (t) \propto a(t)$. Hence, in a Universe
without dark matter, the growth factor for baryon density perturbations
would be at most 
\be
   \frac{a(t_0)}{a(t_{\rm rec})} = 1+z_{\rm rec} = \frac{T_{\rm rec}}{T_0} \approx 10^3 \; .
\label{bgrow}
\ee
Because of the presence of dark energy,
the growth factor is even somewhat smaller. The initial amplitude of density perturbations is very well known from
the CMB anisotropy measurements, $(\delta \rho /\rho)_{\rm i} = 5 \times 10^{-5}$.
Hence, a Universe without dark matter would still be pretty homogeneous:
the density contrast would be in the range of a few per cent. No structure
would have been formed, no galaxies, no life. No structure would be formed
in future either, as the accelerated expansion
 due to dark energy will soon terminate
the growth of perturbations.

Since dark matter particles decoupled from plasma much earlier
than baryons,
perturbations in dark matter started to grow much earlier.
The corresponding growth factor is larger than (\ref{bgrow}),
so that the dark matter density contrast at galactic and
subgalactic scales
becomes of order one, perturbations  enter the non-linear regime and form
dense dark matter clumps at $z = 5$--10. Baryons fall into
potential wells formed by dark matter, so dark matter and baryon
perturbations develop together soon after recombination. Galaxies get
formed in the regions where dark matter was overdense originally.
For this picture to hold, dark matter particles
must be non-relativistic early enough, as relativistic particles
fly through gravitational wells instead of being trapped there.
This means, in particular, that neutrinos cannot
constitute a considerable part
of dark matter.

\subsection{Cold and warm dark matter}
\label{subsec:C-and-W}

Currently, the most popular dark matter scenario is  cold
dark matter, CDM. It consists of particles which get out of
{\it kinetic} equilibrium when they are non-relativistic.
For dark matter particles Y which are initially in thermal equilibrium
with cosmic plasma, this means that their scattering off other particles
switches off at $T=T_{\rm d} \ll m_{\rm Y}$. Since then the dark matter
particles move freely, their momenta decrease due to red shift,
and they remain non-relativistic until now.
Note that the {\it decoupling}
temperature $T_{\rm d}$ may be much lower than the {\it freeze-out}
temperature $T_{\rm f}$ at which the dark matter particles get out of {\it chemical}
equilibrium, i.e., their number in the comoving volume freezes out
(because, e.g., their creation and annihilation processes switch off).
This is the case for many models with weakly interacting massive particles (WIMPs), a class of
dark matter particles we discuss in some detail below. Note also that
dark matter particles may never be in thermal equilibrium; this is the case,
e.g., for axions.

An alternative to CDM is {\it warm dark matter}, WDM, whose particles decouple,
being relativistic. Let us assume for definiteness
that they are in kinetic equilibrium
with cosmic plasma at temperature $T_{\rm f}$
when their number density freezes out
(thermal relic). After
kinetic equilibrium breaks down at temperature $T_{\rm d} \leq T_{\rm f}$,
their spatial momenta decrease as $a^{-1}$,
i.e., the momenta are  of order $T$ all the time after decoupling.
Warm dark matter particles become non-relativistic at $T\sim m$, where $m$ is their mass.
Only after that do the WDM perturbations start to
grow:
 as we mentioned above,
relativistic particles escape from gravitational potentials, so
the gravitational wells get smeared out instead of getting deeper.
Before becoming non-relativistic, WDM particles travel the distance of the
order of the horizon size; the WDM perturbations therefore are suppressed
at those scales.
The horizon size at the time $t_{\rm nr}$ when $T\sim m$ is of order
\[
   l_{\rm H}(t_{\rm nr}) \simeq
H^{-1} (T\sim m) = \frac{M_{\rm Pl}^*}{T^2} \sim  \frac{M_{\rm Pl}^*}{m^2}
\; .
\]
Due to the expansion of the Universe,
the corresponding length at present is
\be
  l_0 =  l_{\rm H}(t_{\rm nr}) \frac{a_0}{a(t_{\rm nr})} \sim
l_{\rm H}(t_{\rm nr}) \frac{T}{T_0} \sim \frac{M_{\rm Pl}}{m T_0} \; ,
\label{l0dwarf}
\ee
where we neglected  (rather weak) dependence on $g_*$.
Hence, in the WDM scenario,
 structures of sizes smaller than $l_0$ are less abundant
as compared to CDM. Let us point out that $l_0$ refers to the
size of the perturbation in the linear regime; in other
words, this is the size of the region from which matter collapses into a
compact object.

There is a hint towards the plausibility of warm, rather than cold,
dark matter. It is the dwarf-galaxy problem.  According
to numerical simulations, the CDM scenario tends to
overproduce small objects---dwarf galaxies:
it predicts hundreds of satellite
dwarf galaxies
in the vicinity of a
large galaxy like the Milky Way,
whereas only dozens of satellites have been observed so far.
This argument is still controversial, but, if correct, it does
suggest that the dark matter perturbations are suppressed
at dwarf-galaxy scales. This is naturally the case in the WDM
scenario.
The present size of a dwarf galaxy is a few kpc, and the density
is about $10^6$ of the average density in the Universe.
Hence, the size $l_0$ for these objects is of order $100~\mbox{kpc}
\simeq 3 \times 10^{23}~\mbox{cm}$. Requiring that perturbations of this
size,
but not much larger, are suppressed, we obtain from (\ref{l0dwarf})
the estimate for the mass of a dark matter particle
\be
 \mbox{WDM}\; : \;\;\;\; m_{\rm DM} = \mbox{3--10 keV} \; .
\label{wdmmass}
\ee
On the other hand, this effect is absent, i.e., dark matter is cold,
for
\be
\mbox{CDM}\; : \;\;\;\; m_{\rm DM} \gg 10~\mbox{keV} \; .
\label{cdmmass}
\ee
Let us recall that these estimates apply to particles that
are initially in kinetic equilibrium with cosmic plasma.
They do {\it not} apply in the opposite case; an example is
axion dark matter, which is cold despite being of very small axion mass.

\subsection{WIMP miracle}

There is a simple mechanism of the dark matter generation in the early
Universe. It applies to {\it cold} dark matter. Because of its simplicity
and robustness, it is considered by many as a very likely one,
and the corresponding dark matter candidates---WIMPs---as the best candidates. Let us describe this mechanism
in some detail.

Let us assume that there exists a heavy stable neutral particle Y, and
that
Y particles can only be destroyed or created via their
pair annihilation or creation, with annihilation products being
the particles of the Standard Model.
The general scenario for the
cosmological behaviour of Y particles is as follows.
At high temperatures, $T \gg m_{\rm Y}$, the
Y particles are in thermal equilibrium with
the rest of the cosmic plasma; there are lots of Y particles
in the plasma, which are continuously created and annihilate.
As the temperature drops below $m_{\rm Y}$, the equilibrium number
density decreases.
 At some `freeze-out' temperature $T_{\rm f}$, the
number density becomes so small that Y particles can no longer
meet each other during the Hubble time, and their annihilation
terminates. After that the number density of  surviving
Y particles decreases like $a^{-3}$, and these relic particles
contribute to the mass density in the present Universe.

Let us
estimate the properties of Y particles such that they really
serve as dark matter.
Elementary considerations of mean free path of a particle in gas
give for the lifetime of a non-relativistic
Y particle in cosmic plasma, $\tau_{\rm ann}$,
\[
      \langle \sigma_{\rm ann}\cdot v \rangle
\cdot \tau_{\rm ann} \cdot n_{\rm Y} \sim 1 \; ,
\]
where $v$ is the relative velocity of Y particles,
$\sigma_{\rm ann}$ is the annihilation cross-section at velocity $v$,
averaging is over the velocity distribution of Y particles
and $n_{\rm Y}$ is the number density. In thermal equilibrium
at $T\ll m_{\rm Y}$,
the latter is given by the Boltzmann
law at zero chemical potential,
\be
   n^{(\rm eq)}_{\rm Y} = g_{\rm Y} \cdot \left(\frac{m_{\rm Y} T}{2\pi} \right)^{3/2}
\mbox{e}^{-\frac{m_{\rm Y}}{T}}\; ,
\label{mar26-15-2}
\ee
where $g_{\rm Y}$ is the number of spin states of a Y particle.
Let us introduce the notation
\[
      \langle \sigma_{\rm ann} \cdot v \rangle = \sigma_0
\]
(in kinetic equilibrium, the left-hand side is the thermal
average). If the
annihilation occurs in an s-wave, then $\sigma_0$ is a constant
independent of temperature; for a p-wave it is somewhat suppressed
at $T \ll m_{\rm Y}$, namely $\sigma_0 \propto v^2 \propto T/m_{\rm Y}$.
A quick way to come to correct estimate
is to compare the lifetime with the Hubble
time, or the annihilation rate $\Gamma_{\rm ann} \equiv \tau^{-1}_{\rm ann}$
with the expansion rate $H$.
At $T \sim m_{\rm Y}$, the equilibrium density is of order
$n_{\rm Y} \sim T^3$, and $\Gamma_{\rm ann} \gg H$ for not too small $\sigma_0$.
This means that annihilation (and, by reciprocity, creation) of
$Y$ pairs is indeed rapid, and Y particles are indeed in complete
thermal equilibrium with the plasma. At very low temperature, on the
other hand, the equilibrium
number density $n_{\rm Y}^{(\rm eq)}$ is exponentially small, and the
equilibrium rate is small too,
$\Gamma_{\rm ann}^{(\rm eq)} \ll H$. At low temperatures we cannot, of course,
make
use of the equilibrium formulas: Y particles no longer annihilate
(and, by reciprocity, are no longer created), there is no thermal
equilibrium with respect to creation--annihilation processes
and the number density $n_{\rm Y}$ gets diluted only because of the
cosmological expansion.

The freeze-out temperature $T_{\rm f}$ is determined by the
relation\footnote{In fact, we somewhat oversimplify the
analysis here. The chemical equilibrium breaks down slightly earlier
than what we find from Eq.~\eqref{mar25-15-1}: the corresponding temperature
is obtained by equating the equilibrium
creation--annihilation rate $\Gamma_{\rm ann}$ to
the rate of evolution of the equilibrium
number density \eqref{mar26-15-2}, rather than to the Hubble parameter $H$.
For $T\ll m_{\rm Y}$, this gives the equation for the temperature
\[
\Gamma_{\rm ann} \simeq \frac{\dot{n}_{\rm Y}}{n_{\rm Y}} \simeq - \frac{m_{\rm Y}}{T}
\frac{\dot{T}}{T} = \frac{m_{\rm Y}}{T} H (T) \; .
\]
This temperature differs by the log--log correction
from $T_{\rm f}$ determined from Eq.~\eqref{mar26-15-5} and, at this temperature,
one has $n_{\rm Y} \gg T^2/(M_{\rm Pl}^* \sigma_0)$, cf. Eq.~\eqref{mar26-15-6}.
However, below this temperature, the annihilation of Y particles
continues, and it terminates at temperature $T_{\rm f}$ determined by
Eq.~\eqref{mar25-15-1}, which gives Eqs.~\eqref{dec1} and \eqref{mar26-15-6}.
All this gives rise to log--log corrections, which we do not calculate anyway.
So, our estimate for the present dark matter mass density remains valid.}
\be
  \tau_{\rm ann}^{-1} \equiv \Gamma_{\rm ann} \simeq H \; ,
\label{mar25-15-1}
\ee
where we  use the equilibrium formulas.
Making use of the relation \eqref{mar25-15-10} between the Hubble
parameter and the temperature at radiation domination, we obtain
\begin{equation}
      \sigma_0 (T_{\rm f}) \cdot n_{\rm Y} (T_{\rm f}) \sim  \frac{T_{\rm f}^2}{M_{\rm Pl}^{*}} \;
\label{dec1}
\end{equation}
or
\be
   \sigma_0 (T_{\rm f})
\cdot g_{\rm Y} \cdot \left(\frac{m_{\rm Y} T_{\rm f}}{2\pi} \right)^{3/2}
\mbox{e}^{-\frac{m_{\rm Y}}{T_{\rm f}}} \sim \frac{T_{\rm f}^2}{M_{\rm Pl}^{*}} \; .
\label{mar26-15-5}
\ee
The latter equation gives the freeze-out temperature, which, up to
log--log corrections, is
\be
   T_{\rm f} \approx \frac{m_{\rm Y}}{\ln (M_{\rm Pl}^{*} m_{\rm Y} \sigma_0)} \;
\label{may10-1}
\ee
(the possible dependence of $\sigma_0$ on temperature
is irrelevant in the right-hand side: we are doing the
calculation in the leading-log approximation anyway).
Note that this temperature is somewhat lower than $m_{\rm Y}$ if the
relevant microscopic mass scale is much below $M_{\rm Pl}$. This means
that Y particles freeze out when they are indeed non-relativistic
and get out of kinetic equilibrium at even lower temperature, hence
the term `cold dark matter'. The fact that the annihilation and
creation of Y particles terminate at a relatively low temperature has
to do with the rather slow expansion of the Universe, which should be
compensated for by the smallness of the number density $n_{\rm Y}$.

At the freeze-out temperature, we make use of Eq.~(\ref{dec1})
and obtain
\be
   n_{\rm Y} (T_{\rm f}) = \frac{T_{\rm f}^2}{M_{\rm Pl}^{*} \sigma_0 (T_{\rm f})} \; .
\label{mar26-15-6}
\ee
Note that this density is inversely proportional to the annihilation
cross-section (modulo a logarithm). The reason is that
for higher annihilation cross-sections, the creation--annihilation
processes are longer in equilibrium, and fewer Y particles survive.

Up to a numerical factor of order 1, the number-to-entropy ratio
at freeze-out
is
\be
   \frac{n_{\rm Y}}{s} \simeq \frac{1}{g_* (T_{\rm f}) M_{\rm Pl}^{*}T_{\rm f} \sigma_0(T_{\rm f})}  \; .
\label{nYs}
\ee
This ratio stays constant until the present time, so
 the present number density of
Y particles is
$   n_{\rm Y, 0} = s_0 \cdot
\left(n_{\rm Y}/s \right)_{\rm \mbox{freeze-out}}$,
and the mass-to-entropy ratio is
\be
  \frac{\rho_{\rm Y, 0}}{s_0} = \frac{m_{\rm Y} n_{\rm Y,0}}{s_0}
   \simeq
\frac{\ln (M_{\rm Pl}^{*} m_{\rm Y} \sigma_0)}{g_*(T_{\rm f}) M_{\rm Pl}^{*} \sigma_0(T_{\rm f})}
\simeq \frac{\ln (M_{\rm Pl}^{*} m_{\rm Y} \sigma_0)}{\sqrt{g_*(T_{\rm f})} M_{\rm Pl} \sigma_0 (T_{\rm f})} \; ,
\nonumber
\ee
where we made use of (\ref{may10-1}).
This formula is remarkable. The mass density depends mostly on
one parameter, the annihilation cross-section $\sigma_0$. The
dependence on the mass
of a Y particle is through the logarithm and through
$g_* (T_{\rm f})$; it is very
mild.
The value of the logarithm here is between 30 and 40, depending on
parameters
(this means, in particular, that freeze-out occurs when the
temperature drops 30 to 40 times below the mass of a Y particle).
Inserting
$g_* (T_{\rm f}) \sim 100$, as well as the numerical factor omitted
in Eq.~(\ref{nYs}), and comparing with (\ref{10p*}), we obtain the estimate
\be
   \sigma_0 (T_{\rm f}) \equiv \langle \sigma v \rangle (T_{\rm f})
= \mbox{(1--2)} \times 10^{-36}~\mbox{cm}^2 \; .
\label{estim}
\ee
This is a weak-scale cross-section,
which tells us that the relevant energy scale
is TeV. We note in passing that the estimate
(\ref{estim}) is quite precise and robust.

If the annihilation occurs in an s-wave,
the annihilation cross-section may be parametrized as
$   \sigma_0 = \alpha^2/ M^2$,
where $\alpha$ is some coupling constant and $M$ is
a mass scale (which may be higher than
$m_{\rm Y}$). This parametrization is suggested by the picture of
Y-pair annihilation via the exchange by another particle of mass
$M$. With $\alpha \sim
10^{-2}$, the estimate for the mass scale is roughly
$ M \sim 1~\mbox{TeV}$.
Thus, with very mild assumptions, we find that the
non-baryonic dark matter may naturally originate from
the TeV-scale physics. In fact, what we have found can be understood as
an approximate equality between the cosmological parameter, the mass-to-entropy
ratio of dark matter and the particle-physics parameters,
\[
\mbox{mass-to-entropy} \simeq \frac{1}{M_{\rm Pl}}
\l \frac{\mbox{TeV}}{\alpha_{\rm W}} \r^2 \; .
\]
Both are of order $10^{-10}~\mbox{GeV}$, and it is very tempting
to think that this `WIMP miracle' is not a mere coincidence.
If it is not, the dark matter particles should be found
at the LHC.

The most prominent candidate
for WIMPs is neutralinos of the supersymmetric extensions of the Standard
Model. The situation with neutralinos is somewhat tense, however.
The point is that the pair annihilation of neutralinos often occurs
in the p-wave, rather than the s-wave. This gives the suppression factor
in $\sigma_0 \equiv \langle \sigma_{\rm ann} v\rangle$
proportional to $v^2 \sim T_{\rm f} / m_{\rm Y} \sim 1/30$.
Hence, neutralinos tend to be overproduced in most of the parameter
space of the Minimal Supersymmetric Standard Model (MSSM) and other models. 
Yet neutralinos remain a good candidate,
especially at high $\tan \beta$.

\begin{figure}[htb!]
\begin{center}
\includegraphics[angle=-90,width=0.7\textwidth]{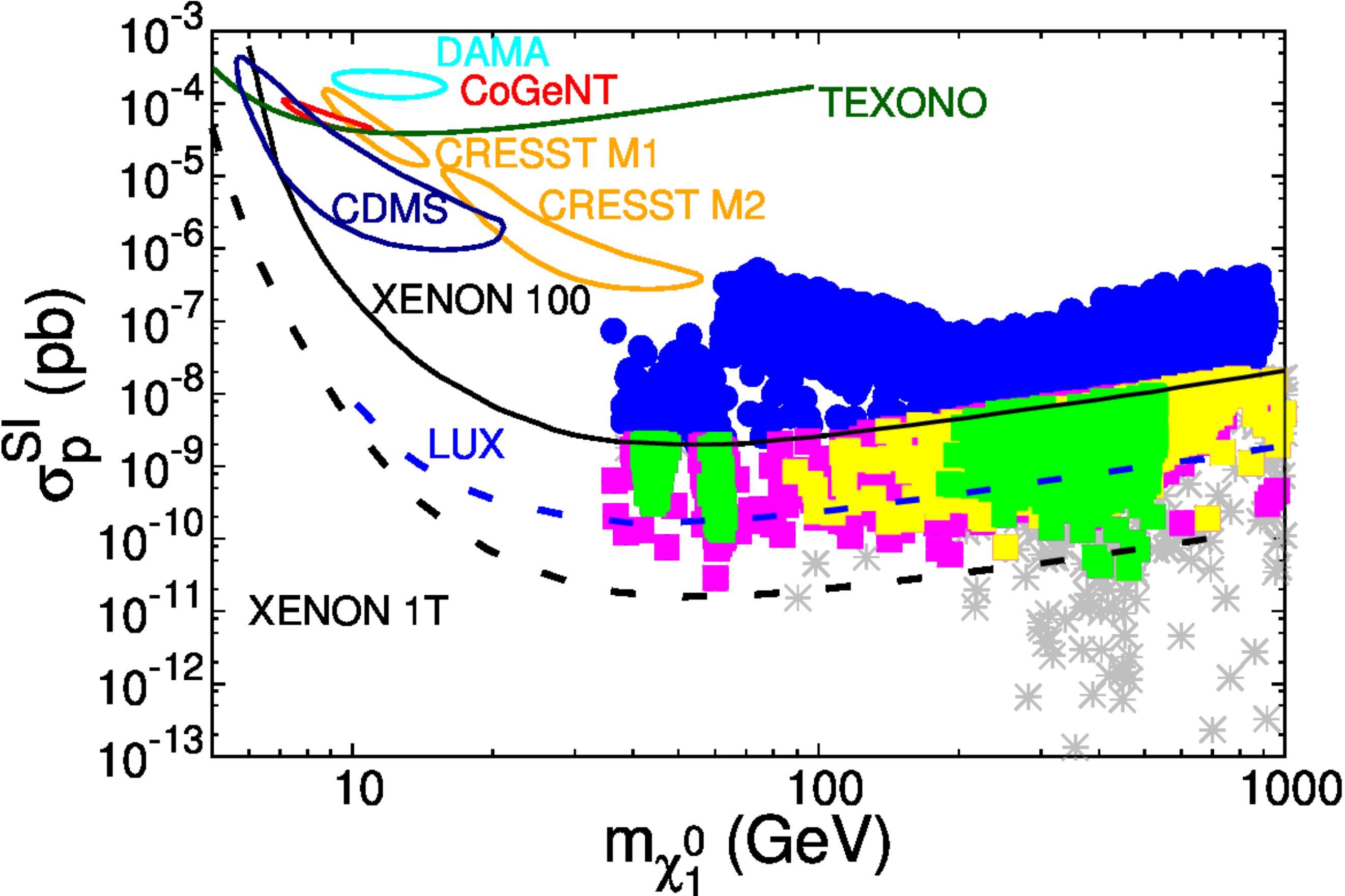}
\end{center}
\caption{
MSSM predictions for spin-independent elastic neutralino--nucleon cross-section versus neutralino mass
and experimentally excluded regions~\cite{Han:2013gba}.
Shaded regions correspond to MSSM parameters consistent
with collider limits and yielding
$\Omega_{\rm DM} \approx 0.25$. Regions above the open
solid lines are ruled out by direct searches, closed solid curves
correspond to regions favoured by experiments indicated.
Dashed lines are sensitivities of future direct search experiments
LUX and XENON 1T.
\label{chap8new-fig2}
}
\end{figure}

\begin{figure}[htb!]
\centerline{\includegraphics[width=0.80\textwidth]{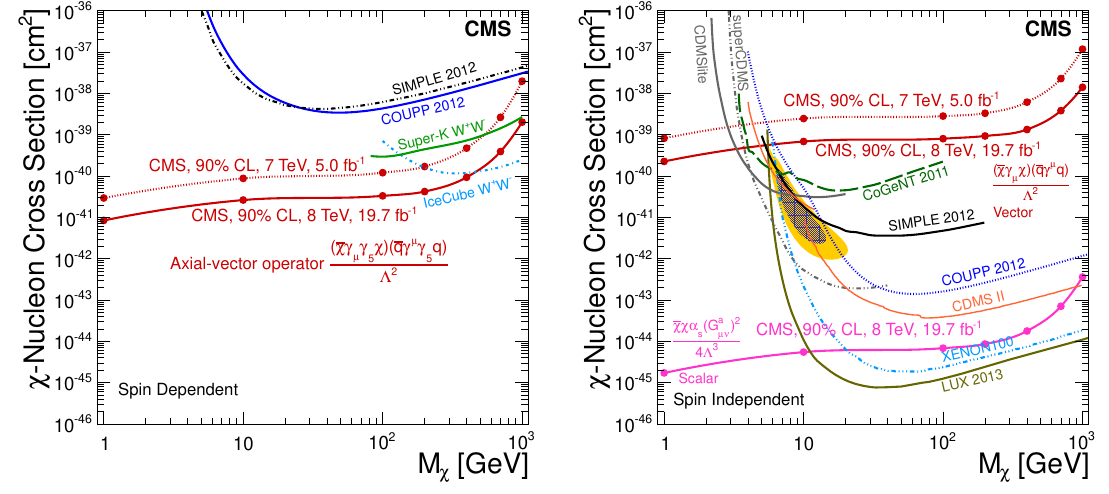}
}
\caption{Excluded regions in the parameter space
$(M_{\rm X},\sigma_{\rm pX})$ \cite{Khachatryan:2014rra} for spin-dependent
(left) and spin-independent (right)
WIMP interactions with nucleons.
Regions above the curves
are ruled out at 90\,\% confidence level.
CMS denotes searches for WIMPs at the LHC
(assuming contact interaction
YYf$_1$f$_2$, where f$_{1,2}$ are Standard Model fermions);
 IceCube and Super-K are searches for neutrinos from
WIMP annihilation in the Sun; others are direct searches.
The shaded region in the middle of the right-hand panel is
favoured by a possible signal at the
CDMS experiment.
\label{chap8-fig1}
}
\end{figure}

A direct search for dark matter WIMPs is underway in underground
laboratories. The idea is that WIMPs orbiting around the centre of our
Galaxy with velocity of order $10^{-3}$ sometimes hit a nucleus in a
detector and deposit a small energy in it. These searches have become
sensitive to neutralinos, as shown in Fig.~\ref{chap8new-fig2}.
Indirect searches for dark matter WIMPs include the search for
neutrinos coming from the centres of the Earth and Sun
(WIMPs may concentrate and annihilate there), see, e.g.,
Ref.~\cite{Avrorin:2014swy} and positrons and
antiprotons in cosmic rays (produced in WIMP annihilations
in our Galaxy), see, e.g., Ref.~\cite{Bergstrom:2013jra}.
Collider searches are sensitive to WIMPs too,
see Fig.~\ref{chap8-fig1}. We conclude that the hunt for WIMPs has
entered the promising stage.

\vspace{0.3cm}
\noindent
{\it Question.} Estimate the energy deposited in the XENON
detector due to elastic scattering of a dark matter WIMP,
for WIMP masses 10~GeV, 100~GeV and 1~TeV. Estimate
the number of events per kilogram per year for the same masses
and elastic cross-sections $10^{-5}$~pb, $10^{-9}$~pb and
$10^{-8}$~pb, respectively (see Fig.~\ref{chap8new-fig2}), assuming
that the WIMP mass density around the Earth is similar to the
average baryon mass density, $\rho_{\rm DM} \sim 0.3~\mbox{GeV cm}^{-3}$,
and that $v_{\rm DM} \sim 10^{-3}$.
\vspace{0.3cm}

\subsection{Light long-lived particles}

Many extensions of the Standard Model contain light
scalar or pseudoscalar particles. In some models these
new particles are  so weakly interacting that
their lifetime exceeds the present age of the Universe. Hence,
they may serve as dark matter candidates. The best motivated of them is
the axion, but there is an entire zoo of axion-like particles.

Let us consider general properties of models with light
scalars or pseudoscalars. These particles should interact
with the usual matter very weakly, so they must be neutral
with respect to the Standard Model gauge interactions.
This implies that interactions of scalars
$S$ and pseudoscalars $P$ with gauge fields are of the
form


\begin{equation}
\label{chap8-axion-2+}
{\cal L}_{SFF}=\frac{C_{SFF}}{4\Lambda}\cdot {S}F_{\mu\nu}F^{\mu\nu}\;,~~~~~
{\cal L}_{PFF}=\frac{C_{PFF}}{8\Lambda}\cdot {P}F_{\mu\nu}F_{\lambda\rho}
\epsilon^{\mu\nu\lambda\rho}
\;,
\end{equation}
where $F_{\mu\nu}$ is the field strength of the
SU$(3)_{\rm c}$, SU$(2)_{\sm W}$ or U$(1)_{\sm Y}$ gauge group.
The parameter
$\Lambda$ has dimension of mass and can be interpreted as
the scale of new physics related to an
$S$ and/or a $P$ particle.
This parameter has to be large; then the interactions
of
$S$ and $P$ with gauge bosons are indeed weak at low energies.
Because of that, the Lagrangians
\eqref{chap8-axion-2+}  contain gauge-invariant
operators of the lowest possible dimension.
Dimensionless constants $C_{SFF}$ and
$C_{PFF}$ are typically
numbers of order 1.
The terms
\eqref{chap8-axion-2+} describe interactions of (pseudo)scalars
with pairs of photons, gluons as well as with Z$\gamma$,
ZZ and W$^+$W$^-$ pairs.

Interactions with fermions can also be written on symmetry grounds.
Since
$S$ and
$P$ are singlets under
SU$(3)_{\rm c}\times {\rm SU}(2)_{\sm
W}\times {\rm U}(1)_{\sm Y}$, no combinations like
${S\bar f f}$ or
${P\bar f\gamma^5f}$ are gauge invariant, so they
cannot appear in the Lagrangian (hereafter $f$ denotes the
Standard Model fermions).
Gauge-invariant operators of the lowest dimension
have the form
$H\bar f f$, where $H$ is the Englert--Brout--Higgs field. Hence,
the interactions with fermions are
\[
{\cal L}{SHff}=\frac{Y_{SHff}}{\Lambda}\cdot SH\bar f f\;,~~~~~
{\cal L}{PHff}=\frac{Y_{PHff}}{\Lambda}\cdot PH\bar f\gamma^5f\;.
\]
It often happens that the couplings
$Y_{SHff}$ and $Y_{SPff}$ are of the order of the Standard Model
Yukawa couplings, so upon electroweak symmetry breaking
the low-energy Lagrangians have the following structure:
\begin{equation}
\label{chap8-axion-3+}
{\cal L}{Sff}=\frac{C_{Sff}m_f}{\Lambda}\cdot S\bar f f\;,~~~~~
{\cal L}{Pff}=\frac{C_{Pff}m_f}{\Lambda}\cdot P\bar f\gamma^5f\;,
\end{equation}
where we assume that the dimensionless couplings $C_{Sff}$ and $C_{Pff}$
are also of order 1.

Making use of
Eqs.~\eqref{chap8-axion-2+} and \eqref{chap8-axion-3+},
we estimate the partial widths of decays of
$P$ and $S$ into the Standard Model particles:
\begin{equation}
\label{chap8-axion-4+}
\Gamma_{P(S)\to AA}\sim \frac{ m^3_{P(S)}}{64\pi \Lambda^2}\;,~~~~~
\Gamma_{P(S)\to ff}\sim\frac{m^2_fm_{P(S)}}{8\pi \Lambda^2}\;,
\end{equation}
where $A$ denotes vector bosons.
By requiring that the lifetime of the new particles exceeds
the present age of the Universe,
$\tau_{S(P)}=\Gamma^{-1}_{S(P)}> H^{-1}_0 $, we find a bound on
the mass
of the dark matter candidates,
\begin{equation}
\label{chap8-saxion-4+}
m_{P(S)}<\l16\pi \Lambda^2H_0\r^{1/3}\;.
\end{equation}
Assuming that the new physics scale is below the Planck scale,
$\Lambda<M_{\rm Pl}$,
we obtain an (almost) model-independent bound,
\begin{equation}
\label{chap8-axion-add-L}
m_{P(S)}<100~\mbox{MeV}\;.
\end{equation}
Hence, the kinematically allowed decays are ${P(S)} \to \gamma \gamma$,
 ${P(S)} \to \nu \bar{\nu}$ and  ${P(S) \to {\rm e}^+ {\rm e}^-}$.
It follows from Eq.~\eqref{chap8-axion-4+} that the two-photon decay
mode dominates,
unless the mass of the new particle is close to that of the electron.

Let us now consider generation of relic (pseudo)scalars
in the early Universe. There are several generation mechanisms;
one of them is fairly generic for the class of models we discuss.
This is generation in decays of condensates (we
will consider  another mechanism later,
in the
model with axions).
The picture is as follows.
Let some scalar field
$\phi$
be in a condensate
in the early Universe. The condensate can be viewed as  a collection of
$\phi$ particles at rest. Equivalently,
the condensate is the homogeneous scalar field that
oscillates at relatively late times, when $m_\phi > H$.
Let both particles,
$\phi$ and $S$, interact with matter so weakly that they
never get into thermal equilibrium, and
let the interaction between $\phi$ and $S$ have the form
$\mu \phi S^2/2$, where
$\mu$ is the coupling constant.
Then the width of the decay
$\phi\to SS$
is estimated as
\begin{equation}
\label{chap8-axion-add-218a*}
\Gamma_{\phi\to SS}\sim \frac{\mu^2}{16\pi m_\phi}\;.
\end{equation}
If the widths of other decay channels do not exceed
the value
\eqref{chap8-axion-add-218a*}, the decay of
the $\phi$ condensate occurs at a temperature
$T_\phi$ determined by
\[
\Gamma_{\phi\to SS}\sim H(T_\phi)=\frac{T_\phi^2}{M_{\rm Pl}^*}\;.
\]
Let the energy density of the $\phi$ condensate at that time
be equal to $\rho_\phi$, so that the number density of decaying
$\phi$ particles is $n_\phi \sim \rho_\phi/m_\phi$.
Immediately after the epoch of
$\phi$-particle decays, the number
density of $S$ particles is of  order $\epsilon \rho_\phi/m_\phi$,
where $\epsilon$ is the fraction of the condensate that
decayed into $S$ particles.
After $S$ particles become non-relativistic,
their mass density is of order
\[
\rho_{S}\sim\epsilon\rho_\phi\cdot\frac{m_{S}T^3}{m_\phi T_\phi^3}\;,
\]
where we omitted the dependence on $g_*$ for simplicity.
In this way we estimate the mass fraction of $S$ particles today,
\begin{equation}
\label{chap8-axion-general-case-3}
\Omega_{S}=\frac{\rho_{S}}{\rho_{\rm c}}\sim
\frac{m_{S}T_0^3}{\rho_{\rm c}}\cdot\frac{\epsilon\rho_\phi}{m_\phi T_\phi^3}
\sim
0.2\cdot
\l\frac{m_{\sm
S}}{1~\mbox{eV}}\r\cdot\frac{\epsilon\rho_\phi}{m_\phi T_\phi^3}\; .
\end{equation}
With an appropriate choice of parameters, the correct value
$\Omega_{S} \simeq 0.2$ can indeed be obtained.
We note that the last factor on the right-hand side of
Eq.~\eqref{chap8-axion-general-case-3} must be small.

\subsection{Axions}
\label{subs:axions}

Let us now turn to a concrete class of models
with Peccei--Quinn symmetry and
axions. This symmetry provides a
 solution to the {\it strong CP-problem, } and the existence
of axions is an inevitable consequence of the construction.

The strong
CP-problem~\cite{'tHooft:1976up,Callan:1976je,Jackiw:1976pf}
emerges in the following way.
One can extend the Standard Model
Lagrangian by adding the following term:
\begin{equation}
\label{chap8-true-axion-1+}
\Delta L=\frac{\alpha_{\rm s}}{8\pi}\cdot\theta_0\cdot G_{\mu\nu}^{\rm a}\tilde
G^{\mu\nu\;{\rm a}}\;,
\end{equation}
where $\alpha_{\rm s}$ is the SU$(3)_{\rm c}$ gauge coupling,
$G_{\mu\nu}^{\rm a}$ is the gluon field strength,
$\tilde
G^{\mu\nu\;{\rm a}} = \frac{1}{2} \epsilon^{\mu\nu\lambda\rho}G^{\rm a}_{\lambda\rho}$
is the dual tensor and
$\theta_0$ is an arbitrary dimensionless parameter
(the factor $\alpha_{\rm s}/(8\pi)$ is introduced for later convenience).
The interaction term
\eqref{chap8-true-axion-1+} is invariant under gauge
symmetries of the Standard Model, but it violates
P and CP.
The term \eqref{chap8-true-axion-1+}
is a total derivative, so it
does not contribute to the
classical field equations, and its contribution to the action
is reduced to the surface integral. For any perturbative
gauge field configurations (small perturbations about
$G^{\rm a}_\mu=0$), this contribution is equal to zero.
However, this is not the case for configurations of
instanton
type.
This means that CP is violated in QCD at the non-perturbative level.

Furthermore, quantum effects due to quarks give rise
to the anomalous term in the Lagrangian,
which has the same form as
Eq.~\eqref{chap8-true-axion-1+} with proportionality coefficient
determined by the phase of the quark mass matrix
$\hat M_{\rm q}$. The latter enters the Lagrangian as
\[
{\cal L}{\rm m} =  \bar{q}_{\rm L} \hat{M}_{\rm q} q_{\rm R} + \mbox{h.c.} \;
\]
By chiral rotation of quark fields, one
makes quark masses real (i.e., physical), but that rotation
induces a new term in the Lagrangian,
\begin{equation}
\label{chap8-true-axion-3+}
\Delta {\cal L}{\rm m}=\frac{\alpha_{\rm s}}{8\pi}\cdot {\rm Arg}\l {\rm Det} \hat M_{\rm q}\r
\cdot G_{\mu\nu}^{\rm a}\tilde G^{\mu\nu\;{\rm a}}\;.
\end{equation}
There is no reason to think that
${\rm Arg}\l {\rm Det}
\hat M_{\rm q}\r=0$. Neither there is a reason to think that
the `tree-level' term
\eqref{chap8-true-axion-1+} and the anomalous contribution
\eqref{chap8-true-axion-3+} cancel each other. Indeed, the former term
is there even in the absence of quarks, while the latter comes
from the Yukawa sector, as the quark masses are due to
their Yukawa interactions with the Englert--Brout--Higgs field.

Thus, the Standard Model Lagrangian should contain the term
 \be
\Delta {\cal L}\theta= \frac{\alpha_{\rm s}}{8\pi}
\l \theta_0+{\rm Arg}\l {\rm Det} \hat M_{\rm q}\r  \r G_{\mu\nu}^{\rm a}\tilde
G^{\mu\nu\;{\rm a}} \equiv \frac{\alpha_{\rm s}}{8\pi}\cdot\theta\cdot
G_{\mu\nu}^{\rm a}\tilde G^{\mu\nu\;{\rm a}} \;.
\label{chap8-true-axion-4+}
\ee
This term violates CP, and off hand the parameter
$\theta$ is of order 1.

The term \eqref{chap8-true-axion-4+} has
non-trivial phenomenological consequences.
One  is that it generates
the electric dipole moment (EDM) of the neutron,
$d_{\rm n}$, which is estimated as~\cite{Kim:2008hd}
\begin{equation}
\label{chap8-true-axion-4++}
d_{\rm n}\sim \theta\times 10^{-16}\, e \mbox{ cm}\;.
\end{equation}
The neutron EDM has not been found experimentally, and the
searches place a strong bound
\begin{equation}
\label{chap8-true-axion-4*}
d_{\rm n}\lesssim 3\times 10^{-26}\, e \mbox{ cm}\; .
\end{equation}
This leads to the bound on the parameter
 $\theta$,
\[
|\theta|<0.3 \times 10^{-9}\;.
\]
The problem to explain such a small value of
$\theta$ is precisely the strong
CP-problem.

A solution to this  problem  does not exist
within the Standard Model.
The solution is offered by models with axions.
These models make use of the following observation.
If at the classical level
the quark Lagrangian is invariant under axial symmetry
U$(1)_A$ such that
\begin{equation}
\label{chap8-true-axion-5+}
q_{\sm L}\to \e^{{\rm i}\beta} q_{\sm L}\;,~~~~~
q_{\sm R}\to \e^{-{\rm i}\beta} q_{\sm R}\;,
\end{equation}
then the $\theta$ term would be rotated away by
applying this transformation. This global
symmetry is called the
Peccei--Quinn (PQ) symmetry~\cite{Peccei:1977hh},
U$(1)_{\rm PQ}$. There is no PQ symmetry
 in
the Standard Model, but
one can extend the Standard Model in such a way that
the classical Lagrangian is invariant under
the PQ symmetry. Quark masses
are not invariant under the PQ transformations
\eqref{chap8-true-axion-5+}, so PQ symmetry is {\it spontaneously
broken}. At the classical level, this leads to the existence
of a massless Nambu--Goldstone field
$a(x)$, an axion. As for any Nambu--Goldstone field, its properties
are determined by its transformation law under the PQ symmetry:
\begin{equation}
\label{chap8-true-axion-R+216a*}
a(x)\to a(x)+\beta\cdot f_{\rm PQ}\;,
\end{equation}
where $\beta$ is the same parameter as in
Eq.~\eqref{chap8-true-axion-5+} and $f_{\rm PQ}$ is a constant of dimension
of mass, the energy scale of
U$(1)_{\rm PQ}$ symmetry breaking. The mass terms
in the low-energy quark
Lagrangian must be symmetric under the transformations
\eqref{chap8-true-axion-5+} and
\eqref{chap8-true-axion-R+216a*}, so the quark and axion fields enter
the Lagrangian in the combination
\begin{equation}
\label{chap8-true-axion-R+216b*}
{\cal L}{\rm m}=\bar q_{\sm R}m_{\rm q} \e^{-2{\rm i}\frac{a}{f_{\rm PQ}}}q_{\sm L}+\mbox{h.c.}
\end{equation}
Making use of
Eq.~\eqref{chap8-true-axion-3+}, we find that at the quantum level
the low-energy Lagrangian contains the term
\begin{equation}
\label{chap8-true-axion-R+}
{\cal L}{\rm a}=C_{\rm g}\frac{\alpha_{\rm s}}{8\pi}\cdot\frac{a}{f_{\rm PQ}}
G_{\mu\nu}^{\rm a}\tilde G^{\mu\nu\;{\rm a}}\;,
\end{equation}
where the constant $C_{\rm g}$ is of order 1; it is determined
by PQ charges of quarks.
Clearly,
PQ symmetry \eqref{chap8-true-axion-5+} and \eqref{chap8-true-axion-R+216a*}
is {\it explicitly} broken by quantum effects of QCD, and an axion
is a {\it pseudo}-Nambu--Goldstone boson.

Hence, the
$\theta$ parameter multiplying the operator
$G_{\mu\nu}^{\rm a}\tilde
G^{\mu\nu\;{\rm a}}$ obtains a shift depending on the space--time point and
proportional to the axion field,
\begin{equation}
\label{chap8-axion-add-216+}
\theta\to\bar\theta(x)=\theta+C_{\rm g}\frac{a(x)}{f_{\rm PQ}}\;.
\end{equation}
Strong interactions would conserve CP provided the
axion vacuum expectation value is such that
$\la\bar\theta\ra=0$. The  QCD effects  indeed do the job.
They generate a non-vanishing quark condensate
 $\la\bar q q\ra\sim\Lambda_{\rm QCD}^3$ at the QCD energy scale
$\Lambda_{\rm QCD}\sim200$~MeV. This condensate breaks chiral
symmetry
and in turn generates  the
axion effective potential
\begin{equation}
\label{chap8-true-axion-eff+}
V_{\rm a}\sim -\frac{1}{2}\bar\theta^2\frac{m_{\rm u}m_{\rm d}}{m_{\rm u}+m_{\rm d}}\la \bar q q\ra +
{\cal O}(\bar\theta^4)\simeq
\frac{1}{8}\bar\theta^2\cdot m_\pi^2 f_\pi^2 +
{\cal O}(\bar\theta^4)\;,
\end{equation}
where $m_\pi = 135$~MeV and $f_\pi= 93$~MeV are pion mass and decay constant.
In fact, the axion potential must be
periodic in $\theta$ with period $2\pi$,
so the expression \eqref{chap8-true-axion-eff+} is valid for small $\theta$
only.
The potential has the minimum at
$\la\bar\theta\ra = 0$, so the strong CP-problem finds an
elegant solution. It follows from
Eqs.~\eqref{chap8-axion-add-216+} and
\eqref{chap8-true-axion-eff+} that the axion has a mass
\begin{equation}
\label{chap8-true-axion-R+216c*}
m_{\rm a}\approx C_{\rm g} \frac{m_\pi f_\pi}{2f_{\rm PQ}}\; ,
\end{equation}
i.e., it is indeed a {\it pseudo}-Nambu--Goldstone boson.

There are various  ways to implement the PQ mechanism.
One is to introduce two Englert--Brout--Higgs doublets
and choose the
Yukawa interaction as
\begin{equation}
\label{chap8-true-axion-6+}
Y^{\rm d}\bar Q_{\sm L} H_1 D_{\sm R}+
Y^{\rm u}\bar Q_{\sm L} {\rm i}\tau^2 H_2^* U_{\sm R}\;.
\end{equation}
The two scalar fields transform under the
U$(1)_{\rm PQ}$ transformation
\eqref{chap8-true-axion-5+} as follows:
\[
H_1\to\e^{2{\rm i}\beta}H_1\;,~~~~~H_2\to\e^{-2{\rm i}\beta}H_2\; .
\]
This ensures U$(1)_{\rm PQ}$ invariance of the Lagrangian
\eqref{chap8-true-axion-6+} and hence the absence of the
$\theta$ term.  Both scalars acquire vacuum expectation values
$v_1$ and $v_2$.
If no  other new fields are added,
we arrive at the Weinberg--Wilczek
model~\cite{Weinberg:1977ma,Wilczek:1977pj}.
In that case,
the axion field $\theta$ is the relative phase of
$H_1$ and $H_2$, and
the PQ scale equals the
electroweak scale:
\[
f_{\rm PQ}= 2\sqrt{v_1^2+v_2^2}= 2v_{\rm SM} =2\times 246~\mbox{GeV}\;.
\]
The axion is quite heavy,
$m_{\rm a}\sim 15~\mbox{keV}$, and its
interaction with quarks, gluons and photons
is too strong. Because of that, the Weinberg--Wilczek  axion
is experimentally ruled out.

This problem is solved in the
Dine--Fischler--Srednicki--Zhitnitsky (DFSZ) model~\cite{Dine:1981rt,Zhitnitsky:1980tq} by adding
a complex scalar field $S$
which is a singlet under the Standard Model gauge group.
Its interactions involve PQ invariants
\[
S^\dagger S\;,~~~H_1^\dagger H_2\cdot S^2 \; .
\]
The field $S$ transforms under
U$(1)_{\rm PQ}$ as
$S\to\e^{2{\rm i}\beta}S$.
The axion field is now a linear combination of the phases
of fields
$H_1$, $H_2$ and $S$ and
\begin{equation}
\label{chap8-true-axion-6*}
f_{\rm PQ}=2\sqrt{v_1^2+v_2^2+v_{\rm s}^2}\;,
\end{equation}
where $v_{\rm s}$ is the vacuum expectation value of the field $S$.
The latter can be large, so it is clear from
Eq.~\eqref{chap8-true-axion-6*} that the  mass of the axion is small and,
most importantly, its couplings to the Standard Model fields
are weak: these couplings
are inversely proportional to
$f_{\rm PQ}\sim v_{\rm s}$.
The DFSZ axion interacts  with  both quarks and leptons.

Another approach is called
the Kim--Shifman--Vainshtein--Zakharov (KSVZ) mechanism~\cite{Kim:1979if,Shifman:1979if}. It does not require more than one
Englert--Brout--Higgs field of the Standard Model.
The  mechanism makes use of additional quark fields
$\Psi_{\sm R}$ and $\Psi_{\sm L}$, which are triplets under
SU$(3)_{\rm c}$ and singlets under
SU$(2)_{\sm W}\times {\rm U}(1)_{\sm Y}$. Only these quarks
transform non-trivially under
U$(1)_{\rm PQ}$, while the usual quarks have zero PQ charge.
One also introduces a complex scalar field
$S$, which is a singlet under the Standard Model gauge group.
One writes the PQ-invariant Yukawa interaction of the new fields,
\[
L = y_\Psi S\bar \Psi_{\sm R} \Psi_{\sm L} + \mbox{h.c.} \; ,
\]
so that
$S$ again transforms under
U$(1)_{\rm PQ}$ as $S\to\e^{2{\rm i}\beta}S$.
PQ symmetry is spontaneously broken by the vacuum expectation
value
$ \la S\ra = v_{\rm s}/\sqrt{2}$. The axion here is the phase of the
field
$S$; therefore,
\begin{equation}
\label{chap8-axion-add-217+}
f_{\rm PQ}=2v_{\rm s}\;.
\end{equation}
The KSVZ model does not contain an explicit interaction of an axion with
the usual quarks and leptons.

To summarize, an axion is a light particle whose interactions with the
Standard Model fields are very weak.  The latter
property  relates to the fact that it is a pseudo-Nambu--Goldstone boson
of a global symmetry spontaneously broken at the high-energy
scale
$f_{\rm PQ}\gg M_{\rm W}$.
As for any Nambu--Goldstone field, the interactions of an axion
with quarks and leptons
are described by
the generalized Goldberger--Treiman formula
\begin{equation}
\label{chap8-true-axion-add-1+RR+}
{\cal L}{\rm af}=\frac{1}{f_{\rm PQ}}\cdot\d_\mu a\cdot J^\mu_{\rm PQ}\; .
\end{equation}
Here
\begin{equation}
\label{chap8-true-axion-add-1+}
J^\mu_{\rm PQ}=\sum_{\rm f} e^{({\rm PQ})}_{\rm f}\cdot\bar f\gamma^\mu\gamma^5 f\;.
\end{equation}
The contributions of fermions to the current
$J^\mu_{\rm PQ}$ are proportional to their PQ charges
$e^{({\rm PQ})}_{\rm f}$;
these charges are model-dependent. In accord with
Eq.~\eqref{chap8-true-axion-R+216b*},
the action \eqref{chap8-true-axion-add-1+RR+}
can be integrated by parts and we obtain instead
\begin{align}
{\cal L}{\rm af}&=-\frac{1}{f_{\rm PQ}}\cdot a\cdot \d_\mu J^\mu_{\rm PQ}
\nonumber \\
&=
-\frac{a}{f_{\rm PQ}}\cdot
\sum_f 2e^{({\rm PQ})}_{f}m_{f}\cdot\bar f\gamma^5 f
\; .
\label{chap8-true-axion-add-1++}
\end{align}
Besides the interaction
\eqref{chap8-true-axion-add-1+RR+}, there are also interactions
of axions with gluons, see
Eq.~\eqref{chap8-true-axion-R+}, and photons,
\begin{equation}
\label{chap8-true-axion-add-1*}
{\cal L}{\rm ag}=C_{\rm g}\frac{\alpha_{\rm s}}{8\pi}\cdot\frac{a}{f_{\rm PQ}}\cdot
G_{\mu\nu}^{\rm a} \tilde G^{\mu\nu\;{\rm a}}
\;,~~~~~
{\cal L}{\rm a\gamma}=C_\gamma\frac{\alpha}{8\pi}\cdot\frac{a}{f_{\rm PQ}}\cdot
F_{\mu\nu} \tilde F^{\mu\nu}
\;,
\end{equation}
where the dimensionless constants
$C_{\rm g}$ and $C_\gamma$ are also model-dependent and, generally speaking,
are of order 1.
The interaction terms \eqref{chap8-true-axion-add-1++} and
\eqref{chap8-true-axion-add-1*}
indeed have the form
\eqref{chap8-axion-2+} and \eqref{chap8-axion-3+},
i.e., models with axions belong to
the class of models with light, weakly interacting pseudoscalars.
The axion mass, however, is not a free parameter: we find from
Eq.~\eqref{chap8-true-axion-R+216c*} that
\begin{equation}
\label{chap8-true-axion-add-2+}
m_{\rm a}\approx m_\pi\cdot\frac{f_\pi}{2f_{\rm PQ}}\approx 0.6~\mbox{eV}~\cdot
\l
\frac{10^7~\mbox{GeV}}{f_{\rm PQ}}
\r
\;.
\end{equation}

The main decay channel of the light axion is decay into two
photons. The lifetime
$\tau_{\rm a}$ is found from
Eq.~\eqref{chap8-axion-4+} by setting $\Lambda=2\pi
f_{\rm PQ}/\alpha$ and using Eq.~\eqref{chap8-true-axion-add-2+},
\[
\tau_{\rm a}=\frac{1}{\Gamma_{{\rm a}\to\gamma\gamma}}
=\frac{64\pi^3 m_\pi^2f_\pi^2}{\alpha^2m_{\rm a}^5}
\simeq 4\times
10^{24}~\mbox{s}~\cdot\l\frac{\mbox{eV}}{m_{\rm a}}\r^5\;.
\]
By requiring that this lifetime exceeds the age of the Universe,
$\tau_{\rm a}>t_0\approx 14$~billion years, we find the bound on the
mass of the axion as a dark matter candidate,
\begin{equation}
\label{chap8-true-axion-add-2*}
m_{\rm a}<25~\mbox{eV}
\;.
\end{equation}
There are astrophysical bounds on the strength of axion interactions
$f_{\rm PQ}^{-1}$ and hence on the axion mass. Axions in theories
with
$f_{\rm PQ}\lesssim 10^{9}$~GeV, which are heavier than
$10^{-2}$~eV, would be intensely produced in stars and supernovae
explosions. This would lead to contradictions with
observations. So, we are left with very light axions,
$m_{\rm a} \lesssim 10^{-2}$~eV.

As far as dark matter is concerned,
thermal production of axions is irrelevant.
There are at least two
mechanisms of axion production in the early Universe
that can provide not only right axion abundance but
also  small initial velocities of axions. The latter
property makes an axion a {\it cold} dark matter
candidate, despite its
very small  mass.
One mechanism has to do with decays of
global strings~\cite{Vilenkin:1982ks}---topological  defects that
exist in theories with spontaneously broken global
U(1) symmetry (U$(1)_{\rm PQ}$ in our case;
for a discussion of this mechanism, see, e.g.,
Ref.~\cite{Battye:1999bd}). Another
mechanism employs an  axion
condensate~\cite{Preskill:1982cy,Abbott:1982af,Dine:1982ah}, an
homogeneous
axion field that oscillates in time after the QCD epoch.
This is called the axion misalignment mechanism.
Let us consider the second mechanism in some detail.

As we have seen in
Eq.~\eqref{chap8-true-axion-eff+}, the axion potential is
proportional to the quark condensate
$\la \bar{q} q \ra$. This condensate breaks chiral symmetry.
The chiral symmetry is in fact restored at high temperatures
Hence, one expects that the
axion potential is negligibly small at
$T \gg
\Lambda_{\rm QCD}$. This is indeed the case: the effective potential
for the field
$\bar\theta=\theta+a/f_{\rm PQ}$
vanishes at high temperatures, and this field can take any value,
\[
\bar\theta_{\rm i} \in\left[ 0\;,2\pi\right)\; ,
\]
where we recall that the field $\bar\theta$ is a phase.
There is no reason to think that the initial value
$\bar\theta_{\rm i}$ is zero. As the temperature decreases,
the axion mass $m(T)$ starts to get generated,  so that
\begin{align}
    m_{\rm a} (T) &\simeq 0 \;\;\;\;\;\; \mbox{at} \;\; T\gg \Lambda_{\rm QCD} \; ,
\nonumber \\
  m_{\rm a} (T) &\simeq m_{\rm a} \;\;\;\;\;\; \mbox{at} \;\; T\ll \Lambda_{\rm QCD} \; .
\nonumber
\end{align}
Hereafter
$m_{\rm a}$ denotes the zero-temperature axion mass.
As the mass increases, at some point the field
$\bar\theta$, remaining homogeneous, starts to roll down
from
$\bar\theta_{\rm i}$ towards its value
$\bar\theta=0$ at the minimum of the potential.
The axion field practically does not
evolve when
$m_{\rm a} (T) \ll H(T)$ and at the time when
$m_{\rm a} (T) \sim H(T)$ it starts to oscillate.
Let us estimate the present energy density
of the axion field in this picture, without using the concrete form
of the function
$m(T)$.

The oscillations start at the time
$t_{\rm osc}$ when
\be
  m_{\rm a} (t_{\rm osc}) \sim H(t_{\rm osc}) .
\label{ch8-addit1}
\ee
At this time, the energy density of the axion field is estimated as
\[
\rho_{\rm a} (t_{\rm osc}) \sim m_{\rm a}^2 (t_{\rm osc}) f_{\rm PQ}^2 \bar{\theta}_{\rm i}^2 \; .
\]
The oscillating axion field
is the same thing as a collection of axions at rest.
Their number density at the beginning of oscillations
is estimated as
\be
 n_{\rm a} (t_{\rm osc}) \sim \frac{\rho_{\rm a} (t_{\rm osc})}{m_{\rm a} (t_{\rm osc})}
\sim  m_{\rm a} (t_{\rm osc}) f_{\rm PQ}^2 \bar{\theta}_{\rm i}^2
\sim  H (t_{\rm osc}) f_{\rm PQ}^2 \bar{\theta}_{\rm i}^2 \; .
\nonumber
\ee
This number density, as any number density of non-relativistic
particles, then decreases as
$a^{-3}$.

The axion-to-entropy ratio at time
$t_{\rm osc}$ is
\[
\frac{n_{\rm a}}{s} \sim   \frac{H (t_{\rm osc}) f_{\rm PQ}^2}{\frac{2\pi^2}{45} g_* T_{\rm osc}^3}
\cdot  \bar{\theta}_{\rm i}^2 \simeq
\frac{f_{\rm PQ}^2}{\sqrt{g_*} T_{\rm osc} M_{\rm Pl}}\cdot  \bar{\theta}_{\rm i}^2 \; ,
\]
where we use the usual relation $H = 1.66 \sqrt{g_*} T^2 /
M_{\rm Pl}$.
The axion-to-entropy ratio remains constant after the beginning of
oscillations, so the present mass density of  axions is
\be
\rho_{\rm a, 0} = \frac{n_{\rm a}}{s} m_{\rm a} s_0 \simeq \frac{m_{\rm a}
f_{\rm PQ}^2}{\sqrt{g_*} T_{\rm osc} M_{\rm Pl}} s_0 \cdot \bar{\theta}_{\rm i}^2 \; .
\label{ch8-addit2}
\ee
In fact, it is a decreasing function of
$m_{\rm a}$. Indeed,
$f_{\rm PQ}$ is inversely proportional to
 $m_{\rm a}$, see Eq.~\eqref{chap8-true-axion-R+216c*};
at the same time, the
axion obtains its
mass near the epoch of QCD transition, i.e., at
$T\sim \Lambda_{\rm QCD}$, so $T_{\rm osc}$ depends on $m_{\rm a}$ rather weakly.

To obtain a simple estimate, let us set
$T_{\rm osc} \sim \Lambda_{\rm QCD}
\simeq 200~\mbox{MeV}$ and make use of Eq.~\eqref{chap8-true-axion-R+216c*}
with $C_{\rm g} \sim 1$. We find
\be
 \Omega_{\rm a} \equiv \frac{\rho_{\rm a, 0}}{\rho_{\rm c}} \simeq
\l \frac{10^{-6}~\mbox{eV}}{m_{\rm a}} \r \bar{\theta}_{\rm i}^2 \; .
\label{ch8-addit3}
\ee
The natural assumption about the initial phase is
$\bar\theta_{\rm i}\sim \pi/2$. Hence, an axion of mass
$10^{-5}$--$10^{-6}$~eV is a good dark matter
candidate. Note that an axion of lower mass
  $m_{\rm a}<10^{-6}$~eV
may also serve as a dark matter particle, if for some reason
the initial phase
$\bar\theta_{\rm i}$ is much smaller than $\pi/2$.
This is {\it cold} dark matter: the oscillating field corresponds to
axions at rest.

A more precise estimate is obtained by taking into account
the fact that
that the axion mass smoothly depends on
temperature:
\begin{equation}
\Omega_{\rm a}\simeq0.2\cdot\bar{\theta}_{\rm i}^2\cdot
\l\frac{4\times10^{-6}~\mbox{eV}}{m_{\rm a}}\r^{1.2}.
\nonumber
\end{equation}
We see that our crude estimate
\eqref{ch8-addit3} is fairly accurate. Interestingly, the string
mechanism of the axion production leads to the same parametric dependence of
$\Omega_{\rm a}$ on the axion mass.

\begin{figure}[!htb]
\begin{center}
\includegraphics[width=0.6\textwidth]{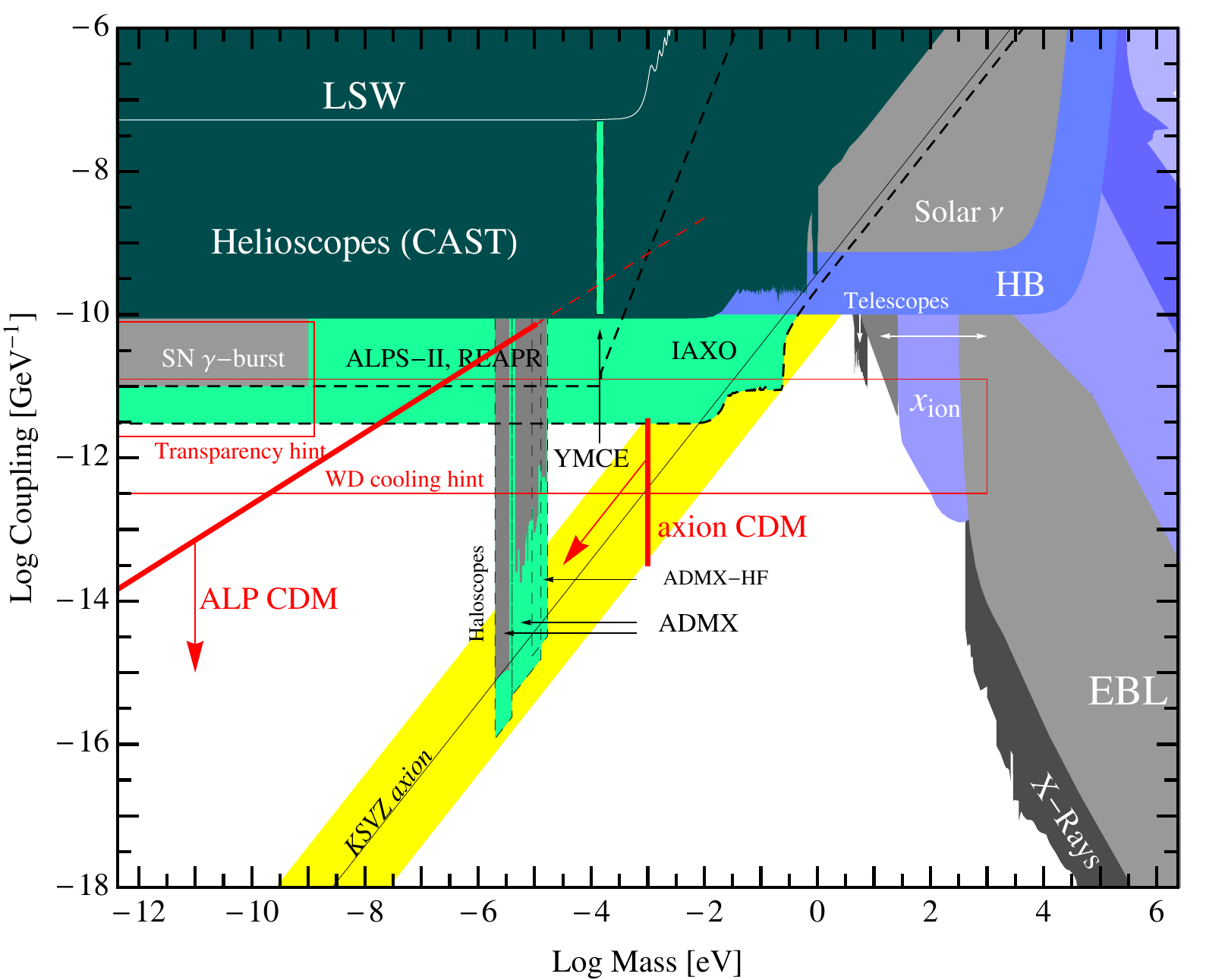}
\end{center}
\caption{
\label{Chap09-New-fig-axion}
Bounds on dark matter axions: axion--photon coupling versus  axion
mass~\cite{Ringwald:2012cu}. Inclined straight line labelled
`KSVZ axion' is the prediction of the KSVZ model, shaded region along
this line is the range of predictions of other axion models.
Region below the line labelled ALP CDM is the range of predictions of other
reasonably motivated models with axion-like particles as dark matter
candidates. Dashed lines show the sensitivities of future experiments.
}
\end{figure}

Search for dark matter axions with mass
$m_{\rm a}\sim10^{-5}$--$10^{-6}$~eV is difficult, but not impossible.
One way is to search for axion--photon conversion
in a resonator cavity filled with a strong magnetic field.
Indeed, in the background magnetic field the axion--photon interaction
(second term in Eq.~\eqref{chap8-true-axion-add-1*}) leads to the conversion
a $\to \gamma$, and the axions of mass $10^{-5}$--$10^{-6}$~eV are converted
to photons of frequency $m/(2\pi) = 2$--0.2~GHz (radio waves).
Bounds on the dark matter axions are shown in Fig.~\ref{Chap09-New-fig-axion}.

\subsection{Warm dark matter: sterile neutrinos and
light gravitinos}

As we discussed in Section~\ref{subsec:C-and-W}, there
are arguments, albeit not particularly strong, that
 favour warm, rather than cold, dark matter.
If WDM particles are thermal relics, i.e., if they
were in kinetic equilibrium at some epoch in the early Universe,
then their mass should be in the range 3--10~keV.
Reasonably well motivated particles of this mass are
sterile neutrinos and gravitinos.

\subsubsection{Sterile neutrinos}

Sterile neutrinos are most probably required for giving masses
to ordinary, `active' neutrinos. The masses of sterile neutrinos
cannot be predicted theoretically. Although sterile neutrinos
of WDM mass $m_{\nu_{\rm s}} = 3$--10~keV are not particularly plausible
from the particle-physics prospective, they are not pathological either.
In the simplest
models
the creation of sterile neutrino
states
$|\nu_{\rm s}\ra$ in the early Universe occurs due to their mixing with
active neutrinos
$|\nu_\alpha\ra$, $\alpha={\rm e},\mu,\tau$. In the approximation of
mixing between two states only, we have
\be
|\nu_\alpha\ra=\cos\theta |\nu_1\ra + \sin\theta
 |\nu_2\ra\;, \;\;\;\;\;
|\nu_{\rm s}\ra=-\sin\theta |\nu_1\ra + \cos\theta
 |\nu_2\ra\;,
\nonumber
\ee
where
$|\nu_\alpha\ra$ and $|\nu_{\rm s}\ra$ are active and sterile neutrino states,
$|\nu_1\ra$ and  $|\nu_2\ra$ are mass eigenstates of masses
$m_1$ and $m_2$, where we order $m_1 < m_2$,
and $\theta$ is the vacuum mixing angle between
sterile and active neutrinos. This  mixing should be weak,
$\theta \ll 1$, otherwise sterile neutrinos would decay too
rapidly, see below.
The heavy state is mostly sterile neutrinos
$|\nu_2\ra \approx
|\nu_{\rm s}\ra$, and
$m_2\equiv m_{\rm s}$ is the sterile neutrino mass.

\begin{figure}[!htb]
\centering
\includegraphics[width=0.6\textwidth]{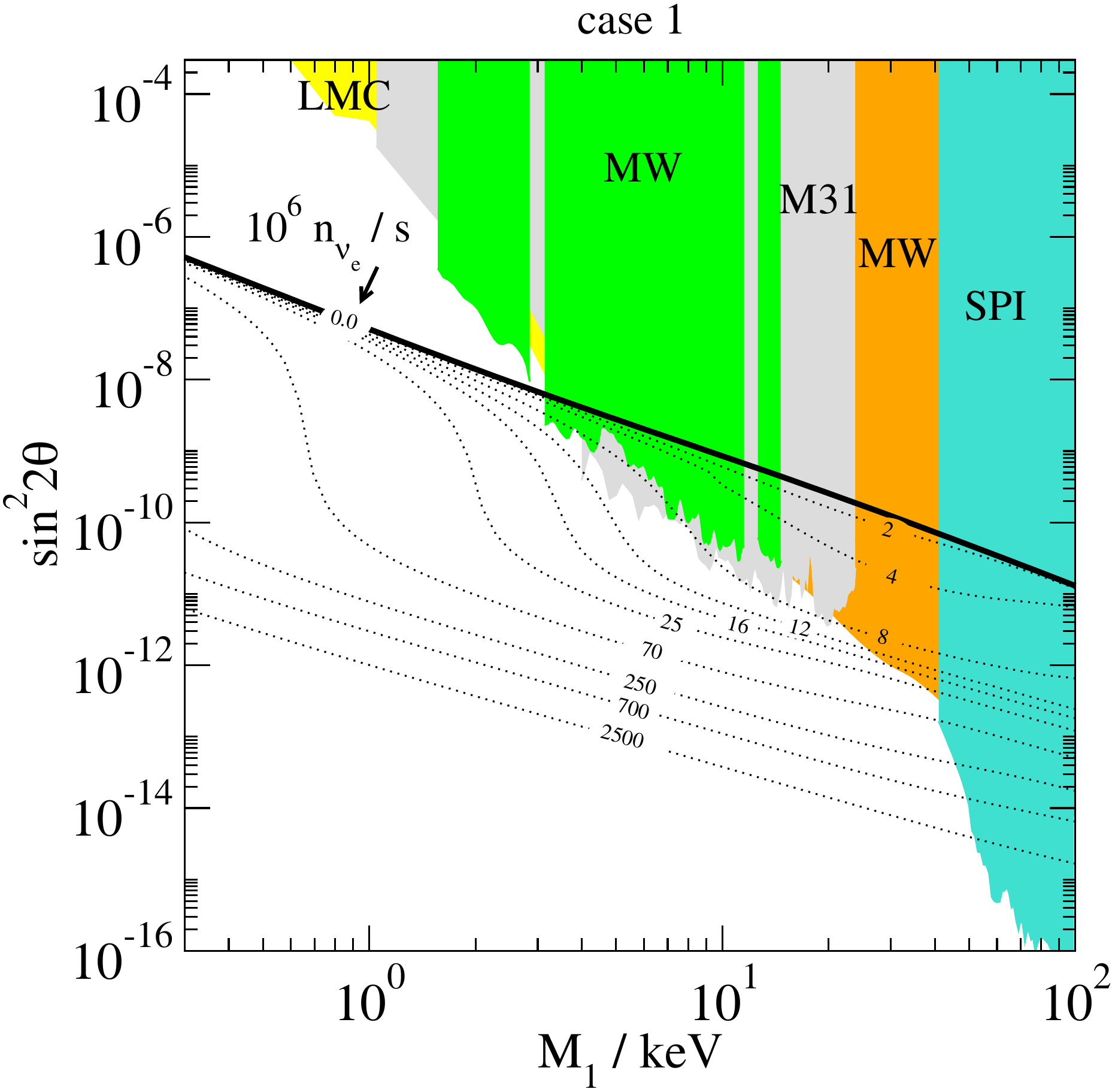}
\caption{Limits on sterile neutrino parameters
(mass $M_1$, mixing angle $\theta$) obtained from
X-ray telescopes. Solid line corresponds to
sterile neutrino dark matter produced in non-resonant
oscillations, Eq.~\eqref{chap7-sterile-4++}.
Dashed lines show the case of resonant oscillations
at non-zero lepton asymmetry; numbers in unit of $10^{-6}$
show the values of lepton asymmetry (lepton-to-photon ratio
$\eta_{\rm L}$)~\cite{Laine:2008pg}.
\label{chap7-newfig-sterile-neutrino}
}
\end{figure}

The calculation of sterile neutrino abundance is fairly
complicated, and we do not reproduce it here. If there is no
sizeable lepton asymmetry in the Universe,
the sterile neutrino production is most efficient at
temperature around
\[
T_* \sim \l\frac{m_{\rm s}}{5G_{\rm F}}\r^{1/3}\simeq 200~\mbox{MeV}\cdot
\l \frac{m_{\rm s}}{1~\mbox{keV}}\r^{1/3}\;.
\]
The resulting number density of sterile neutrinos
is estimated as
\begin{equation}
\label{chap7-sterile-add-4+}
\frac{n_{\nu_{\rm s}}}{n_{\nu_\alpha}}
\sim
T_*^3M_{\rm Pl}^*G_{\rm F}^2\cdot\sin^22\theta
\sim 10^{-2} \cdot \l\frac{m_{\rm s}}{1~\mbox{keV}}\r
\cdot \l \frac{\sin 2\theta}{10^{-4}}\r^2
\;.
\end{equation}
The number density of relic active neutrinos today is about
$110~\mbox{cm}^{-3}$, so
we find from Eq.~\eqref{chap7-sterile-add-4+} the estimate for the
present contribution of sterile neutrinos into energy density,
\begin{equation}
\label{chap7-sterile-4++}
\Omega_{\nu_{\rm s}}\simeq 0.2 \cdot \l\frac{\sin2\theta}
{10^{-4}}\r^2
\cdot \l\frac{m_\nu}{1~\mbox{keV}}\r^2\;.
\end{equation}
Thus, a sterile neutrino of mass
$m_\nu\gtrsim 1$~keV and small mixing angle
$\theta_\alpha\lesssim 10^{-4}$ would serve as a dark matter candidate.
However, this range of masses and mixing angles is ruled out.
The point is that due to its mixing with
an active neutrino,
a sterile neutrino can decay into an active neutrino and a photon,
\[
\nu_{\rm s} \to \nu_\alpha + \gamma \; .
\]
The sterile neutrino decay
width is proportional to $\sin^2 2\theta$.
If sterile neutrinos are dark matter particles, their decays
would produce a narrow line in X-ray flux from the cosmos
(orbiting velocity of dark matter particles in our Galaxy
is small, $v \sim 10^{-3}$, hence the photons produced in their
two-body decays are nearly monochromatic). Such a line has not been
observed, and there exist quite strong limits.
These limits, translated into limits on $\sin^2 2\theta$ as a function
of sterile neutrino mass, are shown in
Fig.~\ref{chap7-newfig-sterile-neutrino};
they rule out the range of masses giving the right mass density
of dark matter, Eq.~\eqref{chap7-sterile-4++}. Recall that
the mass of a sterile neutrino should exceed 3~keV (in fact, a more precise limit
is $m_{\rm s} > 5.7$~keV~\cite{Gorbunov:2008ka}).

A (rather baroque)
way out~\cite{Shi:1998km} is to assume that there is a fairly large
lepton asymmetry in the Universe. Then the oscillations
of active neutrinos into sterile neutrinos may be enhanced
due to the Mikheyev-Smirnov-Wolfenstein (MSW) effect, as at some temperature they occur in the
Mikheev--Smirnov resonance regime. In that case the right abundance of sterile neutrinos is obtained at smaller $\theta$,
and may  be consistent with X-ray bounds. This is also shown in
Fig.~\ref{chap7-newfig-sterile-neutrino}.

\subsubsection{Light gravitino}

A gravitino---a superpartner of a graviton---is necessarily present
in supersymmetric (SUSY) theories. It acquires mass as a result of
SUSY breaking (super-Higgs mechanism).
The gravitino mass is of order
\[
m_{3/2} \simeq \frac{F}{M_{\rm Pl}} \; ,
\]
where $\sqrt{F}$ is the supersymmetry breaking scale. Hence, gravitino
masses are in the right WDM ballpark
for rather low supersymmetry breaking
scales,  $\sqrt{F} \sim 10^{6}$--$10^{7}~\mbox{GeV}$. This is the case,
e.g., in the gauge-mediation scenario. With so low mass, a gravitino
is the lightest supersymmetric particle (LSP), so it is stable in
many supersymmetric extensions of the Standard Model.
From this viewpoint
gravitinos can indeed serve as dark matter particles.
For what follows, important parameters are the
widths of decays of other superpartners into  gravitinos and the Standard
Model particles. These are of order
\be
\Gamma_{\rm \tilde S} \simeq \frac{M_{\rm \tilde S}^5}{F^2} \simeq
\frac{ M_{\rm \tilde S}^5}{ m_{3/2}^2 M_{\rm Pl}^2} \; ,
\label{swidth}
\ee
where $M_{\rm \tilde S}$ is the mass of the superpartner.

One mechanism of the gravitino production in the early Universe
is decays of other superpartners. A gravitino interacts with everything
else so weakly that once produced, it moves freely, without
interacting with cosmic plasma. At production, gravitinos are
relativistic and hence they are indeed {\it warm} dark matter candidates.
Let  us assume that production in decays is the dominant mechanism
and consider under what circumstances the present mass density of gravitinos
coincides with that of dark matter.

The rate of gravitino production in decays of superpartners of the type
$\rm \tilde S$ in the
early Universe is
\[
 \frac{{\rm d} (n_{3/2}/s)}{{\rm d}t} = \frac{n_{\rm \tilde S}}{s} \Gamma_{\rm \tilde S} \; ,
\]
where $n_{3/2}$ and $n_{\rm \tilde S}$ are number densities of gravitinos
and superpartners, respectively,
and $s$ is the entropy density.
For superpartners in thermal equilibrium,
one has  $n_{\rm \tilde S}/s =\mbox{const} \sim g_*^{-1}$
for $T \gtrsim M_{\rm \tilde S}$,
and $n_{\rm \tilde S}/s \propto \mbox{exp} (- M_{\rm \tilde S}/T)$
at  $T \ll M_{\rm \tilde S}$. Hence, the production is most efficient
at $T \sim M_{\rm \tilde S}$, when the number density of superpartners
is still large, while the Universe expands most slowly. The density of
gravitinos produced in decays of the ${\rm \tilde S}$ is thus given by
\[
 \frac{n_{3/2}}{s}
\simeq
\frac{\Gamma_{\rm \tilde S}}{g_*} H^{-1}(T \sim M_{\rm \tilde S})
\simeq \frac{1}{g_*} \cdot
\frac{ M_{\rm \tilde S}^5}{ m_{3/2}^2 M_{\rm Pl}^2} \cdot
\frac{M_{\rm Pl}^*}{M_{\rm \tilde S}^2} \; .
\]
This gives the mass-to-entropy ratio today,
\be
   \frac{m_{3/2} n_{3/2}}{s} \simeq \sum_{\rm \tilde S}
\frac{M_{\rm \tilde S}^3}{g_*^{3/2} M_{\rm Pl} m_{3/2}} \; ,
\label{sum-ino}
\ee
where the sum runs over all  superpartner species {\it
which have ever been relativistic
in thermal equilibrium}. The correct value
(\ref{10p*}) is obtained for gravitino masses in the range
(\ref{wdmmass}) at
\be
M_{\rm \tilde S} = 100\mbox{--}300~\mbox{GeV} \; .
\label{msup}
\ee
Thus, the scenario with a gravitino as a warm dark
matter particle requires light superpartners~\cite{Gorbunov:2008ui},
which are to be
discovered at the LHC.

A few comments are in order. First, decays of superpartners is not
the only mechanism of gravitino production: gravitinos may also be
produced in scattering of superpartners~\cite{deGouvea:1997tn}.
To avoid overproduction
of gravitinos in the latter processes, one has to assume that the
maximum temperature in the Universe (e.g., reached after the post-inflationary
reheating stage) is quite low, $T_{\rm max} \sim 1$--10~TeV.
This is not a particularly
 plausible assumption, but it is
consistent with everything else in cosmology
and can indeed be realized
in some models of inflation.
Second,
existing constraints on masses of
strongly interacting superpartners (gluinos and squarks)
suggest that their masses exceed (\ref{msup}). Hence, these particles
should not contribute to the sum in
(\ref{sum-ino}), otherwise WDM gravitinos would be overproduced.
This is possible if masses of
squarks and gluinos are larger
than $T_{\rm max}$, so that they were never abundant in the
early Universe.
Third, a gravitino produced in decays of superpartners is {\it not}
a thermal relic, as it was never in thermal equilibrium
with the rest of the cosmic plasma. Nevertheless, since gravitinos are
produced at $T \sim M_{\rm \tilde{S}}$ and at that time have energy
$E  \sim M_{\rm \tilde{S}} \sim T$, our estimate \eqref{l0dwarf}
does apply.

\vspace{0.3cm}
\noindent
{\it Question.} Let $\rm \tilde{S}$ be the next-to-lightest superpartner
which decays into a gravitino of mass $m_{3/2}=5$~keV and a Standard
Model particle. Let  $\rm \tilde{S}$ be produced at the LHC at
subrelativistic velocity. How far is the decay vertex of
 ${\rm \tilde{S}}$ displaced from the proton collision point? Give
numerical estimates for $M_{\rm \tilde{S}} = 100$~GeV and $M_{\rm \tilde{S}} = 1$~TeV.
\vspace{0.3cm}

\subsection{Discussion}

If dark matter particles are indeed WIMPs, and the relevant
energy scale is of order 1~TeV, then the hot Big Bang theory will be
probed experimentally up to a temperature of $(\mbox{a~few})\cdot
\mbox{(10--100)}~$GeV
and down to an age of $10^{-9}$--$10^{-11}~$s in the relatively near future (compare
to 1~MeV and 1~s accessible today through BBN).
With microscopic physics to be
known from collider experiments, the WIMP density will be reliably
calculated and checked against the data from observational cosmology.
Thus, the WIMP scenario offers a window to
a very early stage of the evolution of the Universe.

Search for dark matter axions and signals from light sterile
neutrinos makes use of completely different methods. Yet there is
a good chance for discovery if either of these particles make dark
matter.

If dark matter particles are gravitinos, the prospect of
probing quantitatively such an early stage of the cosmological
evolution is not so bright: it would be very hard, if at all possible,
to get an experimental handle on the gravitino mass;
furthermore, the present gravitino mass
density depends on an unknown reheat temperature
$T_{\rm max}$. On the other hand, if this scenario is realized in nature,
then the whole picture of the early Universe will be quite different from
our best guess on the
early cosmology.   Indeed, the gravitino
scenario requires a low reheat temperature, which in turn calls for
a rather exotic mechanism of inflation.

The mechanisms discussed here are by no means the only ones
capable of producing dark matter, and the particles we discussed
are by no means
the only candidates for dark matter particles. Other dark matter
candidates include
axinos, Q-balls,
very heavy relics produced
towards the end of inflation (wimpzillas)
etc. Hence, even though there are grounds to hope
that the dark matter problem will be solved soon, there is no guarantee
at all.

\section{Baryon asymmetry of the Universe}
\label{sec:bau}

As we discussed in Section~\ref{sub:RD}, the baryon asymmetry of the
Universe is characterized by the baryon-to-entropy ratio, which at high
temperatures is defined as follows:
\[
       \Delta_{\rm B} = \frac{n_{\rm B} - n_{\bar{\rm B}}}{s} \; ,
\]
where $n_{\rm \bar{B}}$ is the number density of antibaryons and $s$ is
the entropy density. If the baryon number is conserved and the
Universe expands adiabatically (which is the case at least after the
electroweak epoch, $T \lesssim 100$~GeV), $\Delta_{\rm B}$ is time-independent
and equal to its present value
$  \Delta_{\rm B} \approx 0.8 \times 10^{-10}$, see Eq.~\eqref{mar28-15-1}.
At early times, at temperatures well above 100~MeV, cosmic plasma
contained many quark--antiquark pairs, whose number density
was of the order of the entropy density,
\[
    n_{\rm q} + n_{\rm \bar{q}} \sim s \; ,
\]
while the baryon number density was related to densities of quarks and
antiquarks as follows (baryon number of a quark equals $1/3$):
\[
  n_{\rm B} = \frac{1}{3} (n_{\rm q} - n_{\rm \bar{q}}) \; .
\]
Hence, in terms of quantities characterizing the very early epoch, the
baryon asymmetry may be expressed as
\[
  \Delta_{\rm B} \sim \frac{n_{\rm q} - n_{\rm \bar{q}}}{n_{\rm q} + n_{\rm \bar{q}}} \; .
\]
We see that there was one extra quark per about 10 billion
quark--antiquark pairs. It is this tiny excess that is responsible for
the entire baryonic matter in the present Universe: as the Universe
expanded and cooled down, antiquarks annihilated with quarks,
and only the excessive quarks remained and formed baryons.

There is no logical contradiction to suppose that the tiny excess of
quarks over antiquarks was built in as an initial condition. This is
not at all satisfactory for a physicist, however. Furthermore, the
inflationary scenario
 does not provide such an initial condition for the hot Big Bang epoch;
rather, inflationary theory
predicts that the Universe
was baryon-symmetric just after inflation.
Hence, one would like to explain the baryon asymmetry
dynamically~\cite{Sakharov:1967dj,Kuzmin:1970nx}, i.e., find
the mechanism of its generation in the early Universe.

\subsection{Sakharov conditions}

The baryon asymmetry may be generated from an initially baryon-symmetric state only if three necessary conditions, dubbed Sakharov conditions, are
satisfied. These are:
\begin{enumerate}
\item baryon number non-conservation;
\item C- and CP-violation;
\item deviation from thermal equilibrium.
\end{enumerate}

All three conditions are easily understood.
(1) If baryon number were conserved, and initial net baryon number in
the Universe was zero, the Universe today would still be symmetric.
(2) If C or CP were conserved, then the rate of
reactions with particles would be the same as the rate of reactions
with antiparticles, and no asymmetry would be generated.
(3) Thermal equilibrium means that the system is stationary (no
time dependence at all). Hence, if the initial baryon number is zero, it is
zero forever, unless there are deviations from thermal equilibrium.
Furthermore, if there are processes that violate baryon number,
and the system approaches thermal equilibrium, then the baryon number
tends to be washed out rather than generated.

At the epoch of the baryon-asymmetry generation, all three Sakharov
conditions have to be met simultaneously. There is a qualification, however.
These conditions would be literally correct if there were no other
relevant quantum numbers that characterize the cosmic medium.
In reality, however, lepton numbers also play a role. As we will see shortly,
baryon and lepton numbers are rapidly violated by anomalous
electroweak processes at temperatures above,
roughly, 100~GeV. What is conserved in the Standard Model
is the combination $B-L$, where $L$ is the total lepton number.
So, there are two options. One is to generate the baryon asymmetry
at or below the electroweak epoch, $T \lesssim 100$~GeV, and make sure
that the electroweak processes do not wash out the baryon asymmetry
after its generation. This leads to the idea of electroweak baryogenesis
(another possibility is Affleck--Dine baryogenesis~\cite{Affleck:1984fy}).
Another is
to generate  $B-L$ asymmetry before the electroweak epoch,
i.e., at $T \gg 100$~GeV:
if the Universe is $B-L$ asymmetric above 100~GeV, the electroweak
physics reprocesses $B-L$ partially into baryon number and partially into
lepton number, so that in thermal equilibrium with conserved $B-L$ one has
\[
B = C\cdot (B-L) \; , \;\;\;\;\;\;\; L  = (C-1) \cdot (B-L) \; ,
\]
where $C$ is a constant of order 1 ($C=28/79$ in the Standard
Model at $T \gtrsim 100$~GeV). In the second scenario,
the first Sakharov condition applies to
$B-L$ rather than baryon number itself.

Let us point out two most common mechanisms of baryon number
non-conservation.
One         emerges in grand unified theories and is
due to the exchange of supermassive particles. It is similar,
say, to the mechanism of charm non-conservation in weak interactions,
which occurs via the exchange of heavy W bosons. The scale of these
new, baryon number violating interactions is the grand unification
scale, presumably of order $M_{\rm GUT} \simeq
10^{16}~\mbox{GeV}$. It is rather unlikely, however,
that the baryon asymmetry was generated due to this mechanism:
the relevant temperature would be of order $M_{\rm GUT}$,
while such a high reheat
temperature after inflation is difficult to obtain.

Another mechanism is non-perturbative~\cite{'tHooft:1976up}
and is related to the triangle
anomaly in the baryonic current (a keyword here is
`sphaleron'~\cite{Klinkhamer-Manton,Kuzmin:1985mm}).
It exists already in the Standard
Model and, possibly with mild modifications, operates in all its
extensions. The two main features of this mechanism, as applied to the
early Universe, is that it is effective over a wide range of
temperatures,
$100~\mbox{GeV} < T < 10^{11}~\mbox{GeV}$, and, as we pointed out above,
that it conserves
$B-L$.

\subsection{Electroweak baryon number non-conservation}

Let us pause here to discuss the physics behind electroweak baryon
and lepton number non-conservation in a little more detail,
though still at a qualitative level. A detailed analysis can be found in
the book~\cite{Rubakov:2002fi} and in references therein.

Let us consider the
baryonic current,
\[
B^\mu = \frac{1}{3} \cdot \sum_{i} \bar{q}_{i} \gamma^\mu q_{i} \; ,
\]
where the sum runs over quark flavours. Naively, it is
conserved, but at the quantum level its divergence is non-zero
because of the triangle anomaly (a similar effect
goes under the name of the axial anomaly in the context
of quantum electrodynamics (QED) and QCD),
 \[
    \partial_\mu B^\mu =
\frac{1}{3} \cdot 3_{\rm colours}
\cdot 3_{\rm generations} \cdot  \frac{g^2}{32 \pi^2}
\epsilon^{\mu \nu \lambda \rho}
F^{\rm a}_{\mu \nu} F^{\rm a}_{\lambda \rho} \; ,
\]
where $F^{\rm a}_{\mu \nu}$ and $g$ are the field strength of the SU$(2)_{\rm W}$
gauge field and the SU$(2)_{\rm W}$ gauge coupling, respectively.
Likewise, each leptonic current ($\alpha = {\rm e}, \mu, \tau$) is anomalous
in the Standard Model (we disregard here neutrino masses and mixings,
which violate lepton numbers too),
\be
 \partial_\mu {\cal L}^\mu_\alpha = \frac{g^2}{32 \pi^2}
 \cdot \epsilon^{\mu \nu \lambda \rho}
 F^{\rm a}_{\mu \nu} F^{\rm a}_{\lambda \rho} \; .
\label{mar31-15-1}
\ee
A non-trivial fact is that there exist
large field fluctuations,  $F^{\rm a}_{\mu \nu} ({\bf x}, t) \propto g^{-1}$,
such that
\be
Q \equiv \int~{\rm d}^3x {\rm d}t~   \frac{g^2}{32 \pi^2}
\cdot \epsilon^{\mu \nu \lambda \rho}
F^{\rm a}_{\mu \nu} F^{\rm a}_{\lambda \rho} \neq 0 \; .
\label{mar31-15-2}
\ee
Furthermore, for any physically relevant
fluctuation the value of $Q$ is an integer (`physically relevant' means that
the gauge field strength vanishes at infinity in space--time).
 In four space--time dimensions such fluctuations exist
only in
{\it non-Abelian} gauge theories.

Suppose now that a fluctuation with non-vanishing $Q$ has occurred.
Then the baryon numbers at the end and beginning of the process are different,
\be
 B_{\rm fin} - B_{\rm in} =
 \int~{\rm d}^3x {\rm d}t~  \partial_\mu B^\mu = 3 Q \; .
\label{may7-2}
\ee
Likewise,
\be
   {\cal L}{\rm \alpha,~fin} - {\cal L}_{\rm \alpha,~in} = Q \; .
\label{may7-3}
\ee
This explains the selection rule mentioned above:
 $B$ is violated, $B-L$ is not.

 At zero temperature, the  field fluctuations that induce baryon
and lepton number violation are vacuum fluctuations, called
instantons~\cite{Belavin:1975fg}.
Since these  are {\it large} field fluctuations, their probability
is exponentially suppressed. The suppression factor in the
Standard Model is
\[
\mbox{e}^{- { \frac{16\pi^2}{g^2}}} \sim 10^{-165} \; .
\]
Therefore,
the rate of baryon number violating processes at zero temperature
is suppressed by this factor, making these processes totally negligible.
On the other hand, at high temperatures
there are large {\it thermal} fluctuations
(`sphalerons') whose rate is not necessarily small.
And, indeed,
$B$-violation in the early Universe
is rapid as compared to the cosmological expansion at
sufficiently high temperatures, provided that
(see Ref.~\cite{ew-rev} for details)
\be
\langle\phi \rangle_T < T \; ,
\label{may7-4}
\ee
where
$ \langle\phi \rangle_T$ is the
 Englert--Brout--Higgs expectation value at temperature $T$.

\begin{figure}[htb!]
\begin{center}
\includegraphics[width=0.17\textwidth]{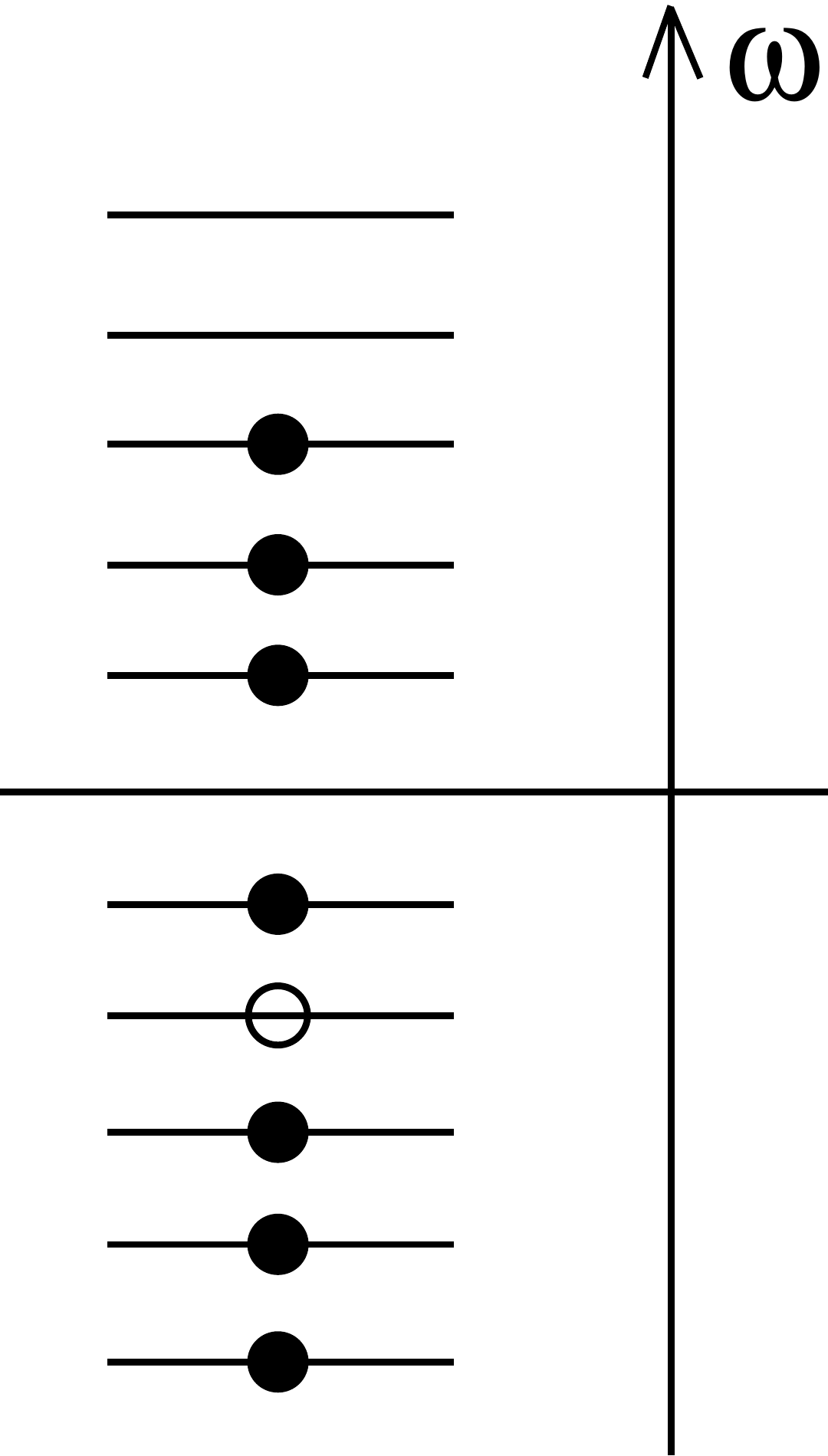}
\end{center}
\caption{Fermion energy levels at zero background gauge field
 \label{a=0}
 }
\end{figure}

One may wonder how baryon number is not conserved in the absence
of explicit baryon number violating terms in the Lagrangian of
the Standard Model. To understand what is going on, let us consider
a massless {\it left-handed} fermion field in the background of the SU(2)
gauge field
${\bf A}({\bf x}, t)$, which depends on space--time coordinates in a
non-trivial way. As a technicality,
 we set the temporal component of the gauge field equal to zero,
$A_0 =0$, by the choice of gauge.
One way to understand the behaviour of the fermion field in
the gauge field background is to study the system of
eigenvalues of the
Dirac Hamiltonian $\{ \omega (t) \}$. The Hamiltonian is defined in the
standard way
\[
  H_{\rm Dirac} (t)
= {\rm i}\alpha^i \left(\d_i - {\rm i}g A_i ({\bf x}, t)\right) \frac{1- \gamma_5}{2} \; ,
\]
where $\alpha^i = \gamma^0 \gamma^i$, so that the Dirac equation has the
Schr\"odinger form,
\[
 {\rm i} \frac{\d \psi}{\d t} = H_{\rm Dirac} \psi \; .
\]
So, let us discuss the eigenvalues $\omega_n (t)$
of the operator $H_{\rm Dirac}(t)$,
treating $t$ as a parameter. These eigenvalues are found from
\[
H_{\rm Dirac} (t) \psi_n = \omega_n (t) \psi_n \; .
\]
At ${\bf A} = 0$, the system
of levels is shown schematically in Fig.~\ref{a=0}. Importantly,
there are both  positive- and negative-energy levels. According to Dirac,
the lowest-energy state (Dirac vacuum) has all negative-energy levels
occupied, and all positive-energy levels empty. Occupied positive-energy
levels (three of them in Fig.~\ref{a=0}) correspond to real fermions,
while empty negative-energy levels describe antifermions (one in
Fig.~\ref{a=0}). Fermion--antifermion annihilation in
this picture is a jump
of a fermion from a positive-energy level to an unoccupied negative-energy level.
As a side remark, this original Dirac picture is, in fact, equivalent
to the more conventional (by now) procedure of the
quantization of the
fermion field, which does not make use of the notion of
negative-energy levels. The discussion that follows can be translated into
the conventional language; however, the original Dirac picture turns out
to be a
lot more transparent in our context. This is a nice example of the
complementarity of various approaches in quantum field theory.

Let us proceed with the discussion of the fermion energy levels
in gauge field backgrounds.
In weak background fields, the energy levels depend on time
(`move'), but nothing dramatic happens. For adiabatically varying background
fields, the fermions merely sit on their levels, while fast-changing fields
generically give rise to jumps from, say, negative- to  positive-energy
levels, that is, creation of fermion--antifermion pairs. Needless to
say, fermion number $(N_{\rm f} - N_{\bar {\rm f}})$ is conserved.

\begin{figure}[htb!]
\begin{center}
\includegraphics[width=0.6\textwidth]{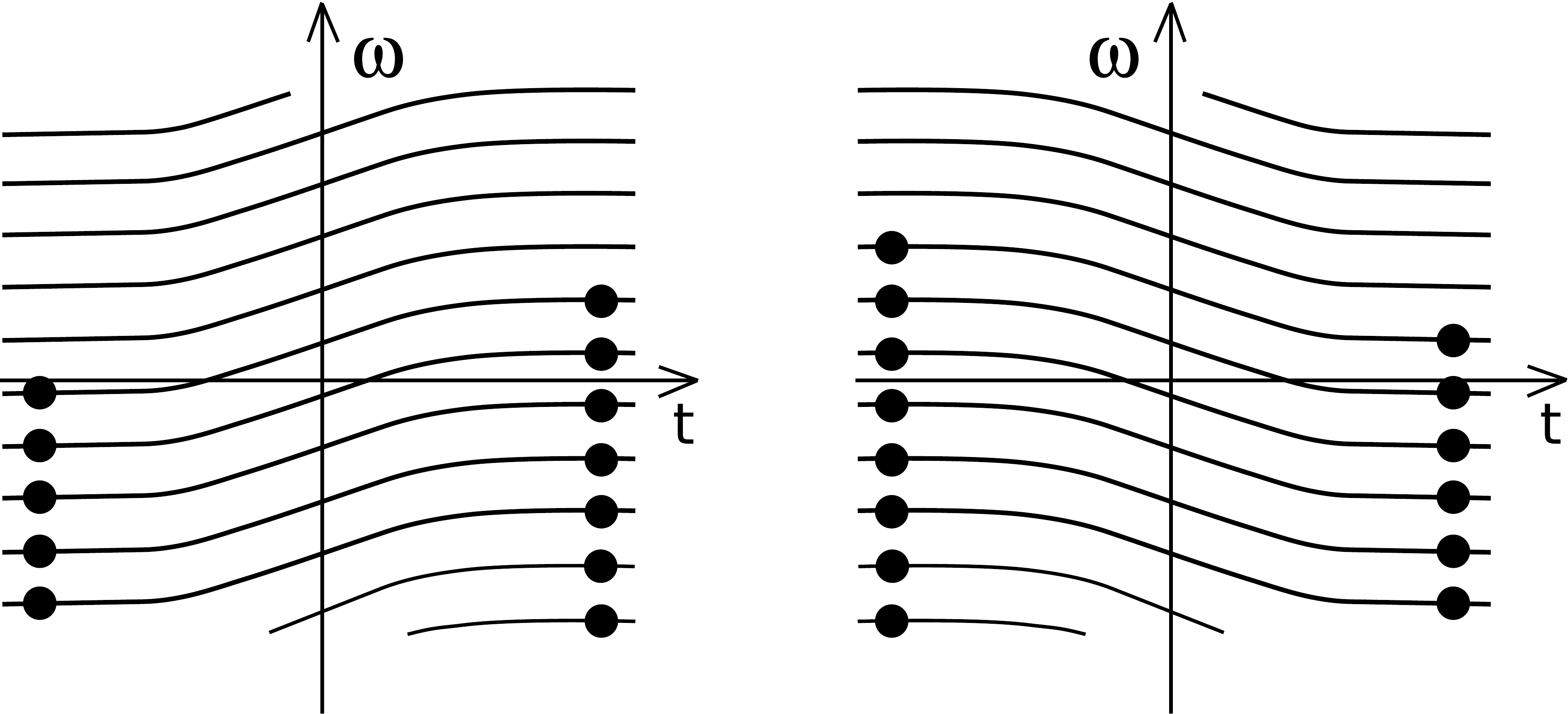}
\end{center}
\caption{Motion of fermion levels in background gauge fields
with non-vanishing $Q$ (shown is the case $Q=2$). Left-hand panel:
left-handed fermions. Right-hand panel: right-handed fermions.
 \label{ane0}}
\end{figure}

The situation is entirely different for the background fields with
non-zero $Q$. The levels of left-handed fermions move as shown
in the left-hand panel of Fig.~\ref{ane0}.
Some levels necessarily cross zero,
and the net number of levels crossing zero from below equals
$Q$. This means that the number of left-handed fermions is not
conserved: for an adiabatically
varying gauge field ${\bf A}({\bf x}, t)$, the motion of levels
shown
in the left-hand panel of Fig.~\ref{ane0} corresponds to the case
in which
the initial state of the
fermionic system is vacuum (no real fermions or antifermions)
whereas the final state
contains $Q$ real fermions (two in the particular case shown).
If the evolution of the gauge field is not adiabatic,
the result for the fermion number non-conservation is the same:
there may be jumps from negative-energy levels to positive-energy levels or vice versa. These correspond to creation or
annihilation of fermion--antifermion pairs, but the net change
of the fermion number (number of fermions minus number of antifermions)
remains equal to $Q$.
Importantly, the initial and final field configurations of the gauge
field may be trivial, ${\bf A} = 0$
(up to gauge transformation), so that fermion number non-conservation
may occur due to a fluctuation that begins and ends in the
gauge field vacuum.
These are precisely instanton-like vacuum fluctuations.
At finite temperatures, processes of this type occur due to
thermal fluctuations, i.e., sphalerons.

If the same gauge field interacts also with right-handed fermions,
the motion of the levels of the latter is opposite to
that of left-handed fermions. This is shown in the right-hand
 panel of Fig.~\ref{ane0}. The change in the number of right-handed
fermions is equal to $-Q$. So, if the gauge interaction is vector-like,
the total fermion number $(N_{\rm left} + N_{\rm right})$ is conserved, while
chirality $(N_{\rm left} - N_{\rm right})$ is violated even for massless fermions.
This explains why there is no baryon number  violation in QCD.
The above discussion implies, instead, that there is
non-perturbative violation of chirality in QCD
in the limit of massless quarks. The latter phenomenon
has non-trivial consequences, which are
indeed confirmed by phenomenology. In this sense anomalous
non-conservation of fermion quantum numbers is an experimentally
established fact.

In electroweak theory, right-handed fermions do not interact with the
SU$(2)_{\rm W}$ gauge field, while left-handed fermions do. Therefore, fermion number is not conserved (the anomalous relations \eqref{mar31-15-1} and
\eqref{mar31-15-2} suggest that this result is valid also in the
presence of the Standard Model Yukawa couplings of quarks and leptons;
this is indeed the case). Since fermions of each SU$(2)_{\rm W}$ doublet
interact with the   SU$(2)_{\rm W}$ gauge bosons
in one and the same way, they are equally created in a process involving
a gauge field fluctuation with non-zero $Q$. This again leads to the
relations (\ref{may7-2}) and  (\ref{may7-3}), i.e., to the selection rules
$\Delta B = \Delta L$ and
$\Delta L_{\rm e} = \Delta L_\mu = \Delta L_{\tau}$.

\subsection{Electroweak baryogenesis}

It is tempting to make use of the electroweak
mechanism of baryon number non-conservation
for explaining the baryon asymmetry of the Universe.
This scenario is known as electroweak baryogenesis.
It meets
two problems, however. One is that CP-violation in the Standard Model
is too weak:  the CKM mechanism
alone is insufficient to generate a realistic value of the
baryon asymmetry. Hence, one needs extra sources of CP-violation.
Another problem has to do with departure from thermal equilibrium
that is necessary for the generation of the baryon asymmetry.
At temperatures well above 100~GeV, electroweak symmetry is restored,
the expectation value of the Englert--Brout--Higgs field
$\phi$ is zero, the relation
(\ref{may7-4}) is valid and the baryon number non-conservation is rapid
as compared to the cosmological expansion. At temperatures of order
100~GeV, the  relation
(\ref{may7-4}) may be violated, but the Universe expands very slowly:
the cosmological time-scale at these temperatures is
\be
H^{-1} = \frac{M_{\rm Pl}^*}{T^2}
\sim 10^{-10}
~\mbox{s} \; ,
\label{may8-1}
\ee
which is very large by the electroweak physics standards. The only
way in which a
strong departure from thermal equilibrium at these
temperatures may occur appears to be   the
first-order phase transition.

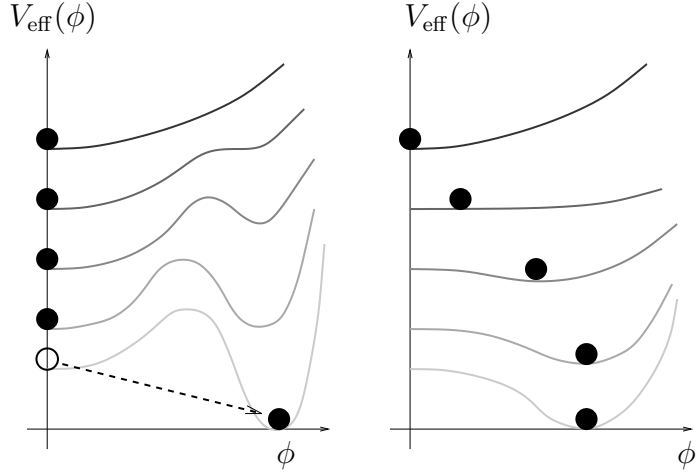
\begin{figure}[htb!]
\begin{center}
\includegraphics[width=0.55\textwidth]{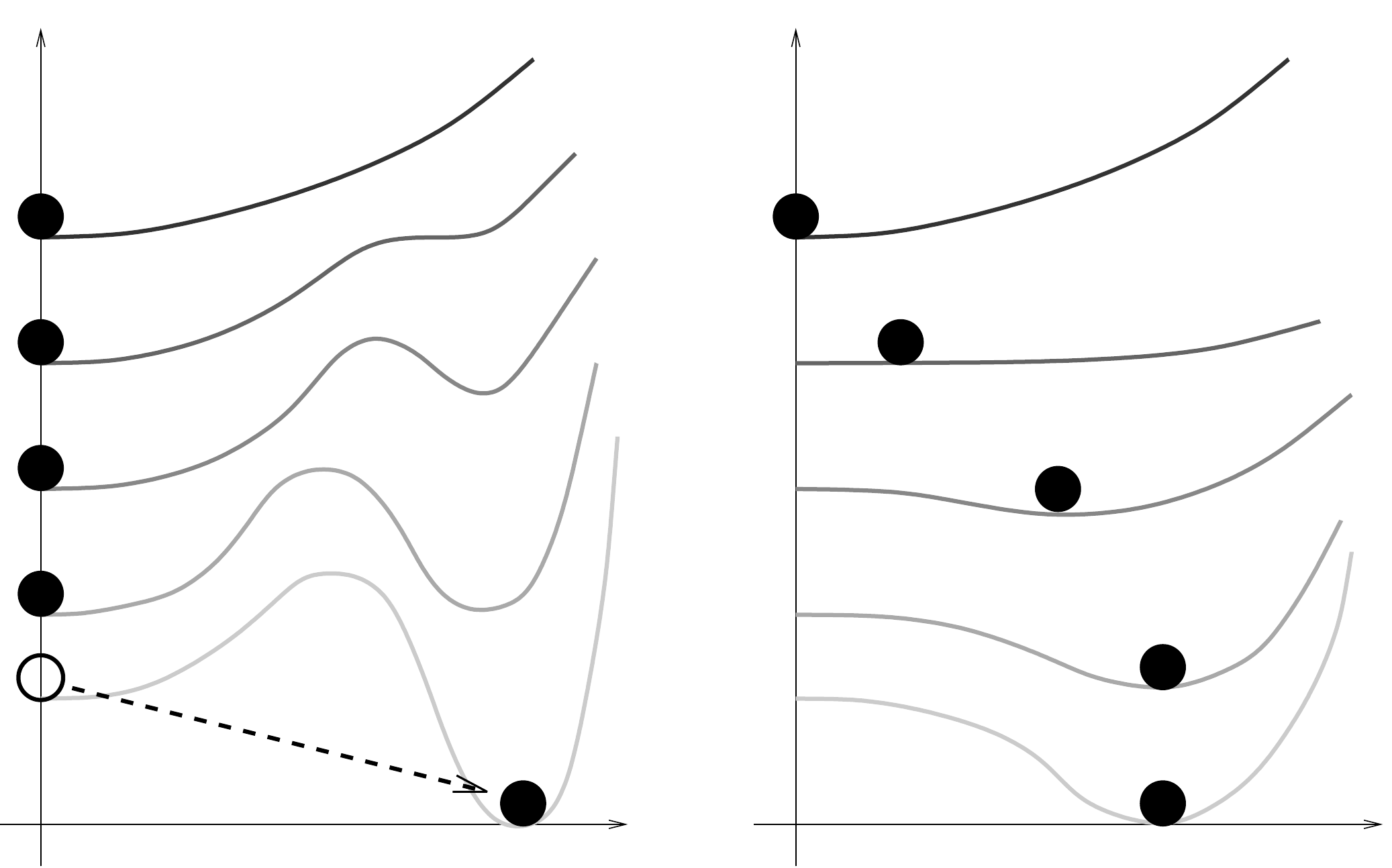}
\begin{picture}(10,10)(0,0)
{\large
\put(-260,160){$V_{\rm eff}(\phi)$}
\put(-112,160){$V_{\rm eff}(\phi)$}
\put(-160,-5){$\phi$}
\put(-10,-5){$\phi$}
}
\end{picture}
\end{center}
\caption{Effective potential as function of $\phi$ at different temperatures.
Left: first-order phase transition. Right: second-order phase transition.
Upper curves correspond to higher temperatures. Black blobs show the
expectation value of $\phi$ in thermal equilibrium. The arrow
in the left-hand panel illustrates the transition from the metastable,
supercooled state to the ground state.
 \label{1-2order}}
\end{figure}

The property that
at temperatures well above 100~GeV the expectation value of
the Englert--Brout--Higgs field is zero, while it is non-zero in vacuo,
suggests that there may be a phase transition
from the phase with $\langle \phi \rangle = 0$ to the phase
 with $\langle \phi \rangle \neq 0$. The situation is pretty
subtle here, as $\phi$ is not gauge invariant and hence cannot serve
as an order parameter, so the notion of phases
with $\langle \phi \rangle = 0$ and
$\langle \phi \rangle \neq 0$
is vague. In fact,  neither electroweak theory
nor most of its extensions have a gauge-invariant order parameter,
so there is no real distinction between these `phases'.
This situation is similar to that in a liquid--vapour
system, which does not have an order parameter and may or may
not experience a vapour--liquid phase transition as temperature decreases,
depending on other, external
parameters characterizing this system, e.g.,  pressure.
In the Standard Model the role of such an `external' parameter is played by
the Englert--Brout--Higgs self-coupling $\lambda$ or, in other words, the
Higgs boson mass.

Continuing to use somewhat sloppy terminology, we recall that
in thermal equilibrium any system is at the global minimum of
its {\it free
energy}. To figure out the expectation value of $\phi$ at a given
temperature, one introduces the temperature-dependent effective potential
$V_{\rm eff}(\phi;T)$, which is equal to the free energy density
in the system where the average field is pinpointed to a prescribed
value $\phi$, but otherwise there is thermal equilibrium.
Then the global minimum of $V_{\rm eff}$ at a given temperature
is at the equilibrium value of $\phi$, while local minima
correspond to metastable states.

The interesting
case for us is the first-order phase transition. In this case,
the system evolves as follows.
At high temperatures, there exists one minimum of $V_{\rm eff}$ at
$\phi =0$, and the expectation value of the Englert--Brout--Higgs field is zero.
As the temperature decreases, another minimum appears at finite
$\phi$, and then becomes lower than the minimum at $\phi =0$; see
left-hand panel of Fig.~\ref{1-2order}.
However, the minima with $\phi =0$ and $\phi \neq 0$ are
separated by a barrier of $V_{\rm eff}$,
the probability of the transition from the phase
$\phi=0$ to the phase $\phi \neq 0$ is very small for some time
and the system gets overcooled. The transition occurs when the
temperature becomes sufficiently low and the transition
probability sufficiently high.
This is to be contrasted
to the case, e.g., of the second-order phase transition,
right-hand panel of Fig.~\ref{1-2order}.
In the latter case, the field slowly evolves,
as the temperature decreases, from zero to non-zero vacuum
value, and the system remains very close to thermal equilibrium
at all times.


During the
first-order phase transition, the field cannot
jump from $\phi=0$ to $\phi \neq 0$ homogeneously
throughout the whole space: intermediate homogeneous configurations
have free energies proportional to the volume of the
system (recall that $V_{\rm eff}$ is free energy {\it density}), i.e.,
infinite.
Instead, the transition occurs just like the first-order
vapour--liquid transition, through boiling. Thermal fluctuations
spontaneously create
bubbles of the new phase inside the old phase. These bubbles
then grow, their walls eventually collide and the new phase finally
occupies the entire space. The Universe boils.
In the cosmological context, this process
happens when the bubble nucleation rate  per Hubble time
per Hubble volume is roughly of order 1,
 i.e., when a few bubbles are created
in Hubble volume in Hubble time.
The velocity of the bubble wall in the relativistic
cosmic plasma is roughly of the order of the speed of light
(in fact, it is somewhat smaller, from $0.1$ to $0.01$),
simply because there are no relevant dimensionless parameters
characterizing the system. Hence, the bubbles grow large before their
walls collide: their size at collision is roughly of the order of the
Hubble size (in fact, one or two orders of magnitude smaller).
While the bubble is microscopic  at nucleation---its size is
determined by the electroweak scale and is roughly of order
$(100~\mbox{GeV})^{-1} \sim 10^{-16}~\mbox{cm}$---its size at collision
of walls is macroscopic,
$R \sim 10^{-2}$--$10^{-3}$~cm, as follows from
(\ref{may8-1}). Clearly,  boiling  is a highly
non-equilibrium process, and one may hope that the baryon asymmetry
may be generated at that time. And, indeed, there exist mechanisms
of the generation of the baryon asymmetry, which have to do with
interactions of quarks and leptons with moving bubble walls.
The value of the resulting baryon asymmetry may well be of order
$10^{-10}$, as required by observations, provided that there is enough
CP-violation in the theory.

A necessary condition for the electroweak generation of the
baryon asymmetry is that the inequality (\ref{may7-4}) must be violated
{\it just after} the phase transition. Indeed, in the opposite case
the electroweak baryon number violating processes are fast after
the transition, and the baryon asymmetry, generated during the
transition, is washed out
afterwards. Hence, the phase transition must be of strong enough
first order. This is {\it not} the case in the Standard Model.
To see why this is so, and to get an idea of in which extensions of
the Standard Model the phase
transition may be of strong enough first order, let us consider
the effective potential in some detail. At zero temperature, the
Englert--Brout--Higgs potential has the standard form,
\[
V(\phi) = - \frac{m^2}{2} |\phi|^2 + \frac{\lambda}{4} |\phi|^4 \; .
\]
Here
\be
   |\phi| \equiv \l \phi^\dagger \phi \r^{1/2}
\label{may8-3}
\ee
is the length of the Englert--Brout--Higgs doublet $\phi$,
$m^2 = \lambda v^2$ and
$v = 246$~GeV is the Englert--Brout--Higgs expectation value in vacuo.
The Higgs boson mass is related to the latter
as follows:
\be
m_{\rm H} = \sqrt{2 \lambda} v \; .
\label{may8-5}
\ee
Now, to the leading order of perturbation theory, the finite-temperature
effects modify the effective potential into
\be
V_{\rm eff} (\phi, T) = \frac{\alpha (T)}{2} |\phi|^2 - \frac{\beta}{3} T |\phi|^3
+ \frac{\lambda}{4} |\phi|^4 \; .
\label{may8-2}
\ee
Here
$\alpha (T) = -m^2 + \hat{g}^2 T^2$,
where $\hat{g}^2$ is a positive linear combination of squares of
coupling constants of all fields to the Englert--Brout--Higgs field (in the
Standard Model, a
linear combination
of $g^2$, $g^{\prime \, 2}$ and $y_i^2$, where $g$ and $g^\prime$
are SU$(2)_{\rm W}$ and U$(1)_{\rm Y}$ gauge couplings and $y_i$ are Yukawa couplings).
The phase transition occurs roughly when $\alpha(T) = 0$.
An important parameter
$\beta$ is a positive linear combination of cubes of
 coupling constants of all {\it bosonic}
fields to the Englert--Brout--Higgs field.  In the
Standard Model, $\beta$ is a
linear combination
of $g^3$ and  $g^{\prime \, 3}$, i.e., a linear combination
of $M_{\rm W}^3/v^3$ and $M_{\rm Z}^3/v^3$,
\be
\beta =  \frac{1}{2 \pi} \frac{2 M_{\rm W}^3 + M_{\rm Z}^3}{v^3} \; .
\label{may8-7}
\ee
The cubic term in (\ref{may8-2})
is rather peculiar:
in view of (\ref{may8-3}), it is not analytic in the original Englert--Brout--Higgs
field
$\phi$. Yet this term is crucial for the first-order phase transition:
for $\beta=0$ the phase transition would be of the second order.

\vspace{0.3cm}
\noindent
{\it Question.} Show that the phase transition is second order
for $\beta = 0$.
\vspace{0.3cm}

The origin of the non-analytic cubic term can be traced back
to the enhancement of the Bose--Einstein thermal
distribution  at low momenta,  $p, m \ll T$,
\[
f_{\rm Bose}(p) = \frac{1}{\mbox{e}^{\frac{\sqrt{p^2 + m_{a}^2}}{T}} - 1}
\simeq \frac{T}{\sqrt{p^2 + m_{a}^2}} \; ,
\]
where $m_{} \simeq g_{a} |\phi|$ is the mass of the boson
$a$ that is generated due to the non-vanishing Englert--Brout--Higgs field, and
$g_{a}$ is the coupling constant of the field $a$ to the Englert--Brout--Higgs field.
Clearly, at $p \ll g_{} |\phi|$ the distribution function is
non-analytic in $\phi$,
\[
f_{\rm Bose}(p) \simeq \frac{T}{g_{a} |\phi|} \; .
\]
It is this non-analyticity that gives rise to the non-analytic cubic term in
the effective potential. Importantly, the Fermi--Dirac distribution,
\[
f_{\rm Fermi}(p) = \frac{1}{\mbox{e}^{\frac{\sqrt{p^2 + m_{a}^2}}{T}} + 1} \; ,
\]
is analytic in $m_{a}^2$, and hence $\phi^\dagger \phi$, so fermions do
not contribute to the cubic term.

With the cubic term in the effective potential, the phase transition
is indeed
of the first order: at high temperatures the coefficient
$\alpha$ is positive and
large, and there is one minimum of the effective potential
at $\phi =0$, while for $\alpha$ small but still positive
there are two minima. The phase transition occurs at $\alpha \approx 0$;
at that moment
\[
V_{\rm eff} (\phi, T) \approx - \frac{\beta T}{3} |\phi|^3
+ \frac{\lambda}{4} |\phi|^4 \; .
\]
We find from this expression that immediately after the phase
transition the minimum of $V_{\rm eff}$ is at
\[
    \phi \simeq \frac{\beta T}{\lambda} \; .
\]
Hence, the necessary condition for successful electroweak
baryogenesis, $\phi > T$, translates into
\be
   \beta > \lambda \; .
\label{may8-6}
\ee
According to (\ref{may8-5}), $\lambda$ is proportional to
$m_{\rm H}^2$, whereas in the Standard Model
$\beta$ is proportional to $(2 M_{\rm W}^3 + M_{\rm Z}^3)$.
Therefore, the relation (\ref{may8-6}) holds for small
Higgs boson masses only; in the Standard Model one makes
use of (\ref{may8-5})
and (\ref{may8-7}) and finds that
this happens
for $m_{\rm H} < 50$~GeV, while in reality   $m_{\rm H} =125$~GeV.
In
fact, in the Standard Model with $m_{\rm H} = 125$~GeV,
there is no phase transition at all; the electroweak transition
is a smooth crossover instead. The latter fact is not visible
from the expression (\ref{may8-2}), but that expression
is the lowest-order perturbative result, while the
perturbation theory is not applicable for describing the transition
in the Standard Model with large $m_{\rm H}$.

This discussion indicates a possible way to make the
electroweak phase transition strong. What one needs
is the existence of new bosonic fields that have
large enough couplings to the Englert--Brout--Higgs field(s), and hence provide large
contributions to $\beta$. To have an effect on the dynamics of
the transition, the new bosons must be present in the cosmic plasma
at the transition temperature, $T \sim 100$~GeV, so their masses
should not be too high, $M \lesssim 300$~GeV. In supersymmetric
extensions of the Standard Model, the  natural candidate for a long time
has been stop (superpartner of top-quark)
whose
Yukawa coupling to the Englert--Brout--Higgs field is the same as that of top,
that is, large. The light
stop scenario for electroweak baryogenesis would indeed work, as has been
shown by the detailed analysis in Refs.~\cite{stop-1,stop-2,stop-3}.

There are other possibilities to make the electroweak
transition strongly first order. Generically, they require an extension
of the scalar sector of the Standard Model and predict new fairly
light scalars which interact with  the Standard Model
Englert--Brout--Higgs field and may or may not participate in
gauge interactions.

Yet another issue is CP-violation, which has to be strong enough
for successful electroweak baryogenesis. As the asymmetry is generated
in the interactions of quarks and leptons
with the bubble walls, CP-violation
must occur at the walls. Recall now that the walls are made of the
scalar field(s). This points towards the necessity of CP-violation
in the scalar sector, which may only be the case in a theory
containing scalar fields other than the Standard
Model  Englert--Brout--Higgs field.

To summarize, electroweak baryogenesis requires a considerable
extension of the Standard Model, with masses of new particles in
the range $100$--300~GeV. Hence, this mechanism will most likely
be ruled out or confirmed by the LHC. We emphasize, however, that electroweak
baryogenesis is not the only option at all: an elegant and well-motivated
competitor is leptogenesis~\cite{Fukugita:1986hr,leptogen-1,leptogen-2}; there are many
other mechanisms proposed in the literature.

\section{Before the hot epoch}
\label{sec:pertu}

\subsection{Cosmological perturbations: preliminaries}

 With BBN theory and observations, and due to
evidence, albeit indirect, for relic neutrinos,
 we are confident of the theory of the early Universe
at temperatures up to
$T\simeq 1$~MeV, which correspond to an age of $t\simeq 1$~s.
With the LHC, we hope to be able to learn the Universe
up to temperature $T\sim 100$~GeV and age $t \sim 10^{-10}$~s.
Are we going to have a handle on an even earlier epoch?

The key issue in this regard is cosmological perturbations.
These are inhomogeneities in the energy density and associated
gravitational potentials, in the first place.
This type of inhomogeneities is called scalar perturbations,
as they are described by 3-scalars. There may exist perturbations
of another type, called tensors; these are primordial gravity
waves. We will mostly concentrate on scalar perturbations, since
they are observed; tensor perturbations are important too,
and we comment on them later on. While
perturbations of the present size of order 10~Mpc and smaller
have large amplitudes today and are non-linear,
amplitudes of all known perturbations
were small in the past, and the perturbations can be described
within the linearized theory. Indeed, CMB temperature anisotropy
tells us that the perturbations at the recombination epoch
were roughly at the level
\be
\delta \equiv \frac{\delta \rho}{\rho} = 10^{-4}\mbox{--}10^{-5} \; .
\label{mar29-15-1}
\ee
Thus, the linearized theory works very well before recombination
and somewhat later.
We will be rather sloppy when talking about scalar perturbations.
In general relativity, there is  arbitrariness in the choice of
reference frame, which can be viewed as a sort of gauge freedom.
In a homogeneous and isotropic Universe, there is a preferred reference frame,
in which quantities like energy density or distribution function of
CMB photons are manifestly homogeneous
and isotropic. It is in this frame that the metric has FLRW
form~\eqref{FRW}. Once there are perturbations,
no preferred reference frame exists any longer.
As an example, one can choose a reference frame such that the
three-dimensional hypersurfaces of constant time are hypersurfaces
of constant total energy density $\rho$. In this frame one has
$\delta \rho =0$, so Eq.~\eqref{mar29-15-1} does not make sense.
Yet the Universe is inhomogeneous in this reference frame,
since there are inhomogeneous metric perturbations
$\delta g_{\mu \nu} ({\bf x},t)$.
We will skip these technicalities and denote the scalar perturbation
by $\delta$ without specifying its gauge-invariant meaning.

Equations for perturbations are obtained by writing for every
variable (including metric) an expression like
$\rho ({\bf x}, t) = \bar{\rho}(t)
+ \delta \rho ({\bf x}, t)$ etc, where $ \bar{\rho}(t)$
is the homogeneous and isotropic background, which we
discussed in Section~\ref{subsec:Friedeq}. One inserts the perturbed
variables into the
Einstein equations and covariant conservation equations
$\nabla_\mu T^{\mu \nu} = 0$
and linearizes this set of equations. In many cases one also has to
use the linearized Boltzmann equations that govern the distribution functions
of particles out of thermal equilibrium; these are necessary for
evaluating the linearized perturbations of the energy--momentum tensor.
In any case,
since the background FLRW metric~\eqref{FRW} does not explicitly depend
on ${\bf x}$, the linearized equations for perturbations do not contain
${\bf x}$ explicitly. Therefore, one makes use of the spatial Fourier
 decomposition
\[
\delta ({\bf x},t)=\int~{\rm e}^{{\rm i}{\bf k}{\bf x}}\delta({\bf k},t)~{\rm d}^3k \; .
\]
The advantage is that modes with different momenta ${\bf k}$ evolve
independently in the linearized theory,
i.e., each mode can be treated separately. Recall that
${\rm d} {\bf x}$ is {\it not} the physical distance between neighbouring
points; the physical distance is $a(t) {\rm d} {\bf x}$. Thus, ${\bf k}$ is
{\it not} the physical momentum (wavenumber); the physical momentum
is ${\bf k}/a(t)$. While for a given mode the comoving (or coordinate)
momentum ${\bf k}$ remains constant in time,
the physical momentum gets red shifted as the Universe expands, see
also Section~\ref{sub:FLRW}. In what follows we set the present value of the
scale factor equal to 1 (in a spatially flat Universe this can always be done by
rescaling the coordinates ${\bf x}$):
\[
    a_0 \equiv a(t_0) = 1 \; ;
\]
then ${\bf k}$ is the {\it present} physical momentum and
$2\pi/k$ is the present physical wavelength, which is also called the comoving
wavelength.

Properties of scalar perturbations are measured in various ways.
Perturbations of fairly large spatial scales (fairly low ${\bf k}$)
give rise to
CMB temperature anisotropy and polarization,
so we have very detailed knowledge of them.
Somewhat shorter wavelengths are studied by analysing
distributions of galaxies and quasars at present and in
relatively near past. There are several other methods, some of which
can probe even shorter wavelengths. There is good overall consistency
of the results obtained by different methods, so we have a pretty good
understanding of many aspects of the scalar perturbations.

The cosmic medium in our Universe has several components that
interact only gravitationally: baryons, photons, neutrinos and
dark matter. Hence, there may be and, in fact, there are
perturbations in each of these components. As we pointed out
in
Section~\ref{sec:dm}, electromagnetic interactions between
baryons, electrons and photons were strong
before recombination,
so to a reasonable approximation
these species made a single fluid, and it is appropriate to talk
about perturbations in this fluid. After recombination, baryons and
photons evolved independently.

By studying the scalar
perturbations,
 we have learned
a number of very important things. To appreciate what they are,
it is instructive to consider first the
baryon--electron--photon fluid before
recombination.

\subsection{Perturbations in the expanding Universe:
subhorizon and superhorizon regimes.}

 Perturbations in the baryon--photon  fluid
before recombination are nothing but sound waves.
It is instructive to compare the wavelength of a perturbation with
the horizon size. To this end, recall (see Section~\ref{sub:RD})
that the horizon size $l_{\rm H}(t)$ is the size of the largest region which
is causally connected by the time $t$, and that
\[
l_{\rm H} (t) \sim
H^{-1}(t) \sim t
\]
at radiation domination and later, see
Eq.~\eqref{mar23-15-7}. The latter relation, however, holds
{\it under the assumption that the hot epoch was the first one in cosmology},
i.e., that the radiation domination started right after the
Big Bang. This assumption is in the heart of what can be called
hot Big Bang theory. We will find that this assumption in fact
is {\it not valid} for our Universe; we are going to see this ad absurdum,
so let us stick to the hot Big Bang theory for the time being.

\begin{figure}[htb!]
\begin{center}
\includegraphics[width=0.7\textwidth]{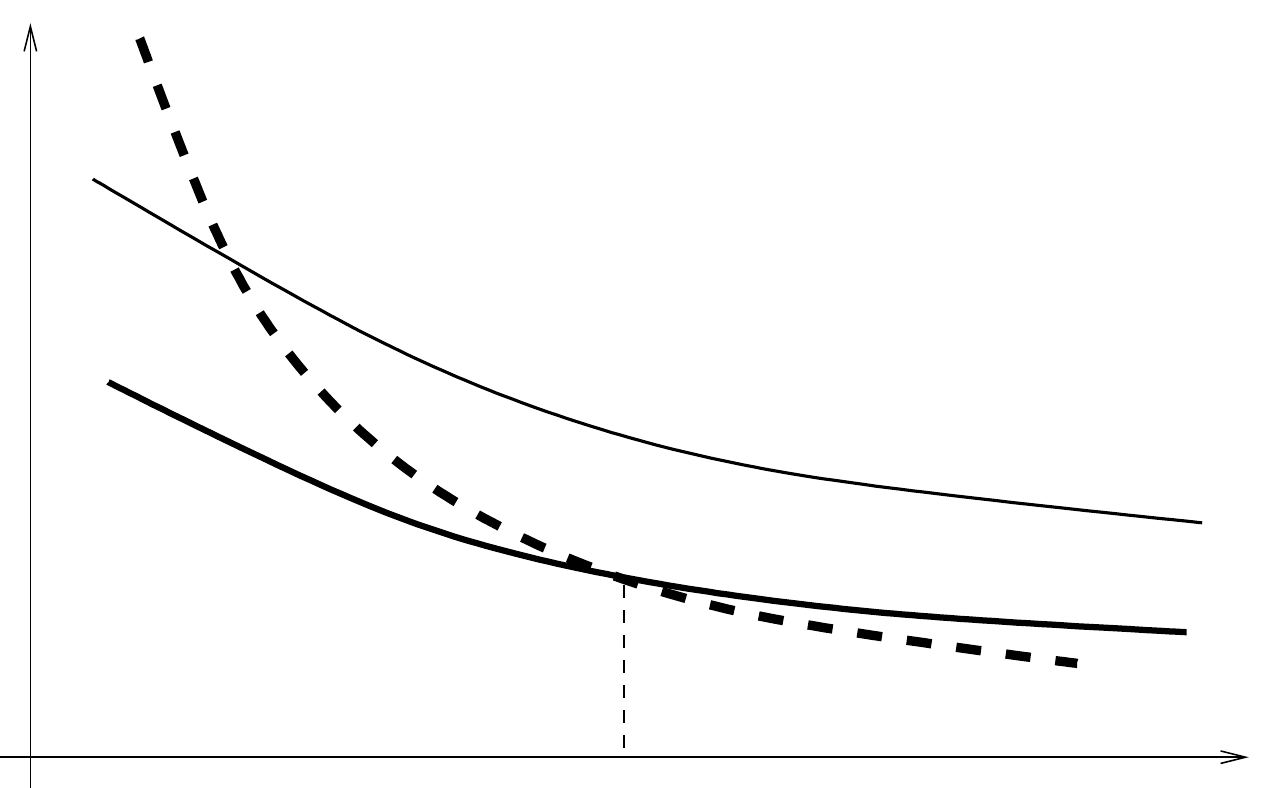}
\begin{picture}(10,10)(0,0)
{
\put(0,0){$t$}
\put(-290,80){$q_2(t)$}
\put(-100,85){$q_1(t) = k_1/a(t)$}
\put(-265,170){$H(t)$}
\put(-170,0){$t_\times$}
}
\end{picture}
\end{center}
\caption{Physical momenta $q(t)=k/a(t)$
(solid lines, $k_2 < k_1$)
and Hubble parameter (dashed line) at radiation-
and matter-dominated epochs. Here $t_\times$ is the horizon entry time.
 \label{length-hubble}
 }
\end{figure}
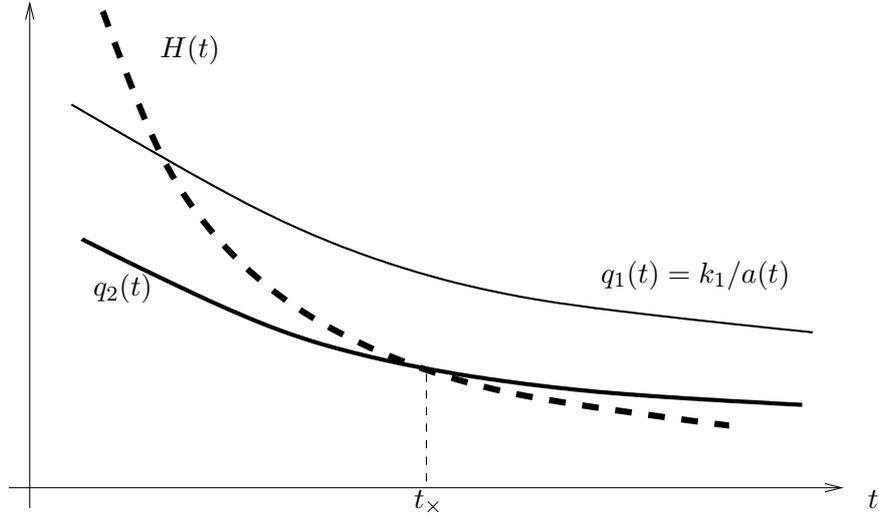

Unlike the horizon size, the physical wavelength of a perturbation
grows more slowly. As an example, at radiation domination
\[
\lambda (t) = \frac{2 \pi a(t)}{k} \propto \sqrt{t} \; ,
\]
while at matter domination $\lambda(t) \propto t^{2/3}$.
For obvious reasons, the modes with $\lambda(t) \ll H^{-1}(t)$
and  $\lambda(t) \gg H^{-1}(t)$ are called subhorizon and superhorizon
at time $t$, respectively. We are able to study the modes which are
subhorizon {\it today}; longer modes are homogeneous throughout the
visible Universe and are not observed as perturbations.
However, {\it the wavelengths which are subhorizon today
were superhorizon at some earlier epoch}. In other words, the
physical momentum $k/a(t)$ was smaller than $H(t)$ at early times;
at time $t_\times$ such that
\[
q(t_\times) \equiv \frac{k}{a(t_\times)} = H(t_\times) \; ,
\]
the mode entered the horizon, and after that evolved in the
subhorizon regime $k/a(t) \gg H(t)$, see Fig.~\ref{length-hubble}.
It is straightforward to see that for all cosmologically interesting wavelengths,
horizon crossing occurs much later than 1~s after the Big Bang, i.e.,
at the time we are confident about. So, there is no guesswork at
this point.

\vspace{0.3cm}
\noindent
{\it Question.} Estimate the temperature at which a perturbation
of comoving size 10~kpc entered the horizon.
\vspace{0.3cm}

\begin{figure}[b!]
\centering
\includegraphics[width=0.4\textwidth,angle=-90]{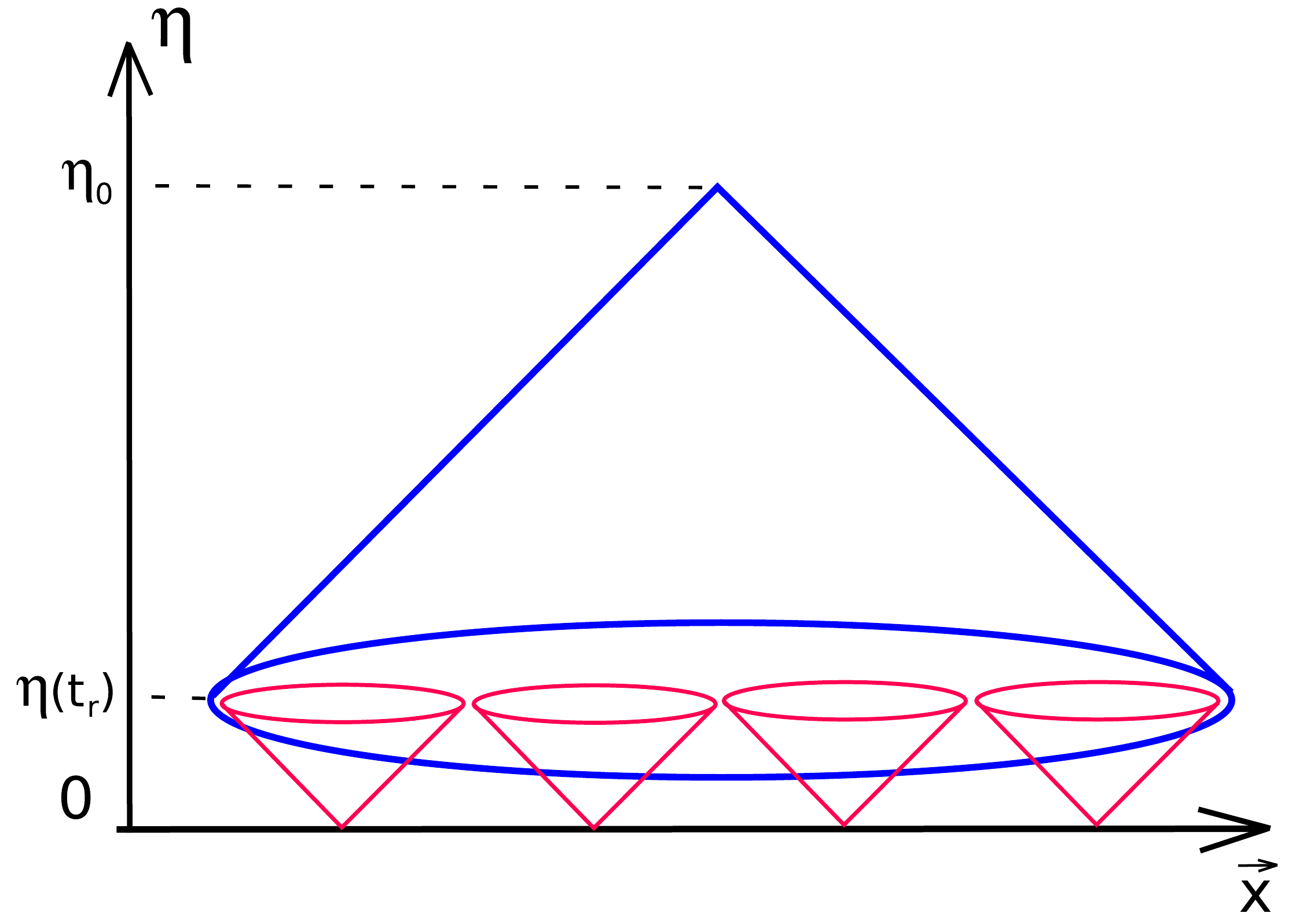}
\caption{Causal structure of space--time in the hot Big Bang theory.
Here $\rm t_{\rm r}$ is the conformal time at recombination.
 \label{noinfl-horizon}
 }
\end{figure}

Another way to look at the superhorizon--subhorizon behaviour
of perturbations is to introduce a new time
coordinate (cf. Eq.~\eqref{sep13-11-6}),
\be
\eta = \int_0^t \frac{{\rm d}t'}{a(t')} \; .
\label{mar29-15-10}
\ee
Note that this integral converges at the lower limit in the hot
Big Bang theory. In terms of this time coordinate, the FLRW metric
\eqref{FRW} reads
\[
{\rm d}s^2 = a^2 (\eta) ({\rm d}\eta^2 - {\rm d}{\bf x}^2) \; .
\]
In  coordinates $(\eta, {\bf x})$, the light cones
${\rm d}s=0$ are the same as in Minkowski space, and $\eta$ is the
coordinate size of the horizon, see Fig.~\ref{noinfl-horizon}.


Every mode of perturbation has a
time-independent coordinate wavelength $2\pi/k$, and
at small $\eta$ it is in superhorizon regime, $2\pi/k \gg \eta$,
and after horizon crossing at time $\eta_\times = \eta(t_\times)$
it becomes subhorizon.

\subsection{Hot epoch was not the first}

\begin{figure}[htb!]
\begin{center}
\includegraphics[width=0.7\textwidth]{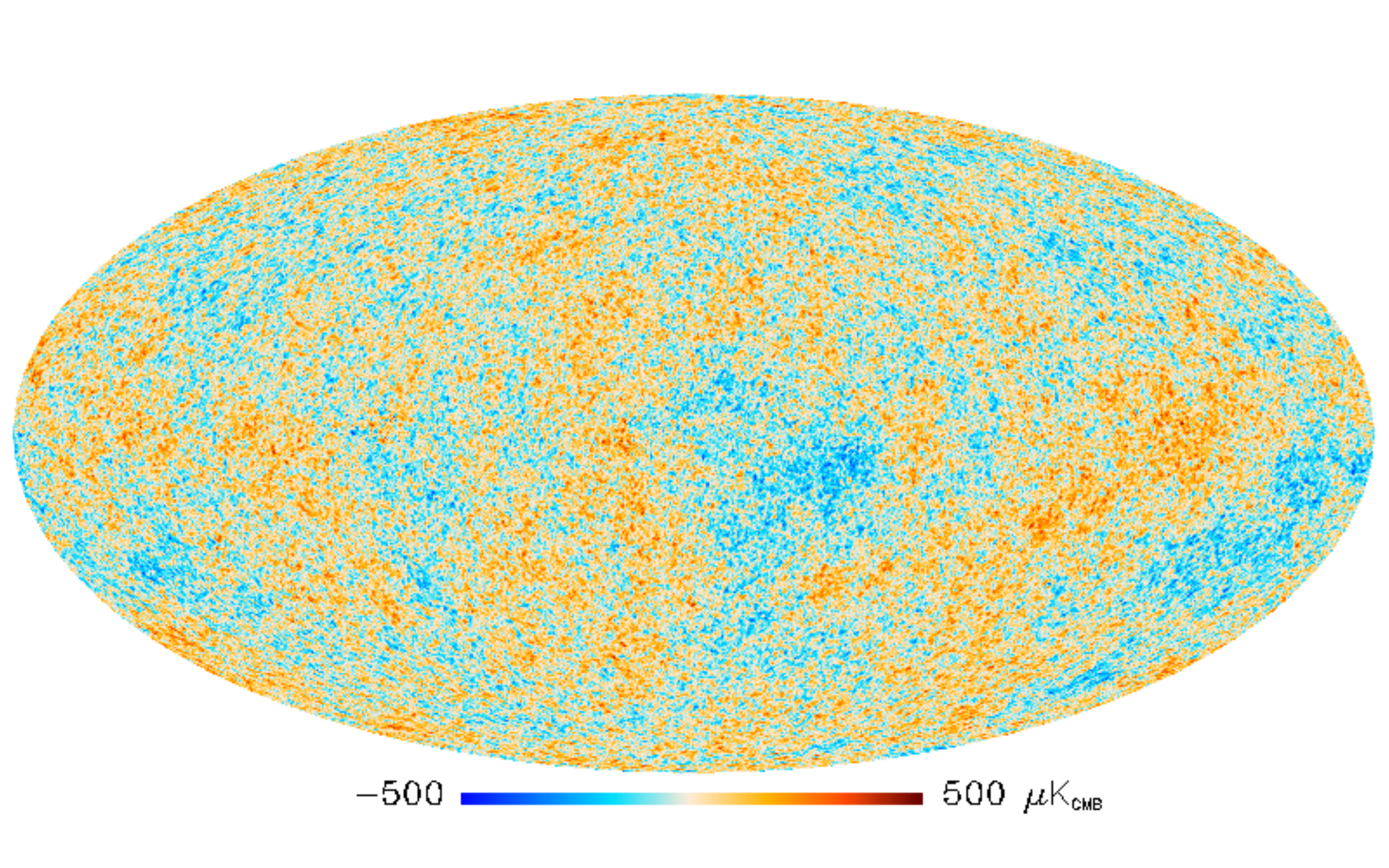}
\end{center}
\caption{\label{Planck-sky}  CMB sky as seen by Planck
 }
\end{figure}

One immediately observes that this picture falsifies the
hot Big Bang theory. Indeed, we see the horizon
at recombination $l_{\rm H}(t_{\rm rec})$ at an angle
$\Delta \theta \approx 2^\circ$, as schematically shown in
Fig.~\ref{noinfl-horizon}. By causality,  at recombination
there should be
no perturbations of larger wavelengths, as
any perturbation can be generated within the causal light cone
only. In other words, CMB temperature must be isotropic when
averaged over angular scales exceeding $2^\circ$; there should be no
cold or warm spots of angular size larger than  $2^\circ$.
Now, CMB provides us with the
photographic picture shown in Fig.~\ref{Planck-sky}.
It is seen by the naked eye that there are cold and warm regions
whose angular size much exceeds $2^\circ$; in fact, there are
perturbations of all angular sizes up to those comparable
to the entire sky. We come to an important conclusion:
{\it the scalar perturbations were built in at the very beginning
of the hot epoch. The hot epoch was not the first, it was preceded
by some other epoch, and the cosmological perturbations were
generated then.}

\vspace{0.3cm}
\noindent
{\it Question.} Assuming (erroneously) that there is
no dark energy, and that recombination occurred deep in the
matter-dominated epoch, estimate the angular scale
of the horizon at recombination.
\vspace{0.3cm}

Another manifestation of the fact that the scalar perturbations were
there already at the beginning of the hot epoch is the
existence of peaks in the angular spectrum of CMB temperature.
In general, perturbations in the baryon--photon medium before recombination
are acoustic waves,
\be
\delta({\bf k}, t)=\delta ({\bf k})
\e^{{\rm i}{\bf kx}}\cos\left[\int_0^t~v_{\rm s}\frac{k}{a(t')}{\rm d}t'+\psi_{\bf k}
\right] \; ,
\label{mar29-15-6}
\ee
where $v_{\rm s}$ is sound speed, $\delta ({\bf k})$ is time-independent amplitude and
$\psi_{\bf k}$ is time-independent phase.
This expression is valid, however, in the subhorizon regime only, i.e.,
at late times.
The two solutions in superhorizon regime at radiation domination
are
\begin{subequations}
\label{mar29-15-5}
\begin{align}
\delta (t) &= \mbox{const} \; ,
\label{mar29-15-5a} \\
\delta(t) &= \frac{\mbox{const}}{t^{3/2}} \; .
\label{mar29-15-5b}
\end{align}
\end{subequations}
Were the perturbations generated in a causal way at radiation domination,
they would be always subhorizon. In that case the solutions~\eqref{mar29-15-5}
would be irrelevant, and there would be no reason for a particular
choice of phase $\psi_{\bf k}$ in Eq.~\eqref{mar29-15-6}. One would rather
expect that $\psi_{\bf k}$ is a random function of ${\bf k}$.
This is indeed the case for specific mechanisms of the generation
of density perturbations at the hot epoch~\cite{Strings-CMB-plots}.

On the other hand, if the perturbations existed at the very beginning
of the hot epoch, they were superhorizon at sufficiently early times,
and were described by the solutions~\eqref{mar29-15-5}. The consistency
of the whole cosmology requires that the amplitude of perturbations was
small at the beginning of the hot stage. The solution~\eqref{mar29-15-5b}
rapidly decays away, and towards the horizon entry the perturbation
is in constant mode~\eqref{mar29-15-5a}. So, the initial condition
for the further evolution is unique modulo amplitude $\delta ({\bf k})$,
and hence the phase $\psi({\bf k})$ is uniquely determined. For modes that
enter the horizon at radiation domination
this phase is equal to zero and,
after entering the horizon, the modes oscillate as follows:
\[
\delta({\bf k},t)=\delta ({\bf k})
\e^{{\rm i}{\bf kx}}\cos\left[\int_0^t~v_{\rm s}\frac{k}{a(t')}{\rm d}t'
\right] \; .
\]
At recombination, the perturbation is
\be
\delta({\bf k},t_{\rm r})=\delta ({\bf k})
\e^{{\rm i}{\bf kx}}\cos(k r_{\rm s}) \; ,
\label{may30-5}
\ee
where
\[
r_{\rm s}=\int_0^{t_{\rm rec}}v_{\rm s}\frac{{\rm d}t'}{a(t')}
\]
 is the comoving size of the sound horizon at recombination,
while its physical size equals $a(t_{\rm rec})r_{\rm s}$. So, we see that
the
density perturbation of the baryon--photon plasma at recombination
{\it oscillates as a function of wavenumber $k$}.
The period of this oscillation is determined by $r_{\rm s}$, which is
a straightforwardly calculable quantity.

So, if the perturbations existed already at the beginning of
the hot stage, they show the oscillatory behaviour in momentum
at the recombination epoch. This translates into an oscillatory pattern
of the CMB temperature angular spectrum.
Omitting details, the
fluctuation of the
CMB temperature is partially due to
the density perturbation in the
baryon--photon medium at recombination. Namely,
the temperature fluctuation
of photons coming from the direction
${\bf n}$ in the sky is, roughly speaking,
\[
\delta T({\bf n}) \propto \delta_\gamma ({\bf x_n}, \eta_{\rm rec})
+ \delta T_{\rm smooth} ({\bf n}) \; ,
\]
where $  T_{\rm smooth} ({\bf n})$ corresponds to the non-oscillatory
part of the CMB angular spectrum and
\[
{\bf x_n}=-{\bf n} (\eta_0 - \eta_{\rm rec}) \; .
\]
 Here $(\eta_0 - \eta_{\rm rec})$
is the coordinate distance to the sphere of photon last scattering,
and ${\bf x_n}$ is the coordinate of the place where the photons
coming from the direction ${\bf n}$ scatter the last time.
 The quantity $  T_{\rm smooth} ({\bf n})$ originates from
the gravitational
potential generated by the dark matter perturbation;
dark matter has zero pressure
at all times, so there are no sound waves in this component,
and there are no
oscillations at recombination as a function of momentum.

\begin{figure}[htb!]
\begin{center}
\includegraphics[width=0.8\textwidth]{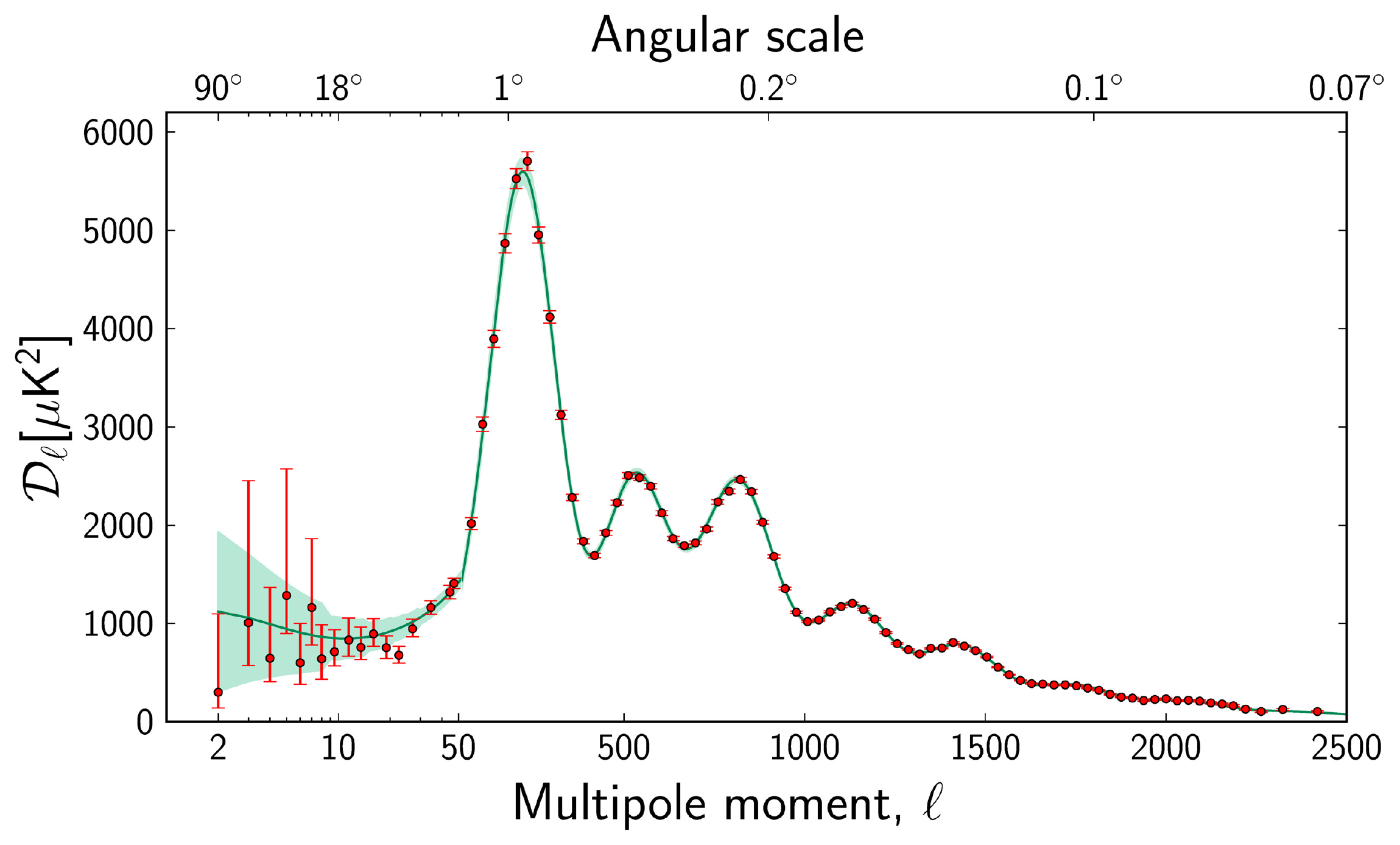}
\caption{The angular spectrum of the CMB temperature
anisotropy~\cite{Ade:2013sjv}. The quantity on the vertical axis  is
$D_l$ defined in (\ref{defdl}). Note the
unconventional scale on the horizontal axis, aimed at showing both
small-$l$ region (large angular scales) and
large-$l$ region.
\label{anisotropyspectrum}}
\end{center}
\end{figure}

One expands
the temperature variation on the celestial sphere
in spherical harmonics:
\[
\delta T({\bf n})=\sum_{lm}a_{lm}Y_{lm}(\theta,\phi).
\]
The multipole number $l$ characterizes the temperature fluctuations
at the angular scale $\Delta \theta = \pi/l$. The sound waves of momentum
$k$ are seen roughly at an angle $\Delta \theta = \Delta x/(\eta_0 - \eta_{\rm rec})$,
where $\Delta x = \pi/k$ is the coordinate half-wavelength.
 Hence, there is the correspondence
\[
  l \longleftrightarrow k (\eta_0 - \eta_{\rm rec}) \; .
\]
Oscillations in momenta in Eq.~\eqref{may30-5} thus translate into
oscillations in $l$, and these are indeed observed, see
Fig.~\ref{anisotropyspectrum}.

To understand what is shown in Fig.~\ref{anisotropyspectrum}, we note that
all
observations today support the hypothesis that
$a_{lm}$ {\it are independent  Gaussian random variables}.
For a hypothetical ensemble of Universes like ours, the
average values
of products of the coefficients $a_{lm}$ would obey
\be
\langle a_{lm} a^*_{l^\prime m^\prime} \rangle = C_l \delta_{l l^\prime}
\delta_{m m^\prime}.
\label{may29-6}
\ee
This gives the expression for
the
temperature fluctuation:
\[
 \langle[\delta T({\bf n})]^2\rangle
=\sum_l\frac{2l+1}{4\pi}C_l\approx
\int~ \frac{{\rm d}l}{l} {\cal D}_l \; ,
\]
where
\be
{\cal D}_l = \frac{l(l+1)}{2\pi} C_l \; .
\label{defdl}
\ee
Of course, one cannot measure the ensemble average
\eqref{may29-6}.
The definition of $C_l$ used in experiments is
\[
C_l = \frac{1}{2l+1} \sum_{m=-l}^l |a_{lm}|^2 \; ,
\]
where $a_{lm}$ are measured quantities.
Since we have only one Universe, this is generically different from
the ensemble average~\eqref{may29-6}: for given $l$, there are only
$2l+1$ measurements, and the intrinsic statistical uncertainty---cosmic variance---is of order $(2l+1)^{-1/2}$. It is this uncertainty,
rather than experimental error, that is shown in Fig.~\ref{anisotropyspectrum}.

We conclude that
the facts that the CMB angular spectrum has oscillatory behaviour
and that there are sizeable temperature fluctuations at $l<50$
(angular scale greater than the angular size of $2^\circ$ of the horizon
at recombination)
unambiguously tell us that the density perturbations were indeed
superhorizon at the hot cosmological stage. The hot epoch has to be
preceded by some other epoch---the epoch of the generation of
perturbations.

\subsection{Primordial scalar perturbations}

There are several things which we already know about the
primordial density perturbations. By `primordial' we mean
the perturbations deep in the superhorizon regime at the
radiation-domination epoch. As we already know, perturbations are
time-independent in this regime, see Eq.~\eqref{mar29-15-5a}.
They set the initial conditions
for further evolution, and this evolution is well understood, at
least in the linear regime.
Hence, using observational data, one is able to measure the properties
of primordial perturbations.
Of course, since the properties we know of
are established by observations, they are valid
within certain error bars. Conversely, deviations from the results
listed below, if observed, would be extremely interesting.

First, density perturbations are {\it adiabatic}.
This means that there are perturbations in the energy density,
but {\it not in composition}. More precisely, the baryon-to-entropy
ratio and dark matter-to-entropy ratio are constant in space,
\be
\delta \left(\frac{n_{\rm B}}{s} \right)= \mbox{const}
\; , \;\;\;\;\;\; \delta \left(\frac{n_{\rm DM}}{s}\right) = \mbox{const} \; .
\label{sep22-11-10}
\ee
This is consistent with the generation of the baryon asymmetry
and dark matter at the hot cosmological epoch: in that case,
all particles were at thermal equilibrium early at the hot
epoch, the temperature completely characterized the whole
cosmic medium at that time and as long as physics
behind the baryon asymmetry and dark matter generation is the
same everywhere in the Universe, the baryon and dark matter
abundances (relative to the entropy density) are necessarily
the same everywhere. In principle,
there may exist {\it entropy} (another term is isocurvature) perturbations,
such that at the early hot epoch energy density (dominated
by relativistic matter) was homogeneous, while the composition was
not. This would give initial conditions for the evolution of
density perturbations, which would be entirely different from
those characteristic of the adiabatic perturbations. As a result,
the angular spectrum of the CMB temperature anisotropy would be entirely
different. No admixture of the
entropy perturbations has been detected so far, but it is worth
emphasizing that even a small admixture will show that many popular
mechanisms for generating dark matter and/or baryon asymmetry
have nothing to do with reality. One will have to think, instead,
that the baryon asymmetry and/or dark matter were generated
before the beginning of the hot stage. A notable example is the
axion misalignment mechanism discussed in Section~\ref{subs:axions}:
in a latent sense, the axion dark matter exists from the very beginning
in that case, and perturbations in the axion field
$\delta \theta_0 ({\bf x})$
(which may be generated
together with the adiabatic perturbations) would show up
as entropy perturbations in dark matter.

Second, the primordial density perturbations are {\it Gaussian random fields}.
Gaussianity means that the three-point and all odd correlation functions
vanish, while the four-point function and all higher order
even correlation functions are expressed through the two-point
function via  Wick's theorem:
\begin{eqnarray}
\langle \delta({\bf k}_1)  \delta({\bf k}_2)  \delta({\bf k}_3) \rangle &=& 0,
\nonumber \\
\langle \delta({\bf k}_1)  \delta({\bf k}_2)  \delta({\bf k}_3)
\delta ({\bf k}_4)\rangle & =
& \langle \delta({\bf k}_1)  \delta({\bf k}_2)
\rangle \cdot \langle \delta({\bf k}_3)  \delta({\bf k}_4)   \rangle
\nonumber\\
&~&{\ +}~
{ \mbox{permutations~of~momenta}} \; .
\nonumber
\end{eqnarray}
We note that this
 property is characteristic of {\it vacuum fluctuations of
non-interacting (linear) quantum fields.} Hence, it is quite likely that
the density perturbations originate from the enhanced vacuum
fluctuations of non-interacting or weakly interacting quantum field(s).
The free quantum field has the general form
\[
\phi ({\bf x}, t) =
\int {\rm d}^3k {\rm e}^{-{\rm i}{\bf kx}} \left(f^{(+)}_{\bf k}(t) a^\dagger_{\bf k}
+  {\rm e}^{{\rm i}{\bf kx}} f^{(-)}_{\bf k}(t) a_{\bf k} \right) \; ,
\]
where $a_{\bf k}^\dagger$ and $a_{\bf k}$ are creation and annihilation operators.
For the field in Minkowski space--time, one has
$f^{(\pm)}_{\bf k} (t) = \mbox{e}^{\pm {\rm i}\omega_k t}$, while enhancement, e.g.,
due to the evolution in time-dependent background, means that
$f^{(\pm)}_{\bf k}$ are large. But, in any case,
Wick's theorem is valid, provided that the state of the system is vacuum,
$a_{\bf k} |0\rangle = 0$.

We note in passing that
{\it non-Gaussianity} is an important topic of current research.
It would show up as a deviation from Wick's theorem.
As an example, the three-point function (bispectrum) may be non-vanishing,
\[\langle
\delta ({\bf k}_1)
\delta ( {\bf k}_2) \delta ({\bf k}_3)
\rangle = \delta( {\bf k}_1 + {\bf k}_2 + {\bf k}_3) ~
G(k_i^2 ;~ {\bf k}_1 {\bf k}_2 ;~
{\bf k}_1  {\bf k}_3) \neq 0 \; .
\]
The functional dependence
 of  $G(k_i^2 ;~ {\bf k}_1 {\bf k}_2 ;~
{\bf k}_1  {\bf k}_3)$ on its arguments
is different in different models of generation of primordial
perturbations, so this shape is a
potential discriminator.
In some models the bispectrum vanishes, e.g., due to symmetries.
In that case the trispectrum
(connected four-point function) may be measurable instead.
It is worth emphasizing that
non-Gaussianity
has not been detected yet.

Another important property is that the
primordial power spectrum of density perturbations
{\it is nearly, but not exactly, flat}.
For a
homogeneous and anisotropic
Gaussian random field, the power spectrum completely determines its
only characteristic, the two-point function.
A convenient definition
is
\be
\langle  \delta  ({\bf k})
      \delta       ({\bf k}^\prime)  \rangle
= \frac{1}{4 \pi k^3} {\cal P} (k) \delta({\bf k} + {\bf k}^\prime) \; .
\label{sep24-11-1}
\ee
The power spectrum ${\cal P} (k) $ defined in this way
determines the fluctuation in a logarithmic
interval of momenta,
\[
\langle \delta^2 ({\bf x}) \rangle
= \int_0^\infty ~\frac{{\rm d}k}{k} ~{\cal P}(k) \; .
\]
By definition, the flat spectrum is such that
${\cal P}$ is independent of $k$. In this case all spatial
scales
are alike; no scale is enhanced with respect to another.
It is worth noting that
the flat spectrum was conjectured by Harrison~\cite{Harrison},
Zeldovich~\cite{Zeldovich} and Peebles and Yu~\cite{Peebles:1970ag}
at the beginning of the
1970s, long before realistic mechanisms of the generation
of density perturbations have been proposed.

In view of the approximate flatness, a natural parametrization is
\be
{\cal P}(k) = A_{\rm s} \left(\frac{k}{k_*} \right)^{n_{\rm s} -1} \; ,
\label{sep24-11-6}
\ee
where
$A_{\rm s} $ is the amplitude, $n_{\rm s}-1 $ is the tilt and $k_*$ is a
fiducial momentum, chosen at one's convenience.
The flat spectrum in this parametrization has $n_{\rm s}=1$.
This is inconsistent with the
cosmological data, which give~\cite{Planck:2015xua}
\[
n_{\rm s} = 0.968 \pm 0.06 \; .
\]
This quantifies what we mean by a nearly, but not exactly flat, power spectrum.

\subsection{Inflation or not?}

\begin{figure}[htb!]
\begin{center}
\includegraphics[width=0.4\textwidth,angle=-90]{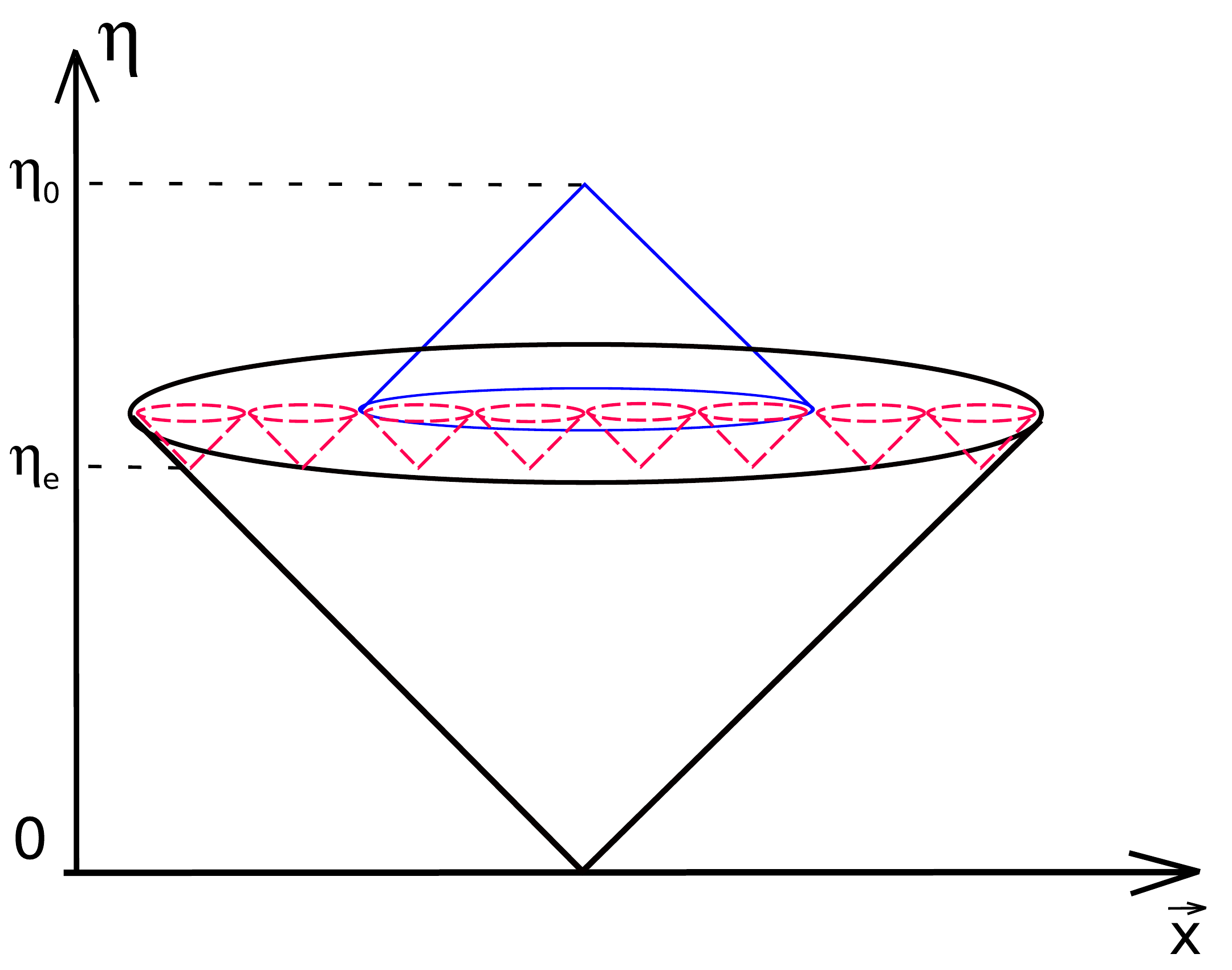}
\end{center}
\caption{Causal structure of space--time in the real Universe
 \label{infl-horizon}
 }
\end{figure}

The pre-hot epoch must be long in terms of the
time variable $\eta$ introduced in Eq.~\eqref{mar29-15-10}.
What we would like to have is that a large part of the Universe
(e.g., the entire visible part) be causally connected towards
the end of that epoch, see Fig.~\ref{infl-horizon}.
A long duration in $\eta$ does not necessarily mean a long duration in
physical time $t$; in fact, the physical duration of the
pre-hot epoch may be tiny.

An excellent hypothesis on the pre-hot stage is inflation,
the epoch of nearly exponential expansion~\cite{inflation-1,inflation-2,inflation-3,inflation-4,inflation-5,inflation-6},
\[
  a(t) = \mbox{e}^{\int H {\rm d}t} \; , \;\;\;\;\;\;\;\;\;
H \approx \mbox{const} \; .
\]
Inflation makes the whole visible Universe, and likely a much greater
region of space, causally connected at very early times.
The horizon size at inflation
is at least
\[
l_{\rm H} (t) = a(t) \int_{t_{\rm i}}^t \frac{{\rm d}t'}{a(t')} = H^{-1} \mbox{e}^{H (t - t_{\rm i})} \; ,
\]
where $t_{\rm i}$ is the time inflation begins, and
we set $H=\mbox{const}$ for illustrational purposes.
This size is huge for $t - t_{\rm i} \gg H^{-1}$, as desired.

\vspace{0.3cm}
\noindent
{\it Question.} Assuming that at inflation
$H \ll M_{\rm Pl}$, show that if the duration of inflation
$\Delta t$
is larger than $100 H^{-1}$, the whole visible Universe
is causally connected by the end of inflation.
What is  $100 H^{-1}$ in seconds for $H=10^{15}$~GeV?
Using the time variable
$\eta$, show that the causal structure of space--time
in inflationary theory with $\Delta t > 100 H^{-1}$
is the one shown in
Fig.~\ref{infl-horizon}.
\vspace{0.3cm}

\begin{figure}[htb!]
\begin{center}
\includegraphics[width=0.6\textwidth]{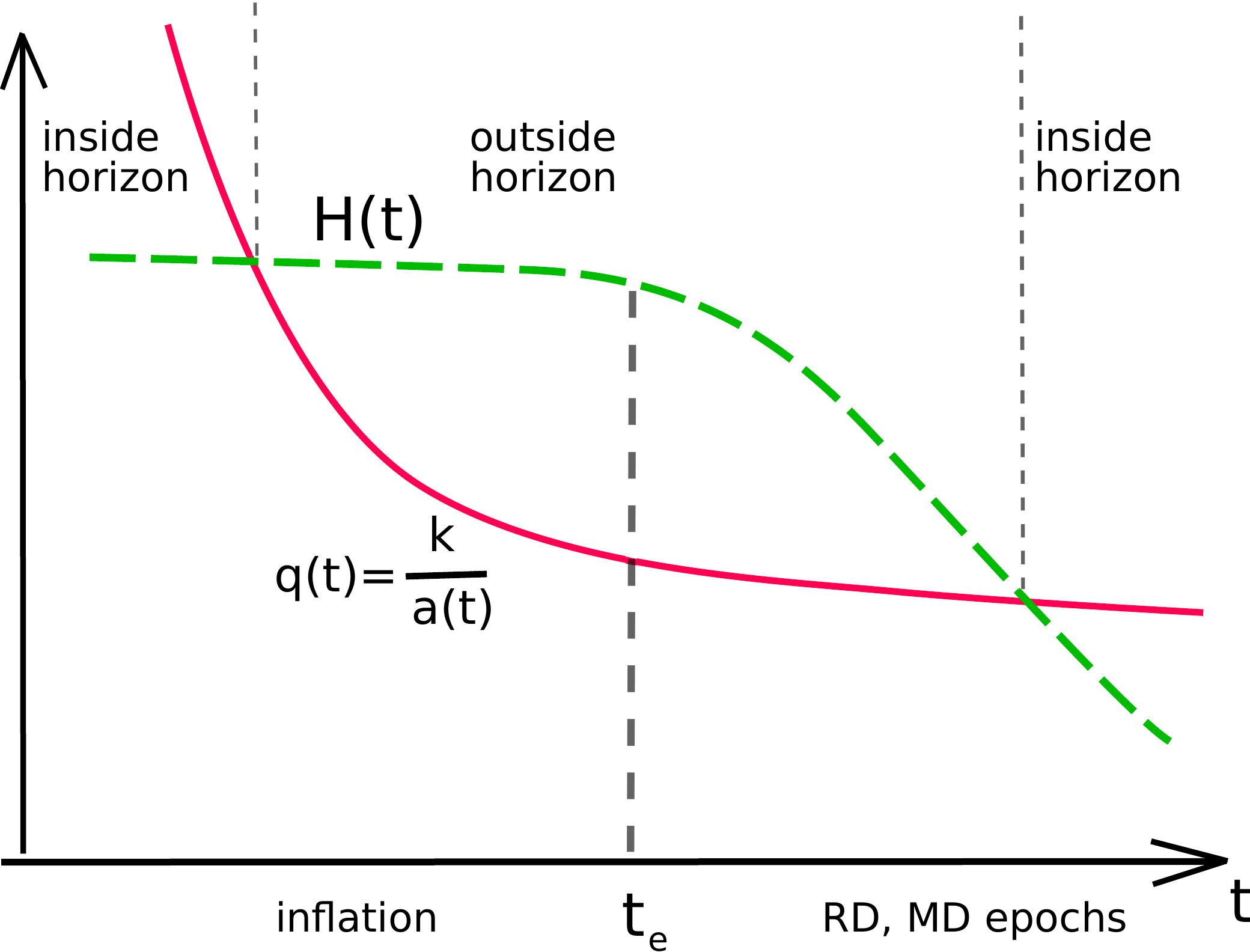}
\caption{Physical momentum and Hubble parameter at inflation and
later: $t_{\rm e}$ is the time of the inflation end
\label{infl3}}
\end{center}
\end{figure}

From the viewpoint of perturbations,
the physical momentum $q(t) = k/a(t)$ decreases
(gets red shifted) at inflation, while the Hubble parameter
stays almost constant. So, every mode is first subhorizon
($q(t) \gg H(t)$) and
later superhorizon ($q(t) \ll H(t)$). This situation is
opposite to what happens at radiation and matter domination, see
Fig.~\ref{infl3};
this is precisely the prerequisite for generating the density
perturbations. In fact, inflation does generate primordial
density
perturbations~\cite{infl-perturbations-1,infl-perturbations-2,infl-perturbations-3,infl-perturbations-4,infl-perturbations-5} whose properties are consistent
with everything we know about them.
Indeed, at the inflationary
epoch, fluctuations of all light fields get enhanced greatly
due to the fast expansion of the Universe.
This is true, in particular, for the field that dominates the energy density
at inflation, called an inflaton. Enhanced vacuum fluctuations of the
inflaton are nothing but
perturbations in the energy density at the inflationary epoch,
which are reprocessed into perturbations in the hot medium after
the end of inflation. The inflaton field is very weakly coupled,
so the non-Gaussianity in the primordial scalar perturbations is
very small~\cite{Maldacena:2002vr}. In fact, it is so small that
its detection is problematic even in the distant future.

The approximate
flatness of the primordial power spectrum
in inflationary theory  is explained by
the symmetry of the de~Sitter space--time,
which is the space--time of constant Hubble rate,
\[
{\rm d}s^2 = {\rm d}t^2 - \mbox{e}^{2Ht} {\rm d}{\bf x}^{2} \; , \;\;\;\;\;\;\;\;
H = \mbox{const} \; .
\]
This metric is invariant under spatial
dilatations
supplemented by time translations,
\[
{\bf x} \to \lambda {\bf x} \;, \;\;\;
t \to t - \frac{1}{2H} \log \lambda \; .
\]
Therefore, all spatial scales are alike, which is also a
defining property of the flat power spectrum.
At inflation, $H$ is almost constant in time and the
de~Sitter symmetry is an approximate symmetry. For this reason,
inflation automatically generates a nearly flat power spectrum.

\begin{figure}[htb!]
\begin{center}
\includegraphics[width=0.8\textwidth]{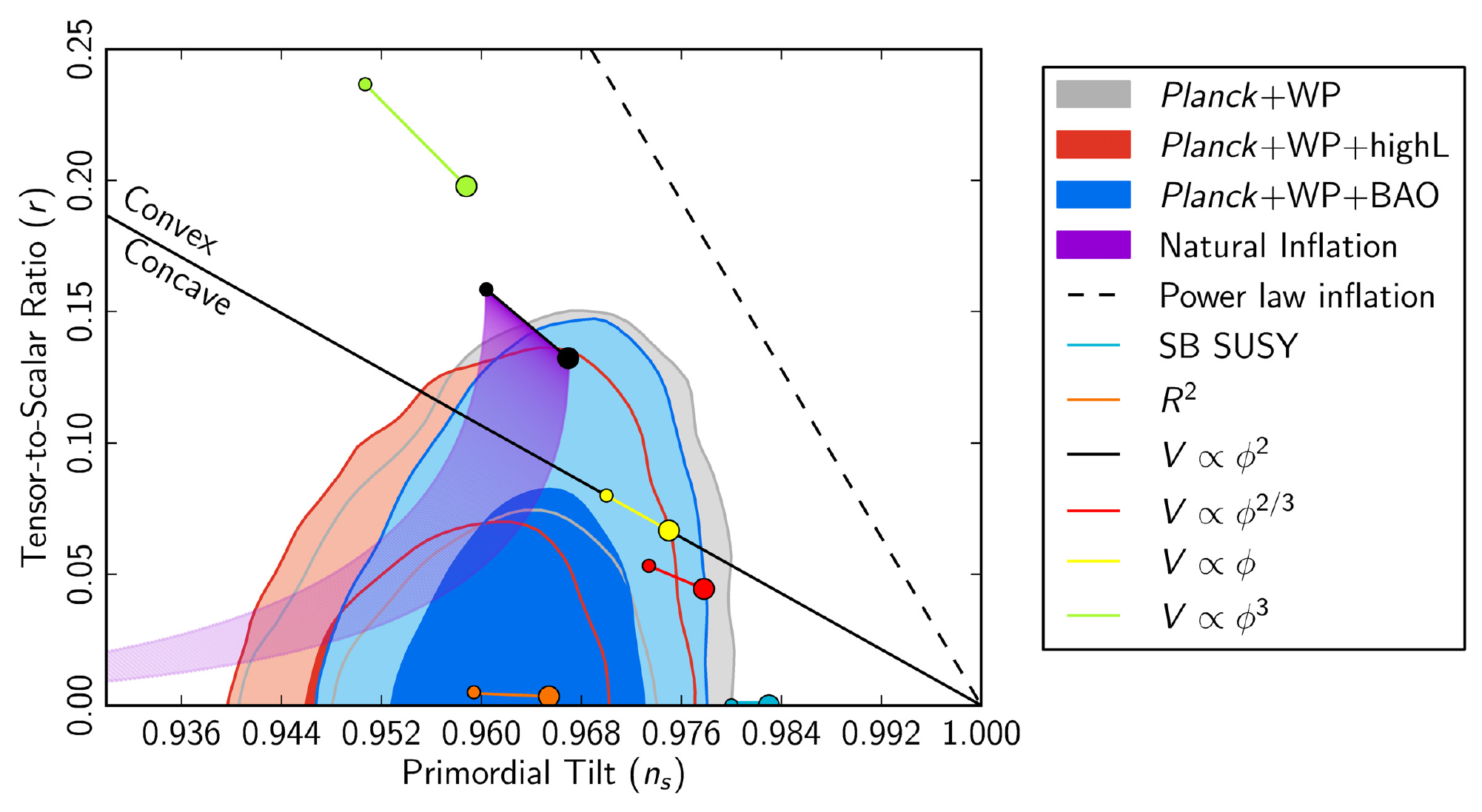}
\caption{Allowed
regions  (at 68\% and 95\% confidence levels) in the plane $(n_{\rm s}, r)$, where $n_{\rm s}$ is the scalar spectral index and $r$ is the tensor-to-scalar
ratio~\cite{Planck:2013jfk}. The right lower corner (the point $(1.0, 0.0)$) is the Harrison--Zeldovich point (flat scalar spectrum, no
tensor modes). Intervals show predictions of popular
inflationary models.
\label{ns-r}}
\end{center}
\end{figure}

The distinguishing property of inflation is  {\it the
generation of tensor modes (primordial gravity waves)}
of sizeable amplitude and nearly flat power spectrum.
The gravity waves are thus a smoking gun for inflation.
The reason for their generation at inflation is that
the exponential expansion of the Universe  enhances vacuum fluctuations
of all fields, including the gravitational field itself.
Particularly interesting are gravity waves whose present
wavelengths are huge, 100~Mpc and larger. Many inflationary
models predict  their amplitudes to be very large,
of order $10^{-6}$ or so. Shorter gravity waves are generated
too, but their amplitudes decay after horizon entry at radiation
domination, and today they have much smaller amplitudes making them
inaccessible  to gravity wave detectors like LIGO or VIRGO,
pulsar timing arrays etc.
A conventional characteristic of the amplitude of
primordial gravity waves is the tensor-to-scalar ratio
\[
r = \frac{{\cal P}_{\rm T}}{{\cal P}} \; ,
\]
where ${\cal P}$ is the scalar power
spectrum defined in Eq.~\eqref{sep24-11-1} and
${\cal P}_{\rm T}$ is the tensor power spectrum defined in a similar
way, but for transverse traceless metric perturbations $h_{ij}$.
The result of the search for effects of the tensor modes on
CMB temperature anisotropy is shown in Fig.~\ref{ns-r}.
This search has already ruled out some of the popular inflationary models.

All the above referred to the simplest, single-field inflationary models.
In models with more than one relevant field, the situation may be different.
In particular, sizeable non-Gaussianity may be generated, while the
amplitude of tensor perturbations
may be very low. So, it would be rather difficult
to rule out the inflationary scenario as a whole.

Inflation is not the only hypothesis proposed so far.
One option is the bouncing Universe scenario, which assumes
that the cosmological evolution begins from contraction,
then the contracting stage terminates
at some moment of time (bounce) and is followed by expansion.
A version is the cycling Universe scenario with many cycles
of contraction--bounce--expansion. See  reviews by~Lehners
and~Brandenberger
in Ref.~\cite{pert-rev-1,pert-rev-2,pert-rev-3,pert-rev-4,pert-rev-5}.
Another scenario is that the Universe starts out from a
nearly flat and static state with nearly vanishing energy density.
Then the energy density increases and, according to the Friedmann
equation, the expansion
speeds up.
This goes under the name of the Genesis scenario~\cite{Creminelli:2010ba}.
Theoretical realizations of these scenarios are more difficult
than inflation, but they are not impossible, as became clear recently.

The generation of the density perturbations
is less automatic in scenarios alternative to inflation.
Similarly to inflationary theory,
the flatness of the scalar power spectrum is likely to be
due to some symmetry.
One candidate symmetry is conformal
invariance~\cite{conf1,conf2-1,conf2-2,conf2-3}. The point is that
the conformal group includes dilatations,
 $x^\mu \to \lambda x^\mu$. This property indicates that
the theory possesses no
scale and has a good chance
for producing the flat spectrum. A
model building along this direction has begun rather recently~\cite{conf2-1,conf2-2,conf2-3}.

\subsection{Hunt continues}

Until now, only very basic facts about the primordial
cosmological perturbations have been
observationally established. Even though very suggestive, these facts
by themselves are not sufficient to unambiguously
figure out what was the Universe at the pre-hot epoch of
its evolution. New properties of cosmological perturbations
will hopefully be discovered in the future and  shed more light on this
pre-hot epoch. Let us discuss some of the potential observables.

\subsubsection{Tensor perturbations~=~relic gravity waves}

As we discussed, primordial tensor perturbations are predicted by
many inflationary models. On the other hand, there seems to be no way of
generating a nearly flat tensor power spectrum in alternatives to
inflation. In fact, most, if not all, alternative scenarios
predict unobservably small tensor amplitudes.
This is why we said that tensor perturbations are
a smoking gun for inflation.
Until recently, the most sensitive probe of the tensor perturbations
has been the CMB temperature anisotropy~\cite{Rubakov:1982df-1,Rubakov:1982df-2,Rubakov:1982df-3,Rubakov:1982df-4}.
However, the most promising tool
is the CMB polarization. The point is that a certain class of
polarization
patterns (called B-mode) is generated by tensor perturbations,
while scalar perturbations are unable to create
it~\cite{Kamionkowski:1996zd-1,Kamionkowski:1996zd-2}. Hence,
dedicated experiments aiming at
measuring the CMB polarization may well discover the tensor perturbations,
i.e., relic gravity waves. Needless to say, this would be a
profound discovery. To avoid confusion, let us note that the CMB
polarization has been already observed, but it belongs to another class of
patterns (so-called E-mode) and is consistent with the
existence of the
scalar perturbations only. The original claim of the BICEP-2
experiment~\cite{Ade:2014xna} to detect the B-mode generated by
primordial tensor perturbations was turned down~\cite{Ade:2015tva}:
the B-mode is there, but it is due to dust in our Galaxy.

\subsubsection{Non-Gaussianity.} As we pointed out already, non-Gaussianity
of density perturbations
is very small in the simplest inflationary models. Hence, its discovery
will signal that either inflation and inflationary generation of
density perturbations
occurred in a rather complicated way, or  an alternative
scenario was realized. Once the non-Gaussianity is discovered,
and its shape is revealed even with modest accuracy,
many concrete models will be ruled out, while at most a few
will get strong support.

\subsubsection{Statistical anisotropy.} In principle, the power spectrum
of density perturbations may depend on the direction of
momentum, e.g.,
\[
{\cal P}({\bf k}) = {\cal P}_0 (k) \left(1 +
w_{ij} (k) \frac{k_i k_j}{k^2}
+ \cdots \right),
\]
where $w_{ij}$ is a fundamental  tensor in our part
of the Universe (odd powers of $k_i$ would contradict
commutativity of the Gaussian random field $\delta ({\bf k})$,
see Eq.~\eqref{sep24-11-1}). Such a dependence would definitely
imply
that the Universe was anisotropic at the
pre-hot stage, when the primordial
perturbations were generated. This statistical anisotropy
is rather hard to obtain in inflationary models, though it is
possible in inflation with strong vector fields~\cite{soda-1,soda-2,soda-3}.
On the other hand, statistical anisotropy is
natural in some other scenarios, including conformal
models~\cite{mlvr+-1,mlvr+-2}.
The statistical anisotropy
would show up in correlators~\cite{aniso-1,aniso-2}
\[
\langle a_{ lm} a_{ l' m'} \rangle \;\;\;\;\;\; \mbox{with}~~{ l'\neq l}
~~
\mbox{and/or}~~{ m' \neq  m}.
\]
At the moment, the constraints~\cite{Kim:2013gka-1,Kim:2013gka-2} on statistical anisotropy
obtained by analysing the CMB data are getting into the region
which is interesting from the  viewpoint of some (though not many)
models of the pre-hot epoch.

\subsubsection{Admixture of entropy perturbations.}
As we explained above,
even a small admixture of entropy perturbations would force us to
abandon the most popular scenarios of the generation of baryon asymmetry
and/or dark matter, which assumed that it happened at the hot epoch.
Once the dark matter entropy mode is discovered,
the WIMP dark matter would no longer be well motivated,
while other, very weakly interacting dark matter candidates, like axions
or superheavy relics, would be preferred. This would redirect
 the experimental
search for dark matter.

\section{Conclusion}

It is by now commonplace that the two fields studying together
the most fundamental properties of matter and the Universe---particle physics and cosmology---are tightly interrelated.
The present situations in these fields have much in common too.
On the particle-physics side, the Standard Model has been completed
by the expected
discovery of the Higgs boson. On the other hand, relatively recently
a fairly unexpected discovery of neutrino oscillations was made,
which revolutionized our view on particles and their interactions.
There are grounds to hope for even more profound discoveries,
notably by the LHC experiments. While in the past there were
definite predictions of the Standard Model, which eventually were confirmed,
there are numerous
hypotheses concerning new physics, none of which is undoubtedly plausible.
On the cosmology side, the Standard Model of cosmology, $\Lambda$CDM,
has been shaped, again not without an unexpected and revolutionary
discovery, in this case
of the accelerated expansion of the Universe. We hope for further
profound discoveries in cosmology too. It may well be that we will soon
learn which is the dark matter particle; again, there is an entire zoo
of candidates, several of which are serious competitors.
The discoveries of new properties of cosmological perturbations
will hopefully
reveal the nature of the pre-hot epoch.
There is a clear best guess, inflation, but it is not excluded that
future observational data will point towards something else.

Neither in particle physics nor in cosmology are new discoveries guaranteed,
however. Nature may hide its secrets. Whether or not it does is
the biggest open issue in fundamental physics.

\section*{Acknowledgement}
This work is supported by the Russian Science Foundation grant
14-22-00161.

\end{document}